\documentclass[a4paper, fleqn, 9pt, onecolumn]{article}

\def\typenumber{0}

\def\typeoddity{0}

\usepackage{jmsn2026v2}

\newcommand{\volumenumber}{1}
\newcommand{\issuenumber}{3}
\newcommand{\manuscriptnumber}{03001}
\newcommand{\yearofpublication}{2026}
\newcommand{\submissiondate}{28 April 2026}
\newcommand{\acceptancedate}{22 June 2026}
\newcommand{\publishingdate}{20 July 2026}
\newcommand{\manuscripttitle}{Photoemission insights into lanthanide-based crystals}
\newcommand{\firstauthorname}{Usachov}
\newcommand{\doi}{10.64214/jmsn.0\volumenumber.\manuscriptnumber}
\newcommand{\authorlist}{
Dmitry\,Yu.\,Usachov,$^{1,2\,\href{https://orcid.org/0000-0003-0390-0007}{\includegraphics[height=2 mm, origin=c]{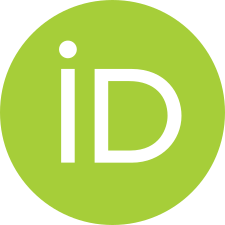}}}$
Georg\,Poelchen,$^{3\,\href{https://orcid.org/0000-0002-1844-2136}{\includegraphics[height=2 mm, origin=c]{orcid_logo_icon.png}}}$
Vasily\,S.\,Stolyarov,$^{1,2\,\href{https://orcid.org/0000-0002-5317-0818}{\includegraphics[height=2 mm, origin=c]{orcid_logo_icon.png}}}$
Kristin\,Kliemt,$^{4\,\href{https://orcid.org/0000-0001-7415-4158}{\includegraphics[height=2 mm, origin=c]{orcid_logo_icon.png}}}$
Cornelius\,Krellner,$^{4\,\href{https://orcid.org/0000-0002-0671-7729}{\includegraphics[height=2 mm, origin=c]{orcid_logo_icon.png}}}$
Denis\,V.\,Vyalikh$^{5-7\,\href{https://orcid.org/0000-0001-9053-7511}{\includegraphics[height=2 mm, origin=c]{orcid_logo_icon.png}}}$}
\newcommand{\affiliationlist}{
$^1$ St. Petersburg State University, St. Petersburg, 199034, Russia \\
$^2$ Moscow Institute of Physics and Technology (State University), Dolgoprudny, Russia \\
$^3$ Institut f\"ur Festk\"orper- und Materialphysik, Technische Universit\"at Dresden, Dresden, Germany \\
$^4$ Kristall- und Materiallabor, Physikalisches Institut, Goethe University Frankfurt, Frankfurt am Main, Germany \\
$^5$ Donostia International Physics Center (DIPC), 20018 Donostia/San Sebastian, Basque Country, Spain \\
$^6$ University of the Basque Country UPV/EHU, 20018 Donostia/San Sebastián, Spain \\
$^7$ IKERBASQUE, Basque Foundation for Science, 48013 Bilbao, Spain
}

\smallskip

\begin{document}

\thispagestyle{firstpagestyle}

\createtitleaffiliations

%

\medskip

\noindent
\lettrine{\textbf{\textsf{T}}}{}\textsf{\textbf{\small \,\,\,he interplay of strongly localized 4f electrons with itinerant spd-valence states gives rise to a wide range of correlated phenomena and properties that place lanthanide materials at the focus of considerable research efforts. Beyond the bulk, their surfaces are of particular interest, where the reduced coordination, a modified crystal electric field, broken inversion symmetry in combination with strong spin-orbit coupling, and the emergence of surface states and resonances considerably reshape 4f-driven electronic and magnetic properties. This, in turn, enables novel functionalities of particular relevance for low-dimensional systems and their applications. This review summarizes how advances in photoelectron spectroscopies, together with improved crystal growth, have enabled detailed insights into bulk and surface phenomena of lanthanide-based crystals. After a brief overview of key developments from the 1970s to the 1990s, we discuss recent progress, focusing on systematic studies by the authors and collaborators that form a coherent line of research. These include the unveiling of ${\boldsymbol k}$-resolved ${\boldsymbol f}$–$\boldsymbol{spd}$ hybridization, the formation and evolution with temperature of ${\boldsymbol f}$-derived Fermi surface in Kondo lattices, layer-dependent 4${\boldsymbol f}$ magnetic anisotropy, the emergence of ferromagnetically ordered surfaces in systems with non-magnetic bulk ground state. This coherent line of research addresses core questions in the physics of 4${\boldsymbol f}$ systems and opens opportunities for engineering novel lanthanide-based architectures, including heterostructures and supramolecular complexes with novel physical properties and functionalities.}}

\thispagestyle{firstpagestyle}

\bigskip

\medskip

\section*{Introduction}

Lanthanide ($Ln$) based materials occupy a unique position in condensed matter physics and materials science due to the coexistence and interaction of highly localized $4f$ electrons and more itinerant valence states. This interplay gives rise to a remarkable diversity of physical phenomena, including complex magnetism, heavy-fermion behavior, valence fluctuation, Kondo physics, \linebreak quantum criticality, and unconventional superconductivity \cite{Coleman2007, Trovarelli_PRL_2000, Stewart2001, Stewart2006, Steglich_1979, Varma_1976, Gschneidner2000RareEarths, Cotton1999LanthanideActinide}. Therefore, from a fundamental perspective, lanthanide systems provide model platforms for studying strong electronic correlations, competing interactions, and the interplay between localized and itinerant degrees of freedom.

At the same time, the distinctive electronic, magnetic, and optical properties of lanthanides provide the basis for a wide range of essential technological applications \cite{Behrsing2024RareEarths, Evans2020, Heffern2014, Amoroso2015, Mathieu2018, Cheng_2024, Li_2025}. These include permanent magnets, magnetic refrigeration, light-emitting devices, lasers and other optical technologies, biomedical applications such as contrast agents for magnetic resonance imaging, catalysis, and emerging quantum and spin-based technologies~\cite{Behrsing2024RareEarths, Evans2020, Heffern2014, Amoroso2015, Mathieu2018, Cheng_2024, Li_2025}.
As a result, lanthanide materials are of both strategic industrial importance and continuing scientific interest, motivating sustained research efforts aimed at investigating, understanding, and controlling their non-trivial and complex properties at a microscopic level.

A central aspect of lanthanide physics is the atomic-like nature of the $4f$ electrons, which are strongly localized and largely shielded from the chemical environment by outer electronic shells~\cite{Gschneidner2000RareEarths, Cotton1999LanthanideActinide}. While this localization stabilizes robust local magnetic moments and well-defined multiplet structures, the interaction of 4$f$ states with more itinerant 5$d$ and 6$s$ electrons plays a crucial role in mediating magnetic exchange, transport properties, and many-body phenomena. The resulting balance between localization and hybridization is highly sensitive to external parameters such as temperature, pressure, chemical substitution and doping, leading to rich and tunable phase diagrams~\cite{Coleman2007, Trovarelli_PRL_2000, Stewart2001, Stewart2006, Steglich_1979, Varma_1976}.

Surfaces and interfaces in lanthanide systems introduce an additional level of complexity on the one hand, but also provide new degrees of freedom for tuning and exploiting their physical properties. Reduced coordination, broken inversion symmetry, surface relaxation and reconstruction, as well as the emergence of surface electronic states and resonances, can substantially modify $4f$-driven properties, altering the electronic structure, magnetism, and correlation effects relative to the bulk properties. These surface-induced changes make surface-sensitive experimental techniques essential for a comprehensive understanding of lanthanide systems.

In this regard, synchrotron-based spectroscopic methods have played a central role in advancing the field. X-ray absorption spectroscopies (XAS), performed in total and partial electron yield (TEY and PEY) as well as fluorescence (TFY) modes at selected absorption edges of lanthanide atoms, together with photoemission spectroscopy (PES), including momentum-resolved (ARPES) and resonant variants, provide direct access to the electronic structure of lanthanide materials. The additional use of X-ray magnetic dichroism techniques employing circularly and/or linearly polarized light (XMCD and XMLD) offers detailed insight into magnetic-related phenomena, including element- and orbital-selective magnetic mo-\linebreak ments and anisotropies. Combined, these techniques provide a comprehensive understanding of both surface and bulk phenomena related to $4f$ physics. When applied to well-characterized single-crystalline samples and interpreted in conjunction with structural, magnetic, and transport measurements, in combination with theoretical modeling, these techniques offer a powerful framework for understanding and linking microscopic electronic structure to macroscopic physical properties.

In this review, we focus on experimental studies of lanthanide-based single crystals, with an emphasis on \linebreak synchrotron-based photoemission investigations \cite{Himpsel_1983, Woodruff_1983,  Hufner_2003, Hufner_2007, King_review_2021, Damascelli_review_2003, Sobota_review_2021}. We discuss both seminal early work on $4f$ systems that established key concepts in the field and more recent advances that highlight the continuing relevance of lanthanide materials for fundamental physics~\cite{Coleman2007, Trovarelli_PRL_2000, Stewart2001, Stewart2006} and applications~\cite{Behrsing2024RareEarths, Evans2020, Heffern2014, Amoroso2015, Mathieu2018, Cheng_2024, Li_2025}. A central theme of this review is the consistent emphasis on surface-related properties and phenomena in lanthanide-based crystals, a topic that, in comparison to bulk studies, has been less explicitly addressed and has received relatively limited attention within the community. We argue that only combined and systematic surface–bulk investigations can provide a more complete understanding of these systems and open clearer pathways toward their practical implementation. We note that certain lanthanide-based crystals are intrinsically composed of well-defined atomic-scale layers that can be regarded as natural building blocks. Once their electronic and magnetic properties are well characterized and understood, these ready-made “Lego-like” modules may serve as structural units for the design and realization of novel low-dimensional or heterostructured systems with diverse functional applications. Finally, our overview also instructively illustrates how a coherent and systematic research line, developed by our team and collaborators, enables us to address key phenomena and questions in the field.

\begin{figure}[t]
    \centering
    \includegraphics[width=0.99\linewidth]{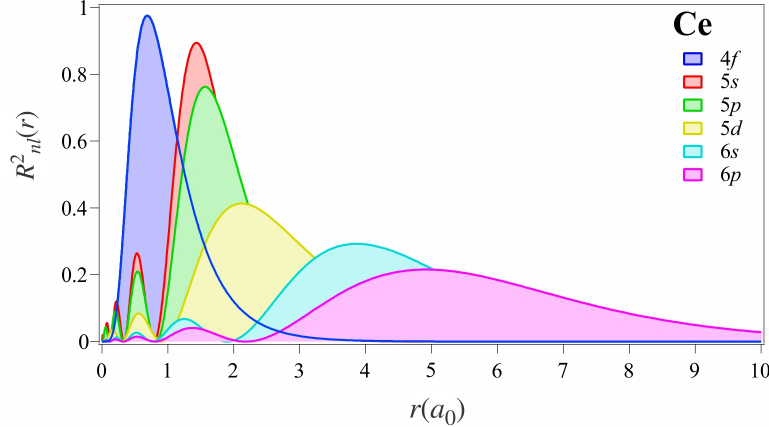}
    \caption{
    Representative radial probability distributions $R_{n\ell}^{2}(r)$ of selected atomic orbitals for a Ce atom are shown as a function of the radial distance $r$ (expressed in units of the Bohr radius $a_0$), with Ce serving as a representative example~\cite{Goldschmidt_1978, Huang_2010_Ln_Book}. The strongly localized $4f$ states (light blue) are confined to small radii, largely inside the closed $5s$ and $5p$ shells, reflecting their atomic-like character and weak contribution to chemical bonding.  In contrast, the $5d$ and $6sp$ orbitals exhibit significantly broader radial distributions extending to larger distances, indicative of their more delocalized nature and enhanced spatial overlap.  This clear separation in radial extent highlights the distinct roles of the electronic states in lanthanides, with localized $4f$ electrons governing magnetic properties, while the more extended $5d$ and $6sp$ states mediate bonding and itinerant electronic behavior.}
    \label{fig:figure_1}
\end{figure}

\pagestyle{otherpagestyle}

\section*{Properties of 4$\boldsymbol{f}$ electrons}

The diverse physical properties of lanthanide materials originate from the strongly localized character of their $4f$ electrons~\cite{Goldschmidt_1978, Huang_2010_Ln_Book}. The $4f$ orbitals are radially confined well inside the filled $5s$ and $5p$ shells, which efficiently screen them from the surrounding crystal environment and limit their spatial extension. As a result, the spatial overlap of $4f$ wave functions with those of neighboring atoms remains small, leading to weak hybridization with ligand orbitals and itinerant $spd$ states. Consequently, the $4f$ electrons largely preserve an atomic-like character even in the solid state. This localized behavior is further enhanced by the increasing effective nuclear charge across the lanthanide series.

The radial probability distributions of the relevant atomic orbitals clearly illustrate this separation of electronic character (figure~\ref{fig:figure_1}). The $4f$ states exhibit pronounced radial maxima at small $r$ and remain strongly localized near the ionic core, resulting in extremely weak intersite wave-function overlap, whereas the more extended $5d$ and $6s$ states penetrate deeply into the interatomic region and therefore show substantial overlap with orbitals on neighboring lattice sites. This separation provides the basis for the distinct roles of the electronic states in lanthanide systems: localized $4f$ electrons govern magnetic properties and multiplet excitations, while the more extended $5d$ and $6s$ states mediate chemical bonding and itinerant electronic behavior.

The limited screening efficiency of the $4f$ electrons plays a central role in the phenomenon known as the lanthanide contraction. As the atomic number increases across the lanthanide series, additional electrons occupy the $4f$ shell; however, their poor shielding only partly compensates the increasing nuclear charge. As a consequence, the effective nuclear attraction acting on the outer electrons progressively increases, leading to a gradual reduction of atomic and ionic radii along the series. The lanthanide contraction has important implications for coordination chemistry, complex stability, and systematic trends in chemical reactivity.

The pronounced localization of $4f$ electrons can already be understood within an effective single-electron description in a spherically symmetric potential. In addition to the central ionic potential, the large orbital angular momentum $l =3$ introduces a strong centrifugal term,
\begin{eqnarray}
\nonumber
V_{\mathrm{cf}}(r) = \frac{\hbar^2 l (l+1)}{2 m r^2},
\end{eqnarray}
which confines the $4f$ electrons to a narrow radial region close to the ionic core. This centrifugal barrier strongly suppresses radial extension and interatomic overlap, helping to preserve the largely atomic-like character of the $4f$ shell in solids.

The strong localization of the $4f$ states implies that their electronic structure is dominated by intra-atomic interactions. Hund’s-rule coupling and spin-orbit interaction set the principal energy scales and give rise to well-separated multiplet states together with robust local magnetic moments. Crystal-electric-field (CEF) effects imposed by the surrounding lattice act as a comparatively weak perturbation. This hierarchy of interactions, in which Hund’s coupling and spin-orbit interaction dominate over CEF splitting, is a defining characteristic of lanthanide-based materials. This situation contrasts with transition-metal ($3d$) systems, where CEF splitting often sets the main energy scale. In lanthanides, CEF energies are typically smaller than Hund’s-rule and spin-orbit interactions, and the atomic multiplet character of the $4f$ shell is therefore largely retained, allowing orbital degrees of freedom to remain active. At the same time, CEF effects often play a decisive role in determining the magnetic ground state and anisotropy.

The stability of specific $4f$ configurations with vanishing total orbital angular momentum, notably $4f^{0}$, $4f^{7}$, and $4f^{14}$, reflects the predominantly atomic character of the $4f$ shell. This underlies the special roles of Ce, Eu, and Yb within the lanthanide series. In Ce, the proximity in energy of the $4f^{1}$ and closed-shell $4f^{0}$ configurations promotes strong hybridization and pronounced valence fluctuations. A closely related situation occurs in Yb, where valence fluctuations involve adding or removing a single $4f$ hole relative to the filled $4f^{14}$ shell.

In Eu systems, the half-filled $4f^{7}$ configuration is \linebreak strongly stabilized by Hund’s-rule exchange, favoring a divalent Eu$^{2+}$ state with a large local spin moment ($J=S=7/2$), as observed, for example, in antiferromagnetic EuRh$_2$Si$_2$~\cite{Chick14, Hoeppner2013}. At the same time, the trivalent $4f^{6}$ configuration with a non-magnetic $J=0$ ground state can be realized in suitable chemical environments, as in EuCo$_2$Si$_2$, while intermediate-valence behavior is also known, for instance in EuIr$_2$Si$_2$~\cite{Seiro2011, Stockert2020}. The close energetic proximity of magnetically active and non-magnetic configurations makes Eu-based materials highly sensitive to external parameters, so that even moderate changes in temperature, pressure, or chemical composition can lead to complex phase behavior. Notably, finite $4f$–$spd$ hybridization may coexist with pronounced magnetic properties, as illustrated by EuRh$_2$Si$_2$~\cite{Hoeppner2013} and EuIr$_2$Si$_2$~\cite{Susanne2019, EIS_2020}.

The differing stability of the $4f$ configurations discussed above has direct consequences for the magnetic response of lanthanide systems. In Eu, the Hund’s-rule half-filled $4f^{7}$ shell supports a large and robust local moment. In contrast, the near-degeneracy of competing $4f$ configurations in Ce and Yb promotes $f$-$spd$ hybridization, resulting in enhanced Kondo screening, partial or complete quenching of local magnetic moments, and pronounced many-body renormalization.

Owing to the predominantly localized character of the $4f$ moments, magnetic ordering in lanthanide compounds typically develops via indirect exchange interactions mediated by itinerant electrons, most notably the Ruderman-Kittel-Kasuya-Yosida mechanism. As a result, ordered $4f$ moments often remain close to their free-ion values, whereas the ordering temperatures are comparatively low and sensitive to details of the electronic structure. The same localization is reflected in spectroscopic experiments, where $4f$ states appear as sharp, only weakly dispersive features with pronounced multiplet structure, providing a characteristic fingerprint of their atomic-like nature.

\section*{Photoemission of lanthanides: early \\ developments}

In the late 1970s and early 1980s, the conceptual basis for photoemission spectroscopy of lanthanide metals was established \cite{Gerken_1983, Gerken_PRL_1981, Gerken_1985}. High-resolution X-ray photoemission spectroscopy and early synchrotron experiments \linebreak demonstrated the highly localized nature of the 4$f$ states and their pronounced atomic-like multiplet structure, requiring a many-body description beyond a band-like picture. A key theoretical advance was the implementation of intermediate coupling calculations combined with the fractional parentage scheme, enabling a quantitative description of 4$f$ photoemission spectra across the lanthanide series and clarifying the breakdown of pure LS coupling, particularly for the heavy lanthanides \cite{Gerken_1983}.

At the same time, surface-sensitive photoemission measurements revealed distinct bulk and surface 4$f$ contributions separated in binding energy by substantial surface-induced shifts (so called surface-core-level energy shifts), an insight made possible by the comprehensive calculation of 4$f$ multiplet intensities in the early 1980s. These results established surface effects as intrinsic to lanthanide electronic structure and laid the groundwork for later single-crystal and ARPES studies.

Between the mid-1980s and the early 2000s, the group led by G.~Kaindl, together with its collaborators, established a coherent research line and produced key results in the field of photoemission studies of lanthanide-based materials.\cite{Laubschat_PhysScr_1990,Laubschat_PRL_1990, Weschke_PRB_1991, Laubschat_SS_1992, Weschke_PRL_1992,Fedorov_PRL_1993,Navas_PRB_1993,Comment_CeRh3_1994,Reply_CeRh3_1994,Kaindl_1995,
Weschke_JESRP_1995,Weschke_PRL_1996,Weschke_Films_1998,Schuessler_PRB_1999, Starke_Tb_1994,Fedorov_1994,Arenholz_1995,Laan_1996, Starke_1997,Laan_1997,Arenholz_1998,Laan_1999_scaling,Hu_1999,Richter_2000,Starke_2001} Their research program combined methodological consistency with conceptual clarity and played a key role in shaping the modern understanding of surface and bulk electronic structure in $4f$ systems. The studies addressed elemental lanthanide metals, Ce-based intermetallic compounds, and epitaxial rare-earth films, employing a broad set of complementary spectroscopic techniques, including core-level photoemission, valence-band photoemission, resonant photoemission, inverse photoemission, and magnetic dichroism. \cite{Laubschat_PhysScr_1990,Laubschat_PRL_1990, Weschke_PRB_1991, Laubschat_SS_1992,Weschke_PRL_1992,Fedorov_PRL_1993,Navas_PRB_1993,Comment_CeRh3_1994,Reply_CeRh3_1994,Kaindl_1995,
Weschke_JESRP_1995,Weschke_PRL_1996,Weschke_Films_1998,Schuessler_PRB_1999, Starke_Tb_1994,Fedorov_1994,Arenholz_1995,Laan_1996, Starke_1997,Laan_1997,Arenholz_1998,Laan_1999_scaling,Hu_1999,Richter_2000,Starke_2001}

A unifying theme throughout this decade of work is the recognition that surface effects strongly influence the photoemission response of lanthanide materials. Early investigations of Ce metal and $\alpha$-like Ce compounds such as CeIr$_2$, CePd$_3$, and CeRh$_3$ conclusively demonstrated that near-surface layers of Ce exhibit enhanced $4f$ localization even when the bulk electronic structure reveals a strong $f$-$d$ hybridization.\cite{Laubschat_PRL_1990,Weschke_PRB_1991,Laubschat_SS_1992} These findings helped to resolve long-standing inconsistencies between information derived from surface-sensitive photoemission spectra and bulk-sensitive probes, showing that many apparent discrepancies originate from surface-dominated contributions rather than intrinsic bulk physics. Our own experience with polycrystalline samples demonstrated that the shape of the $4f$ PE spectrum depends sensitively on the surface preparation, in particular on the cleaning procedure using a diamond file under UHV conditions. More importantly, pronounced variations in the lineshape were observed when spectra were taken from different locations on the same sample surface. Under these conditions, it was difficult to assess the reliability of such results and to disentangle intrinsic electronic properties from surface-related effects.

\begin{figure}[t]
\centering
\includegraphics[width=7cm]{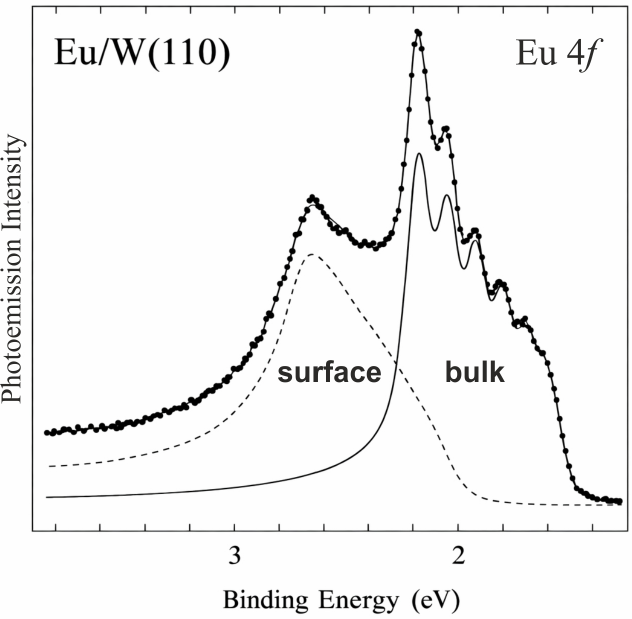}
\caption{High-resolution photoemission spectrum of a thin Eu(110) film prepared on W(110), taken with He II radiation at 20 K. Individual $^{7}F_J$ multiplet components of the bulk Eu spectrum are clearly resolved. The contribution from the topmost surface layer (dashed subspectrum) is less well resolved and shifted toward higher binding energies due to the surface core-level shift. Data taken from Ref.~[\citeonline{Kaindl_1995}]. }
\label{fig:Eu_Kaindl}
\end{figure}

By systematically exploiting variable probing depth and resonant enhancement at the $Ln$ $4d\rightarrow4f$ and $3d\rightarrow4f$ thresholds, the group succeeded in separating surface and bulk spectral components with unprecedented clarity~\cite{Laubschat_PhysScr_1990}. This approach established resonant photoemission as a key methodological tool for lanthanide systems, enabling nearly background-free access to $4f$-derived spectral weight and revealing strong multiplet- and channel-selective enhancement effects.

The importance of surface-related phenomena was further demonstrated by a series of accurate studies of $4f$ surface core-level shifts in heavy lanthanide metals.\cite{Navas_PRB_1993, Kaindl_1995} In parallel, the first highly resolved $4f$ photoemission (PE) spectra of lanthanides became available. Figure~\ref{fig:Eu_Kaindl} presents a high-resolution PE spectrum of the (110) face of a monocrystalline bcc Eu metal film, in which the individual $^{7}F_J$ multiplet components of the bulk spectrum are clearly resolved. In contrast, the spectral contribution from the topmost surface layer (dashed subspectrum) is less well resolved and shifted toward higher binding energies, reflecting the presence of a surface core-level shift.\cite{Kaindl_1995}

These investigations demonstrated that, for well- defined close-packed surfaces, the magnitude of the surface core-level shift is systematic and significantly smaller than values reported earlier for polycrystalline samples. The results showed that exaggerated shifts often originate from low-coordination sites such as steps or defects, emphasizing the necessity of controlled surface preparation. Thermochemical models were shown to account consistently for the observed shifts, providing a physically transparent interpretation in terms of modified local bonding and screening.

A second major contribution from the group of G. Kaindl is the identification and detailed characterization of a localized, $d$-like surface state that appears to be a universal feature of close-packed lanthanide metal surfaces.\cite{Weschke_JESRP_1995} This surface state, typically located near the Fermi level and partially occupied, emerged as a sensitive probe of surface magnetism. Angle-resolved photoemission and inverse photoemission studies of Gd(0001) revealed a temperature-dependent exchange splitting inconsistent with simple spin-mixing scenarios, instead supporting a Stoner-like description with enhanced magnetic robustness at the surface.\cite{Weschke_PRL_1996} Related studies of oxygen-induced surface oxides on Gd demonstrated that magnetically split surface-related states can persist even above the bulk Curie temperature, highlighting the stability of surface and interface magnetism in lanthanide systems.\cite{Schuessler_PRB_1999}

The decade also saw important advances in the combined use of photoemission and inverse photoemission to probe both occupied and unoccupied electronic states.\cite{Hufner_2003, Hufner_2007, Woodruff_1983} Inverse photoemission experiments on La metal revealed a clear surface shift of the unoccupied $4f$ electron-addition state, quantitatively consistent with thermochemical expectations.\cite{Fedorov_PRL_1993} This result provided compelling evidence that surface-induced energy shifts affect both sides of the Fermi level and are a general property of lanthanide surfaces.

A particularly instructive aspect of this research program is the explicit discussion of the boundary between localized and itinerant descriptions of $4f$ electrons. In strongly hybridized CeRh$_3$, combined photoemission and inverse photoemission measurements indicated that the unoccupied spectral weight is more consistently described by band-structure calculations than by a single-impurity Anderson model~\cite{Weschke_PRL_1992}. This interpretation was critically examined in a Comment and Reply communications~\cite{Comment_CeRh3_1994, Reply_CeRh3_1994} published in Physical Review Letters. The Comment questioned the band-like interpretation on the basis of possible energy-referencing ambiguities,\cite{Comment_CeRh3_1994} while the Reply clarified that the discrepancy corresponds to a rigid shift of the entire inverse-photoemission spectrum and does not invalidate the underlying physical conclusion.\cite{Reply_CeRh3_1994} Taken together, this discussion emphasized the sensitivity of BIS (bremsstrahlung isochromat spectroscopy) measurements to energy calibration, while leaving unresolved the broader question of whether the $4f$ states in CeRh$_3$ are predominantly bandlike or best described in terms of localized states.

In the later part of the 1990s, the focus expanded toward epitaxial lanthanide films, enabling band-resolved photoemission studies under well-controlled structural conditions.\cite{Weschke_Films_1998} Investigations of Ce and La films revealed similar valence-band dispersions in agreement with density-functional-theory (DFT) band-structure calculations. By combining resonant photoemission with deliberate suppression of surface contributions, it became possible to extract bulk $4f$ features directly and to quantify surface-to-bulk energy shifts with high reliability.

\begin{figure}[t]
    \centering
    \includegraphics[width=7cm]{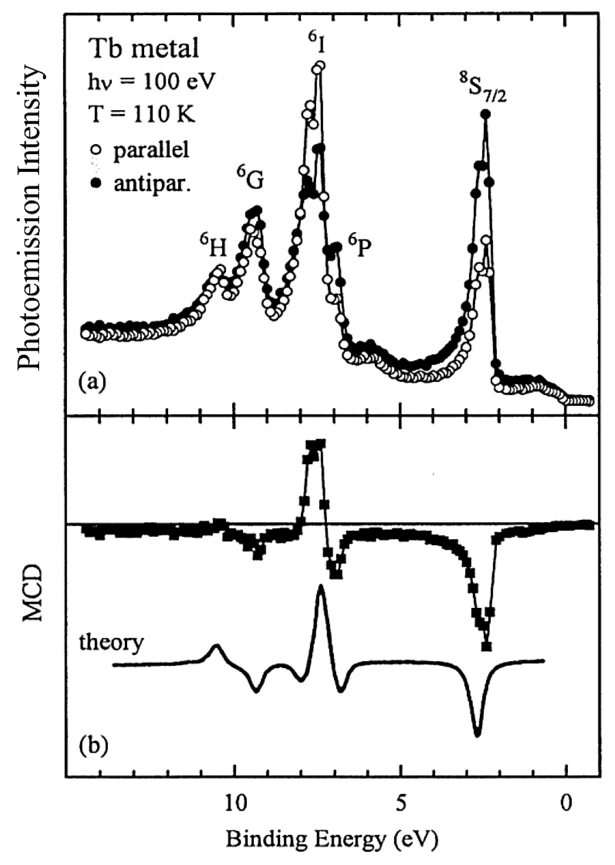}
    \caption{High-resolution Tb $4f$ photoemission spectra measured at $h\nu=100$~eV from a remanently magnetized Tb(0001)/W(110) film (150~\AA, $T=110$~K). Open (filled) symbols correspond to nearly parallel (antiparallel) orientations of photon spin and sample magnetization. (b) MCD signal obtained as the intensity difference; the solid line shows the calculated atomic-multiplet MCD spectrum. Data taken from Ref.~[\citeonline{Arenholz_1995}].}
    \label{fig:Tb_Kaindl}
\end{figure}

Let us focus more precisely on the achievements of Kaindl's group and their partners in the study of magnetic properties of lanthanide materials using \linebreak photoemission-based techniques, including magnetic \linebreak dichroism. Through a series of pioneering investigations, this work established magnetic dichroism in photoemission as a quantitative and element-specific probe of localized $4f$ moments. It was demonstrated that, despite the largely atomic character of $4f$ electrons, key magnetic properties, such as magnetic order, exchange interactions, and anisotropy, are directly reflected in photoemission spectra.

Early experiments provided the first systematic observation of magnetic circular dichroism (MCD) in the $4f$ photoemission of lanthanides, demonstrating that dichroic asymmetries directly reflect the magnetic polarization of the $4f$ shell, as exemplified for Tb~\cite{Starke_Tb_1994,starke_2000_book}. These findings also opened direct perspectives for magnetic imaging applications based on photoemission electron microscopy. In contrast to techniques such as neutron scattering, resonant inelastic X-ray scattering, or bulk magnetometry, which predominantly probe collective excitations or spatially averaged magnetic responses, magnetic circular \linebreak dichroism in photoemission offers an element- and shell-specific view of magnetism on a local, atomic scale~\cite{starke_2000_book}.

Subsequent work revealed that the exchange splitting of unoccupied $5d$-derived states, observed in inverse photoemission spectra of Gd(0001), originates from the local, intra-atomic exchange interaction between the localized $4f$ moments and the itinerant $5d$ states of the same lanthanide ion\cite{Fedorov_1994}, providing one of the first spectroscopic determinations of the $4f$–$5d$ exchange field in lanthanide metals.

A major conceptual advance was achieved by demonstrating that the complex multiplet structure of core-level photoemission spectra from lanthanide materials, exemplified by Gd, Tb (see figure~\ref{fig:Tb_Kaindl}), and Dy, contains detailed information on magnetism. By resolving and analyzing individual $4f$ photoemission multiplet components, it was shown that different final states exhibit distinct magnetic circular dichroism responses, both in magnitude and sign, reflecting the underlying magnetic moment configuration~\cite{Arenholz_1995}. In the case of Tb, in particular, the well-defined $^8S_{7/2}$ $4f^7$ final state  component of the $4f$ photoemission spectrum at a binding energy of about 2–3 eV shows a pronounced dichroic behavior, with the emission intensity strongly enhanced for one orientation of the sample magnetization relative to the photon spin and nearly quenched for the opposite orientation. These results established multiplet-resolved $4f$ photoemission as a highly sensitive, element-specific probe of local magnetic order and laid the basis for magnetic imaging and magnetometry based on photoemission techniques.

\begin{figure}[t]
    \centering
    \includegraphics[width=0.94\linewidth]{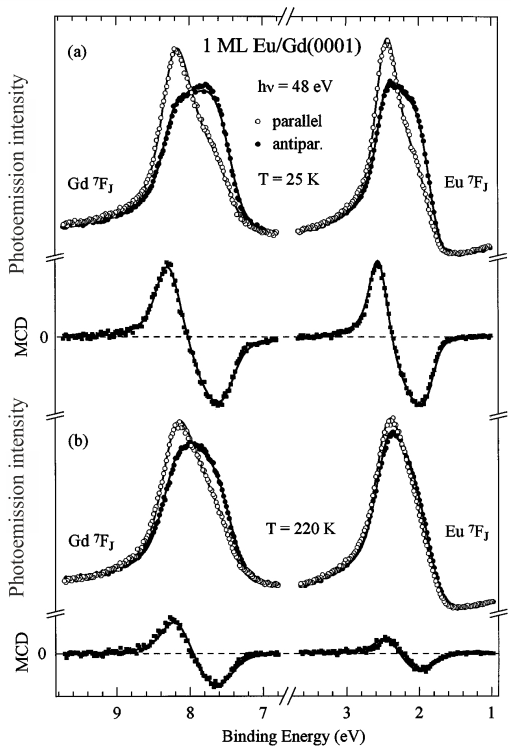}
    \caption{Magnetic circular dichroism in $4f$ photoemission from a monolayer Eu/Gd(0001) interface at 25~K and 220~K, showing element-specific Eu and Gd responses for parallel and antiparallel alignment of photon spin and sample magnetization. Data taken from Ref.~[\citeonline{Arenholz_1998}].}
    \label{fig:Eu_Gd}
\end{figure}

In particular, magnetic circular dichroism in core-level photoemission was shown to be strongly sensitive to spin-orbit interaction and to the coupling between spin and orbital angular momenta in the photoemission final states. This sensitivity enables the spin character of the individual multiplet components to be identified within an atomic many-body description, without the need for explicit spin-resolved detection~\cite{Laan_1996}.

Based on this understanding, the surface sensitivity of photoemission, when combined with magnetic circular dichroism, enabled detailed shell-specific investigations of lanthanide magnetism in thin films. In resonant $4d \rightarrow 4f$ photoemission from magnetized Gd films, the magnetic circular dichroism line shape was shown to vary strongly with photon energy across the giant-resonance region and could be consistently described within an atomic multiplet framework. In particular, the sign and magnitude of the dichroic signal were found to be governed primarily by the total angular momentum of the intermediate $4d^{9}4f^{8}$ state~\cite{Starke_1997}. Complementary studies of $5p$ core-level photoemission from Gd(0001) and Tb(0001) demonstrated large dichroic effects that are well reproduced by intermediate-coupling calculations, allowing a quantitative determination of Coulomb and exchange interactions in the photoemission final state.\cite{Laan_1997}

The broader significance of magnetic dichroism in photoemission was further emphasized by its application to chemically and magnetically well-defined heterostructures. In particular, Arenholz \emph{et al.} investigated an atomically sharp interface formed by a single monolayer of Eu on ferromagnetic Gd(0001)~\cite{Arenholz_1998}. As illustrated in figure~\ref{fig:Eu_Gd}, the multiplet-resolved $4f$ photoemission spectra of Eu and Gd exhibit pronounced and consistent dichroic intensity changes upon reversal of the photon spin, reflecting the parallel alignment of their magnetic moments. Despite the antiferromagnetic ground state of bulk Eu metal, magnetic circular dichroism in photoemission \linebreak (MCD-PES) from the $4f$ states revealed a sizable Eu magnetization at low temperatures, aligned with the underlying Gd film. This behavior was attributed to a strong positive interlayer exchange coupling across the Eu/Gd interface, which overcomes the weak antiferromagnetic intralayer interactions within the two-dimen- sional Eu overlayer. By analyzing multiplet-resolved $4f$ photoemission spectra of both elements, the study demonstrated that MCD-PES provides direct, element-specific access to local magnetic moments in atomically thin layers and at interfaces.

Subsequent studies on Tb(0001)/W(110), using Tb $3d \rightarrow 4f$ resonant photoemission, established a quantitative description of magnetic circular dichroism based on atomic multiplet theory in intermediate coupling, incorporating spin--orbit interaction, Coulomb and exchange interactions, angular-momentum coupling, and core-hole lifetime and decay effects, with detailed agreement between theory and experiment~\cite{Laan_1999_scaling}. At the same time, using thin Gd(0001) films on W(110), detailed investigations clarified the origin of spin-flip transitions in resonant photoemission, demonstrating that they arise predominantly during the photoexcitation step, where spin is no longer a good quantum number due to spin--orbit coupling in the intermediate core-hole state, rather than during the decay process~\cite{Hu_1999}.

Investigating the magnetically ordered surface oxide on Gd(0001), it was shown that hybridized valence states exhibit a pronounced magnetic exchange splitting governed by localized $4f$ moments, providing direct insight into $4f$--valence coupling~\cite{Schuessler_PRB_1999}. Subsequent work on epitaxial thin films of the heavy lanthanide metals Gd, Tb, Dy, and Ho showed that the magnetic splitting of valence states exhibits a systematic Stoner-like temperature dependence and scales linearly with the $4f$ spin moment, even in antiferromagnetic systems, demonstrating the dominant role of exchange coupling between localized $4f$ moments and itinerant electrons~\cite{Richter_2000}.
The next key advance was the realization of element-specific magneto-optics in the soft x-ray regime, where resonant $4d \rightarrow 4f$ reflectivity provides direct access to the $4f$ magnetization~\cite{Starke_2001}.

The papers discussed above demonstrate that photoelectron spectroscopy, when combined with magnetic dichroism, provides a powerful and quantitative instrument for probing exchange interactions, magnetic \linebreak anisotropy, and the coupling between localized $4f$ moments and itinerant electronic states in lanthanide systems. Through a series of systematic and conceptually linked studies, the Kaindl group and their collaborators clearly established that surface and bulk regions of lanthanide materials can exhibit markedly different magnetic properties, reflecting the enhanced role of reduced coordination and modified exchange interaction at surfaces. At the same time, these studies demonstrated that, once surface effects are properly taken into account, photoemission spectroscopy yields results fully consistent with bulk electronic-structure concepts, even in strongly correlated $4f$ systems~\cite{Arenholz_1998,Laan_1999_scaling}. These results formed the basis for subsequent developments in ARPES and modern studies of surface and interface magnetism in lanthanide-based and other correlated-electron materials.

Significant contributions to the theoretical description of lanthanide XAS at the 3$d$ edge and core-level photoemission were made by Thole and van der Laan \cite{Thole_PRB_1991, Thole_3dXAS_1985, XMCDsum_1992, LaanII_PRB_1993}, and became particularly important for interpreting such spectra in systems exhibiting intermediate-valence behavior. They demonstrated that both photoemission and XAS line shapes are governed by multiplet-resolved spectral contributions associated with different electronic configurations and established a quantitative approach for analyzing and interpreting these spectra. In later work, they derived sum rules for magnetic circular dichroism that relate the integrated dichroic intensity to the orbital and spin magnetic moments \cite{XMCDsum_1992}. In addition, detailed calculations of 4$f$ core-level photoemission showed that the multiplet structure exhibits a pronounced polarization dependence, leading to strong magnetic dichroism effects \cite{LaanII_PRB_1993}.

In addition, important contributions to the understanding of rare-earth photoemission and XAS were made by Kotani and Ogasawara. For instance, it was shown that core-level spectra of rare-earth oxides can be consistently described within an impurity Anderson-model framework that explicitly includes hybridization between localized 4$f$ states and ligand $2p$ bands.\textcolor[rgb]{1.00,0.00,0.00}{\cite{ Kotani_1992}} Their analysis of $3d$ and $4d$ X-ray photoemission and absorption spectra demonstrated that the spectral line shapes are governed by the interplay of covalent $4f$–ligand mixing, charge-transfer processes, and intra-atomic multiplet interactions. In particular, they clarified systematic trends across the lanthanide series and emphasized that both initial-state and final-state hybridization effects contribute to the observed core-level splitting, while multiplet effects become increasingly important for higher-energy core excitations. This work established a microscopic basis for extracting parameters such as the hybridization strength $V$, charge-transfer energy $\Delta$, and effective 4$f$ occupancy $n_f$, thereby providing a quantitative framework for understanding intermediate-valent behavior and the electronic structure of rare-earth oxides. Below, we present the results of our modeling of the $Ln$ 4$f$ spectra, with particular emphasis on the contributions of individual $M_J$ components, as well as the $Ln$ XAS spectra at the 4$d$ and 3$d$ edges and the corresponding $Ln$ 4$d$ photoemission spectra. In that regard, the concepts and methodology\cite{ Kotani_1992} have played an essential role and were used in our analysis.

\section*{Early studies and debates on the near-$\boldsymbol{E_\mathrm{F}}$ \\ feature in Kondo systems}

Following these contributions from the group led by \linebreak G.~Kaindl and collaborators, we now turn to earlier key results and developments that emerged already in the late 1970s. In particular, photoemission experiments led by J.~W.~Allen and his collaborators played a central role in elucidating the fine electronic structure of lanthanide systems and in establishing a framework for interpreting the observed spectra in terms of $4f$ physics.

\begin{figure}[t]
\centering
\includegraphics[width=6cm]{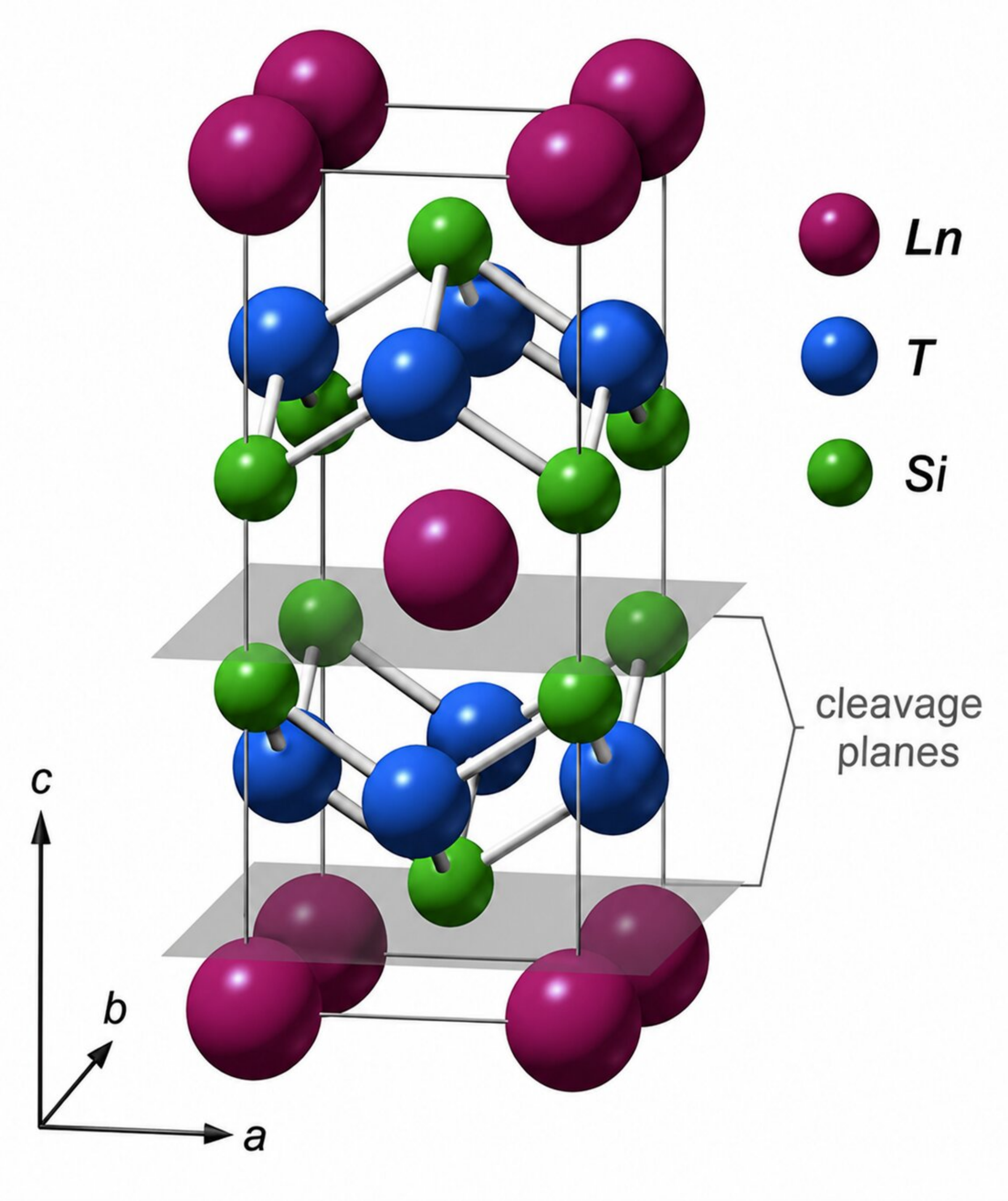}
\caption{View of the body-centered tetragonal ThCr$_2$Si$_2$-type structure (space group $I4/mmm$, No.~139), the so-called 122 phase. The structure is formed by stacks of alternating atomic layers with the sequence -$T$-Si-$Ln$-Si-$T$- \cite{Krellner2012,Kliemt_CRT_2020}, where $Ln$ denotes a lanthanide atom and $T$ represents a transition-metal atom. The $Ln$ atomic layers are well separated from each other by the strongly bonded Si-$T$-Si trilayer blocks. The cleavage planes between the Si and $Ln$ layers, along which such crystals are commonly cleaved, are indicated in gray.}
\label{fig:Cryst_struct}
\end{figure}

\begin{figure*}[t]
\centering
\includegraphics[width=\textwidth]{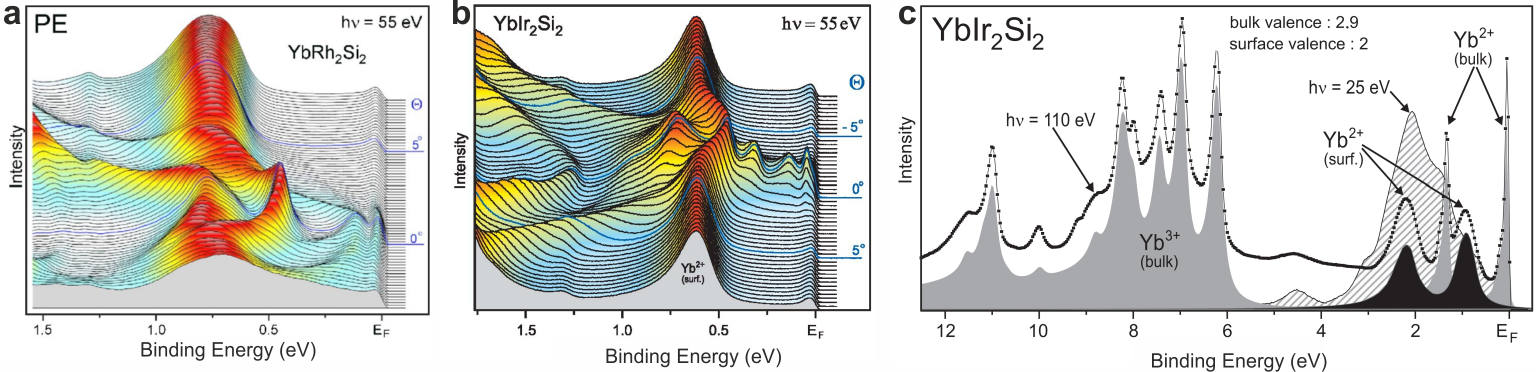}
\caption{ARPES data taken from YbRh$_2$Si$_2$ (a) and YbIr$_2$Si$_2$ (b) using 55 eV photons, and (c) survey photoemission spectra recorded at 25 eV (hatched) and 110 eV (dots). All data were taken at a temperature of $\sim 20$ K. The coexistence of bulk-related Yb $4f^{12}$ and $4f^{13}$ emission features indicates an intermediate Yb valence ground state.~\cite{Danzenb_PRL_2006, Danzenb_PRB_2007}}
\label{fig:YRS_1}
\end{figure*}

In pioneering experiments on Sm (and Tm) metal films, performed using photon energies in the range \linebreak 50–250~eV, a regime covering the 4$d \rightarrow 4f$ excitation threshold and not previously explored for rare-earth metals, it was shown that 4$f$-derived spectral features cannot be understood within simple band-like or integer-valence models. Instead, the spectra reflect strong electron correlations and hybridization effects, with indications of differences between surface- and bulk-related contributions to the photoemission signal~\cite{Allen_1978}. Subsequent experiments including polycrystalline SmB$_6$, performed using the same photon-energy range, revealed pronounced bulk-to-surface shifts of the Sm 4$f$ core level. These shifts were shown to imply an inhomogeneous valence mixing at the surface, distinct from the bulk electronic configuration. This work demonstrated that surface-related effects are intrinsic to photoemission spectra from lanthanides and must be carefully accounted for when interpreting 4$f$-derived features~\cite{Allen_PRB_1980}. Further studies on Yb revealed pronounced differences between clean and oxidized Yb metal, particularly in the vicinity of the $4d \rightarrow 4f$ absorption threshold. The results demonstrated that while clean Yb metal has a 4$f^{14}$ ground-state configuration, oxidation stabilizes a 4$f^{13}$ configuration, leading to a strong enhancement of 4$f$ photoemission. This oxidation-induced resonant enhancement provided a clear experimental realization of a Fano line shape arising from the interference of direct photoemission with autoionization decay of a discrete excited state in a solid~\cite{Johansson_1980}. Further photoemission studies focused on the Ce-based superconductors CeRu$_2$ and CeCo$_2$, with superconducting transition temperatures of approximately 6~K and 1~K, respectively. Resonant photoemission measurements revealed clear Ce~4$f$ spectral weight in both compounds, demonstrating that Ce is in a valence fluctuating state rather than purely tetravalent. This result revised the prevailing view of these materials as tetravalent Ce systems and established the participation of 4$f$ states in their low-energy electronic structure~\cite{Allen_1982a}. Essential contributions to the understanding of Ce-based systems were made through spectroscopic studies of compounds such as CeRu$_2$, CeNi$_2$, CeIr$_2$, and CeAl, as well as CeSi$_2$, CeOs$_2$, CePd$_3$, CeCo$_2$, and CeNi$_5$, as comprehensively discussed in Ref.~[\citeonline{Allen_1986}], where results from valence-band photoemission, inverse photoemission, and core-level spectroscopy were interpreted within the impurity Anderson Hamiltonian, with first-principles calculations providing the parameters required to account for correlation effects beyond mean-field descriptions.

At this point, it is essential to mention several key works and the intense discussions they stimulated. In particular, during the mid-1980s and early 1990s the interpretation of near-$E_\mathrm{F}$ 4$f$ photoemission features in Ce- and Yb-based systems became the subject of an explicit debate between different experimental groups, most notably those led by A.~J.~Arko and by J.~W.~Allen~\cite{Allen_1986, Patthey_1987, Arko_PRL_1}.

In brief, while J.~W.~Allen and co-workers argued that the observed temperature-dependent 4$f$ intensity near the Fermi level reflects the Kondo resonance within the Anderson impurity framework, J.~J.~Joyce and A.~J.~Arko emphasized that the line shape, width, and apparent temperature dependence could be accounted for by phonon broadening, surface sensitivity, and analysis procedures, thereby questioning a direct identification; this position was hotly debated within the community.\cite{Tjeng_1993, Joyce_1994, Tjeng_1994_1, Murani_1994, Tjeng_1994_2, Patthey_1993}

In retrospect, these discussions clarified that reliable insight into Kondo-related physics of Ce- and Yb-based systems from photoemission requires experiments on high-quality single crystals, a careful separation of surface and bulk contributions to the measured signal, close monitoring of sample aging during measurements, strict control of surface cleanliness, and a critical analysis of the measured observables in order to ensure a meaningful interpretation.

In this regard, we discuss below our temperature-dependent ARPES measurements on the AFM Kondo lattice CeRh$_2$Si$_2$ obtained exclusively from well-defined Ce- and Si-terminated surfaces~\cite{patil_16, Georg_2020}. Such control enables a reliable separation of surface and bulk contributions to the PE signal and establishes a clear spectroscopic picture of the Kondo-related electronic structure and its evolution with temperature at the surface and in the bulk~\cite{Georg_2020}.

\section*{YbRh$_2$Si$_2$: the 4$\boldsymbol{f}$ bands and the $\boldsymbol{f}$-like Fermi surface}

We now turn to an overview of our own contributions to the field, beginning in the 2000s and focusing mainly on lanthanide materials crystallizing in the ThCr$_2$Si$_2$ structure, which is schematically illustrated in figure~\ref{fig:Cryst_struct}. At the end of our Review, we additionally provide a section entitled \emph{Crystal growth}, where we discuss in detail our know-how and experience in the synthesis of high-quality single crystals that were subsequently investigated using the synchrotron-based spectroscopic techniques discussed below, beginning with the Yb-based systems.

In general, our work combines the development of high-quality single crystals, their detailed characterization, and their investigation using bulk-sensitive methodologies that provide insight into thermodynamic, magnetic, and transport properties, together with advanced synchrotron-based spectroscopic techniques. This combined and synergistic approach addresses the experimental challenges outlined above and enables reliable and conclusive insight into $4f$-driven physics. The coherent and systematic research line followed in presenting the results below has allowed us to address key aspects of lanthanide physics, in particular through the continuous disentangling of bulk and surface phenomena and their corresponding temperature scales.

We begin with YbRh$_2$Si$_2$, a prototypical heavy-fermion system that has provided deep insights into Kondo physics and quantum criticality and enabled the acquisition of high-quality ARPES data, which we overview below. Since the first observation of pronounced non-Fermi-liquid behavior at ambient pressure and zero field, in proximity to an exceptionally low AFM ordering temperature $T_\mathrm{N} \simeq 70$ mK~\cite{Trovarelli_PRL_2000}, YbRh$_2$Si$_2$ has emerged as a benchmark system attracting broad attention from the community~\cite{Custers_2003,Paschen_2004}. Its delicate magnetic order can be suppressed by vanishingly small magnetic fields, giving rise to a field-induced quantum critical point, at which transport and thermodynamic properties indicate a divergence of the quasiparticle effective mass and a crossover to Landau Fermi-liquid behavior at higher fields~\cite{Gegenwart_PRL_2002}.

Figure~\ref{fig:YRS_1} presents the first reliable ARPES data obtained from YbRh$_2$Si$_2$ (with a Kondo temperature $T_\mathrm{K} \sim 25$ K) and YbIr$_2$Si$_2$ ($T_\mathrm{K} \sim 40$ K)~\cite{Danzenb_PRL_2006, Danzenb_PRB_2007}. While results from bulk measurements on these materials were already available around 2000, their investigation by photoemission experiments was challenged by material- and experiment-specific issues. Only after systematic optimization of sample preparation, in-situ cleavage under UHV conditions, and accurate identification of surface terminations did ARPES measurements become reproducible and \linebreak reliable~\cite{Danzenb_PRL_2006, Danzenb_PRB_2007}.

The spectral patterns shown in figure~\ref{fig:YRS_1}a,b probe a similar region near the Brillouin-zone center for both systems. The spectra were taken mainly from Yb-terminated surfaces, as evidenced by the strong divalent Yb $4f_{7/2}$ surface signal at binding energies of $\sim0.6–0.7$~eV, which clearly dominates the bulk Yb contribution near $E_\mathrm{F}$. Two parabolic valence bands with hole-like dispersion are observed, one of which intersects the $4f_{7/2}$ surface emission close to $\bar{\Gamma}$. Near this intersection, the $4f$ surface-related feature splits into at least two dispersive components, separated by up to $\sim 0.25$~eV. This spectral pattern provides a particularly clear view of the $f$–$d$ hybridization, although it is predominantly associated with the surface electronic structure of these systems.

\begin{figure}[t]
\includegraphics[width=0.99\linewidth]{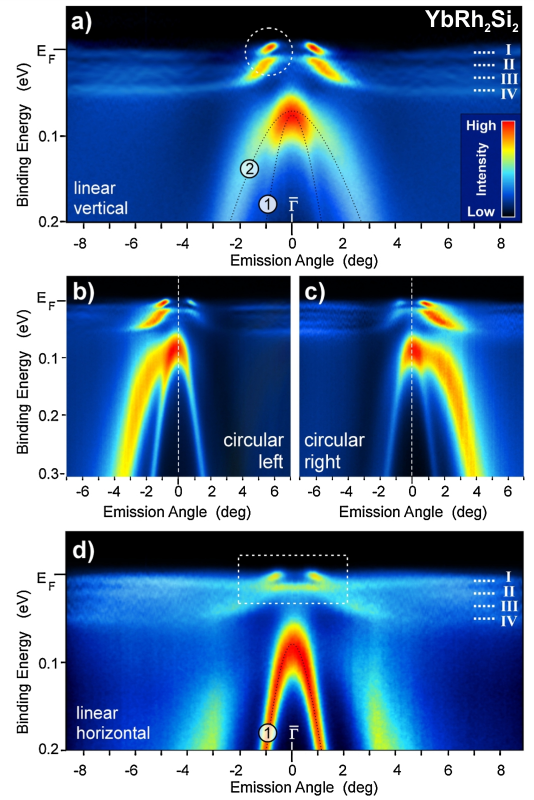}
\caption{ARPES spectra taken from the Si-terminated surface of YbRh$_2$Si$_2$ along the $\bar{\Gamma}$-$\bar{{\rm X}}$ direction of the surface Brillouin zone, measured using 45~eV photons with linear vertical (a), circular left (b), circular right (c), and linear horizontal (d) polarizations~\cite{Vyalikh_PRL_2010}. The color scale represents the photoemission intensity, with red and yellow corresponding to high intensity and blue to low intensity. }
\label{fig:YRS_2}
\end{figure}

Thus, the next step was to resolve the fine electronic structure near the Fermi level, which posed a substantial challenge.
The Yb-terminated surface was not well suited for this purpose, since, as seen in figures~\ref{fig:YRS_1}a,b, its intense surface Yb $4f$ signal outshines and thereby obscures the most relevant features near $E_\mathrm{F}$. It was therefore essential to find and explore a surface with an alternative termination, which was the Si-terminated surface. In this case, the spectral weight near the Fermi level is expected to arise predominantly from electron emission from Yb layers beneath the Si surface (primarily the fourth and deeper atomic layers), consistent with the probing depth, thereby allowing the fine $E_\mathrm{F}$-related features to be discussed in direct connection with bulk electronic properties.

Regarding the evaluation of the surface termination, we note the following. By performing DFT calculations for these materials, in which the $4f$ states were frozen and treated as core states, we found that the charge density between Si and Yb is significantly depleted compared to that within the silicide Si–$T$–Si blocks. This implies that the chemical bonding between Si and Yb is weaker than within the Si–$T$–Si part of the crystal. Consequently, we expected that the crystal would predominantly cleave between the Si and Yb planes, resulting in either Si- or Yb-terminated surfaces~\cite{Vyalikh_2010_Elspec}. Indeed, based on our experimental experience, the cleaved surface typically consists of a mosaic of crystallites with different (Si- and Yb-) terminations. In favorable cases, it was possible to identify a sufficiently large crystallite with the desired termination, selectively illuminate it with the photon beam, and perform ARPES measurements.

After several further attempts, we gradually succeeded in obtaining the first clear ARPES spectral patterns reflecting the interaction between Yb $4f$ and itinerant states near the Fermi level in YbRh$_2$Si$_2$~\cite{Vyalikh_PRL_2008,Vyalikh_PRL_2010}. In figure~\ref{fig:YRS_2}, we present high-resolution ARPES data taken from YbRh$_2$Si$_2$ at a temperature slightly below 1~K, measured at the One-Cubed ARPES endstation at the BESSY~II facility~\cite{Vyalikh_PRL_2010}. These results provide the first direct experimental evidence that crystal-field–split $4f$ states in lanthanide systems can exhibit a pronounced momentum dependence. The data further demonstrate that dispersive valence \linebreak bands induce an effective dispersion of the $4f$ crystal-field levels, leading to substantial variations of their energy splittings across the Brillouin zone and even to an interchange of the ground and excited state sequence near the $\bar{\Gamma}$ point~\cite{Vyalikh_PRL_2010}.

In the framework of this study, a simple approach for the analysis of the ARPES data was proposed. \textit{Ab~initio} band-structure calculations were performed both for the surface (using a slab setup) and for the bulk, where the Yb $4f$ states were frozen and treated as core states. Based on the resulting theoretical band structure, a simple hybridization model was then constructed, and comparison of the modeled spectra with ARPES data allowed the symmetry of the four dispersive Kramers doublets near the Fermi level in the vicinity of the $\bar{\Gamma}$ point to be established~\cite{Vyalikh_PRL_2010}. While the proposed approach is clearly oversimplified, we did not find a more useful or practical method to model the ARPES data. From today’s perspective, although these results have been available since 2010, no DMFT-based studies providing deeper insight into these observations have been reported so far.

As we have already seen, the question of surface-related $4f$ properties versus bulk contributions is a traditional and well-known issue in ARPES measurements, and in the context of the discussed data for YbRh$_2$Si$_2$ we also critically evaluated this point. In fact, even keeping in mind that the signal shown in figure~\ref{fig:YRS_2} was taken from a Si-terminated surface, which should to some extent reflect bulk contributions, it was not a priori obvious that surface effects do not play a significant role here. Therefore, we decided to perform an additional simple experiment by depositing adatoms on the Si surface in order to explore how the underlying spectral structure of YbRh$_2$Si$_2$ is modified. As adatoms, we have chosen Ag.

\begin{figure}[h!]
\includegraphics[width=0.99\linewidth]{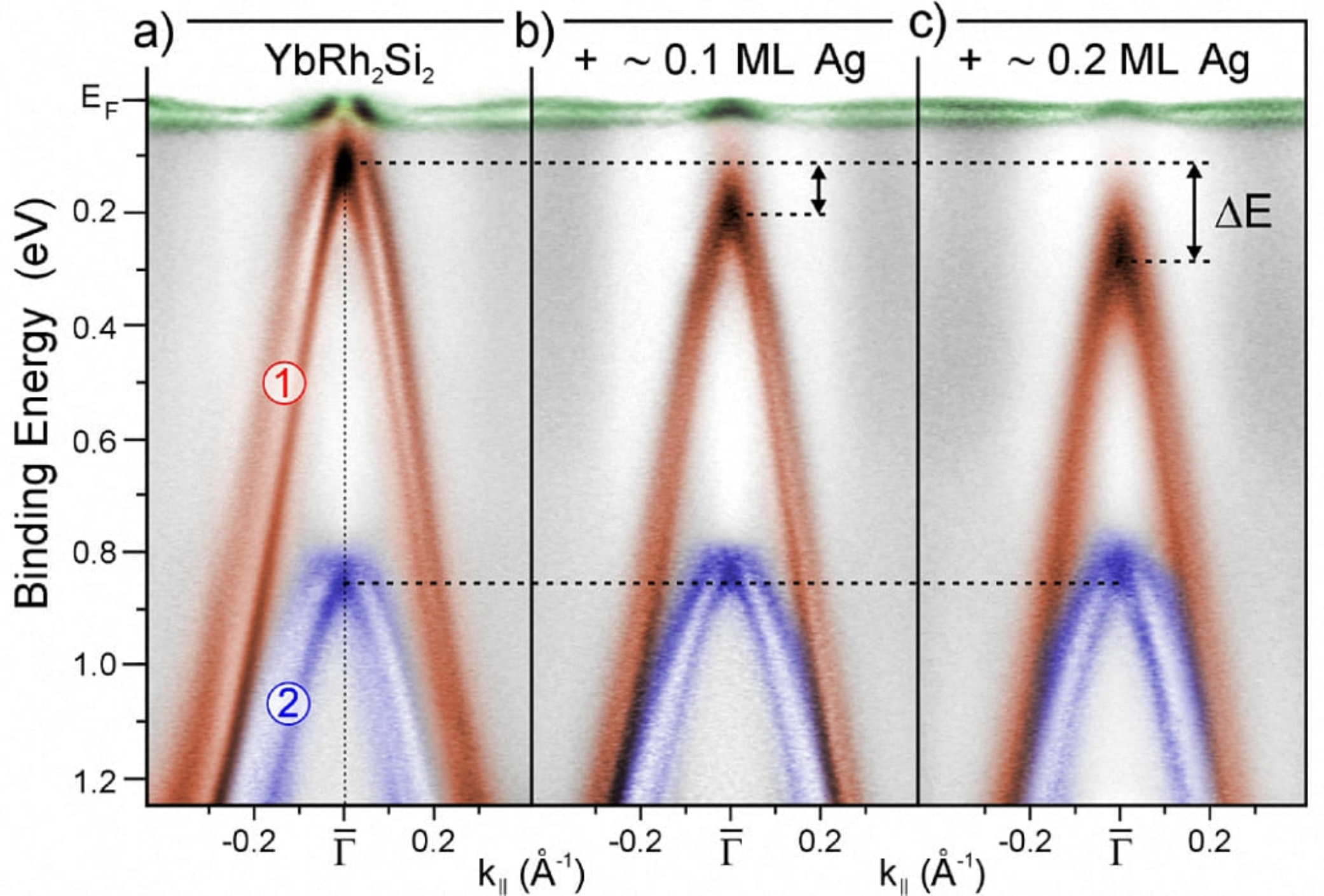}
\caption{Experimental ARPES patterns taken at $h\nu = 45$~eV along the $\bar{\Gamma}$--$\bar{{\rm X}}$ direction of YbRh$_2$Si$_2$ for (a) the Si-terminated surface and after deposition of (b) $\sim$0.1~monolayer (ML)  and (c) $\sim$0.2~ML of Ag, respectively. High photoemission intensity is represented by dark colors.\cite{Molodtsov_YbRhSi_2009} The distinct bands are highlighted using different colors for better visualization.}
\label{fig:YRS_Ag}
\end{figure}

In figure~\ref{fig:YRS_Ag}, we present the results of Ag deposition on the atomically clean, Si-terminated surface of YbRh$_2$Si$_2$. It is clearly seen that gradual deposition of Ag atoms leads to a transfer of Ag $5s$ charge into the Rh $4d$ bands. This substantially modifies the energy overlap, and hence the hybridization strength, between the interacting Yb $4f$ and Rh $4d$ states in the near-surface region of the Si-terminated surface~\cite{Vyalikh_2010_Elspec, Molodtsov_YbRhSi_2009}.

This approach not only enables a controlled separation of surface and bulk contributions, making it evident that, to a certain extent, the observed $f$–$d$ hybridization pattern near the $\bar{\Gamma}$ point at the Fermi level reflects surface-related properties, but also provides important insight into the mechanisms underlying Kondo behavior. It offers guidance for future studies of bulk materials and opens promising perspectives for investigations of heavy-fermion and Kondo physics in thin films, multilayers, clusters, and other low-dimensional systems.

These findings further demonstrate that electronic hybridization at the surface of heavy-fermion systems can be finely tuned by the deposition of adatoms. As a consequence, heavy-fermion and Kondo-related surface properties may strongly deviate from their bulk counterparts.

Comparing the ARPES patterns shown in figure~\ref{fig:YRS_2}a and in the left panel of figure~\ref{fig:YRS_Ag}, which both reflect ARPES data taken from freshly cleaved Si-terminated surfaces, one can see that in the experiment with Ag only two CEF states are observed instead of four. This also explains the challenges faced in ARPES measurements: the quality of the samples used in the early experiments~\cite{Vyalikh_PRL_2008} and with Ag deposition~\cite{Vyalikh_2010_Elspec, Molodtsov_YbRhSi_2009} was not sufficiently high and did not allow us to resolve the spectral pattern with the anticipated four Kramers doublets~\cite{Vyalikh_PRL_2010}.

Once we understood how to synthesize high-quality single crystals and how to handle them properly in order to obtain clear ARPES data, the next task was to determine the Fermi surface from ARPES measurements. This issue had already been intensively discussed within the community, and corresponding data were available from de Haas–van Alphen measurements~\cite{Rourke_PRL_2008}, although these data were taken in magnetic field.

In pursuing this goal, we faced certain challenges, as the ARPES instruments available at that time did not offer an electron-deflection mode.
Consequently, mapping the full Fermi surface required rotating the sample while keeping the photon beam on the same crystallite, which was difficult to control given the small Si-terminated terraces on the samples. Nevertheless, we succeeded in obtaining reliable Fermi-surface maps from ARPES measurements performed at $T = 10$~K, i.e., below the Kondo temperature of $\sim 25$~K~\cite{Danzenbacher_PRL_2011, Kummer_PRX}.

\begin{figure}[h!]
\centering
\includegraphics[width=0.85\linewidth]{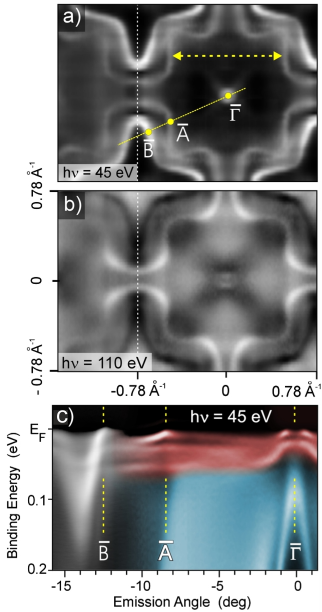}
\caption{Fermi-surface patterns of YbRh$_2$Si$_2$ derived from ARPES measurements using photon energies of 45~eV (a) and 110~eV (b), respectively. The line connecting $\bar{\Gamma}$,  $\bar{{\rm A}}$, and $\bar{{\rm B}}$ indicates the momentum direction along which the energy–momentum map shown in (c) was obtained.\cite{Danzenbacher_PRL_2011}. The $4f$ and itinerant bands in (c) are highlighted using red and blue shading, respectively, to facilitate their visualization. }
\label{fig:YRS_FS}
\end{figure}

In figure~\ref{fig:YRS_FS}, we show the acquired Fermi surface maps obtained from ARPES measurements. To emphasize the contribution of Yb $4f$ states relative to the emission from Rh $4d$ valence-band states, it was particularly useful to combine photon energies of $h\nu = 45$~eV, where both types of states contribute to the measured signal, and $h\nu = 110$~eV, at which the emission from Rh $4d$ states is strongly suppressed due to a Cooper minimum in the Rh $4d$ photoionization cross section~\cite{Vyalikh_2010_Elspec, Molodtsov_YbRhSi_2009, Danzenbacher_PRL_2011}.

The essential aim of these measurements was therefore: (i) to disentangle surface states from bulk-derived spectral features, (ii) to visualize the Fermi surface of YbRh$_2$Si$_2$ in the Kondo regime, and (iii) to explore its evolution as a function of temperature~\cite{Danzenbacher_PRL_2011, Kummer_PRX, Monika_Compton}. Let us first discuss the features observed in figure~\ref{fig:YRS_FS}. An analysis of the observed spectral pattern allows one to conclusively identify the sharp and highly intense feature at $\bar{{\rm B}}$, which forms a diamond-like contour around the $\bar{{\rm M}}$ point at the corner of the Brillouin zone, as a Shockley-type surface state. This state is localized within the surface-related Si–Rh–Si–Yb block and clearly interacts with the Yb $4f$ states within this block. The presence of this coupling implies the existence of a surface-related Kondo scale in YbRh$_2$Si$_2$, and essentially, this Shockley state is an intrinsic feature of the Si-terminated surface~\cite{Danzenbacher_PRL_2011}.

As discussed above, the spectral pattern in the vicinity of the $\bar{\Gamma}$ point predominantly reflects surface-related properties, as independently confirmed by experiments involving Ag deposition~\cite{Vyalikh_2010_Elspec, Molodtsov_YbRhSi_2009}. However, DFT calculations suggest that the hole-like parabolic bands observed near $\bar{\Gamma}$ exhibit a surface-resonance character and therefore penetrate deeper into the bulk compared to the Shockley state centered at $\bar{{\rm M}}$. Consequently, the intrinsic bulk-like Fermi surface derived from these ARPES measurements corresponds to the square-like feature exhibiting open necks near the $\bar{{\rm X}}$ point, commonly referred to as the Doughnut ($\mathcal{D}$) Fermi surface of YbRh$_2$Si$_2$. Figure~\ref{fig:YRS_FS} illustrates the formation of this feature at an arbitrary $\bar{\mathrm{A}}$ point. The weakly dispersive 4$f$ bands observed close below $E_\mathrm{F}$ are seen to interact with a waterfall-like intensity pattern that stems from a bulk-projected state, where the resulting 4$f$-hybrid states exhibit a pronounced bending and cross the Fermi level~\cite{Danzenbacher_PRL_2011}. Notably, the ARPES-derived bulk Doughnut ($\mathcal{D}$) Fermi surface of YbRh$_2$Si$_2$ closely resembles the Fermi surface \cite{Rourke_PRL_2008}, indicating a consistent picture across different experimental probes.

At this point, we mention the essential issue of the Fermi surface in Ce- and Yb-based Kondo lattices. Within the Kondo scenario, below a characteristic temperature $T_\mathrm{K}$ the localized $4f$ magnetic moments become screened through their interaction with itinerant conduction electrons, giving rise to heavy-fermion behavior. This raises the fundamental question of how the $4f$ degrees of freedom participate in the formation and topology of the Fermi surface, and what the relevant temperature scale for this evolution is.

Originally, the concept of the Fermi surface was formulated for non-interacting fermions, and later on it was generalized by Luttinger, who showed that Fermi surface volume remains conserved in interacting Fermi liquids~\cite{Luttinger_1960}. In 1982, R. Martin extended this fundamental principle suggesting that it can also be applied to strongly correlated fermionic systems~\cite{Martin_Richard_1, Martin_Richard_2}. Its application to 4$f$ materials revealing strong interaction between local 4$f$ and itinerant states implies that if the local magnetic moments are quenched due to the Kondo effect, then the underlying local electrons must be counted in the Fermi surface and then it becomes “large”. Otherwise they are excluded from the Fermi surface, and the latter is “small”.

In spite of its conceptual simplicity, experimental evidence of how, and on which temperature scale, the Luttinger sum rule is realized in $f$-electron systems has remained elusive. At the time these issues were discussed, there were no reports that fully visualized the working of Luttinger’s theorem in strongly correlated $f$ materials. In particular, an experimental demonstration of a temperature-driven transition between the “large” and “small” Fermi surface was still lacking.

Turning back to the ARPES-derived Fermi surface of YbRh$_2$Si$_2$, we note that the measurements were performed at 10~K, well below the Kondo temperature $T_\mathrm{K} \approx 25$~K, i.e., in the Kondo regime. Consequently, the observed Fermi surface is expected to correspond to a “large” Fermi surface and, in fact, it agrees well in shape with  the “large” Fermi surface case (also known as “4$f$ itinerant”) reported in Ref.~[\citeonline{Rourke_PRL_2008}] for the respective case. Motivated by this result, the next challenge was to access, using ARPES, the complementary case of a “small” Fermi surface, where the $4f$ moments remain localized and do not interact with the itinerant electrons. In this context, we consider YbCo$_2$Si$_2$~\cite{YbCoSi, Monika_YbCoSi_2014}, for which bulk measurements do not indicate any signatures of Kondo behavior down to very low temperatures. Consequently, YbCo$_2$Si$_2$ can be regarded as a stable-valent Yb$^{3+}$ system and serves as an isoelectronic reference for the heavy-fermion system YbRh$_2$Si$_2$.

\begin{figure}[h!]
\centering
\includegraphics[width=8cm]{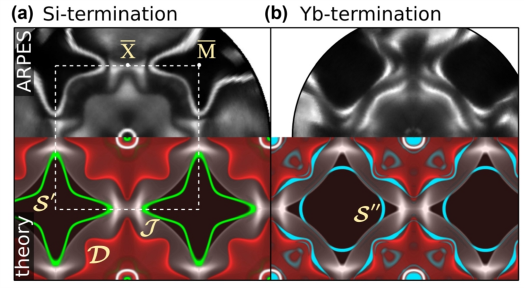}
\caption{ARPES-derived Fermi surfaces from (a) Si- and (b) Yb-terminated surfaces of YbCo$_2$Si$_2$, obtained at a photon energy of 45~eV and a temperature of 11~K~\cite{Monika_YbCoSi_2014}. The colored FSs in the lower panels were derived from band structure calculations employing the open-core approach. Bulk FS sheets are labeled $\mathcal{D}$ (red) and $\mathcal{J}$ (gray). Surface states obtained from the slab supercells are labeled $\mathcal{S'}$ (green) and $\mathcal{S''}$ (cyan) for Si- and Yb-termination, respectively. The first surface BZ is indicated by a dotted square. For further details, see Ref.~[\citeonline{Monika_YbCoSi_2014}].}
\label{fig:YbCoSi}
\end{figure}

Figure~\ref{fig:YbCoSi} shows ARPES-derived Fermi-surface maps obtained from both Si- and Yb-terminated surfaces of YbCo$_2$Si$_2$ at a temperature of 11~K~\cite{Monika_YbCoSi_2014}. The differences between both terminations can be consistently attributed to surface states and resonances. For instance, as already mentioned for YbRh$_2$Si$_2$, a sharp diamond-shaped surface state lies within the bulk-projected gap around the $\bar{{\rm M}}$ point for the Si-terminated surface, while it is absent for the Yb-terminated surface. This surface state is highlighted in green in the panel displaying the results of the open-core calculations. For our discussion, the main feature of interest is the square, ``Doughnut''-shaped bulk Fermi surface sheet ($\mathcal{D}$), highlighted in red. The topology of this sheet plays a significant role in clarifying the nature of the AFM quantum phase transition in YbRh$_2$Si$_2$. As seen, this Fermi surface sheet forms disjoint hole pockets in $k$ space and is ideally suited to track changes of the Fermi surface volume. In YbCo$_2$Si$_2$, the ``Doughnut'' lies entirely inside the first Brillouin zone. Its size and topology perfectly match the theoretical Fermi surface sheet obtained within the open-core approximation, which neglects any hybridization between the localized $4f$ shell and itinerant states.

\begin{figure}[t]
\includegraphics[width=8cm]{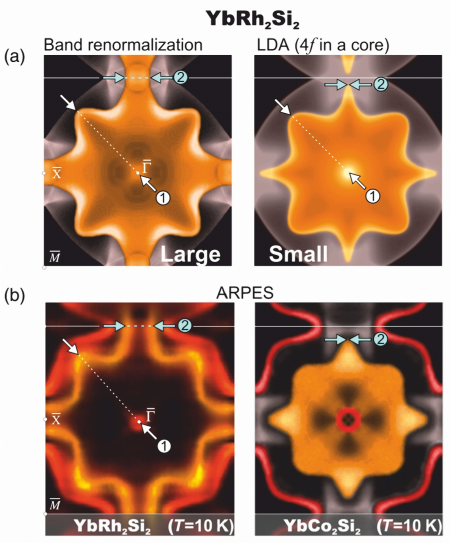}
\caption{(a) The “large” and “small” Fermi surface of YbRh$_2$Si$_2$ calculated within the renormalized band structure approach and the LDA scheme, respectively. (b) Fermi surface of YbRh$_2$Si$_2$ ($T_\mathrm{K} = 25$~K) and YbCo$_2$Si$_2$ ($T_\mathrm{K} < 1$~K) as seen in ARPES at $T \sim 10$~K \cite{Kummer_PRX}. The bulk and surface states are highlighted using orange and red shading, respectively.}
\label{fig:YRS_YBC}
\end{figure}

Thus, the bulk Fermi surface of YbCo$_2$Si$_2$ can be identified as a “small” Fermi surface that does not include the degrees of freedom of the Yb $4f$ hole. The ARPES-derived Fermi surface of YbRh$_2$Si$_2$ exhibits a similar topology, however, in contrast to the Co system, the $\mathcal{D}$ Fermi surface is enlarged and extends into the neighboring Brillouin zone via open necks at the $X$ point~\cite{Danzenbacher_PRL_2011}. Then, taking the Fermi surface of YbCo$_2$Si$_2$ as a reference for the “small” Fermi surface, the enlargement of the ``Doughnut'' in YbRh$_2$Si$_2$, compared to the unhybridized case, is clearly confirmed and can be attributed to the formation of a large Fermi surface that includes the Yb $4f$ hole. In figure~\ref{fig:YRS_YBC}, we summarize our theoretical and experimental findings for YbRh$_2$Si$_2$ and YbCo$_2$Si$_2$, which provide the basis for temperature-dependent measurements aimed at unveiling the temperature scale at which the transition from a small to a large Fermi surface occurs~\cite{Kummer_PRX}.

At the time when the respective experiments were planned, ARPES measurements on such small crystals were technically very demanding, largely due to the absence of the so-called deflection mode in the ARPES analyzer. Modern analyzers employ this mode, using electric fields within the lens system to steer photoelectrons toward the detector. This makes it possible to study even very small samples and acquire the Fermi surface without extensive sample rotation and the need to continuously keep the photon spot on the same region of the tiny crystal.

Therefore, it was essential to elaborate the experimental plan and carefully select the temperatures at which the ARPES measurements should be performed and at which changes in the Fermi surface shape could be expected. In discussions within our team and with our partners, it became clear that there was no consensus on this point. Some experts suggested focusing on temperatures below and moderately above $T_\mathrm{K}$ (for example, 10~K and 40~K for $T_\mathrm{K} \approx 25$ K), assuming that the relevant changes would occur in its vicinity. Others argued in favor of exploring a much broader temperature range, since across the Kondo crossover the evolution of the Fermi surface might be gradual and not necessarily abrupt or clearly discernible close to $T_\mathrm{K}$.

\begin{figure}[t]
\includegraphics[width=0.98\linewidth]{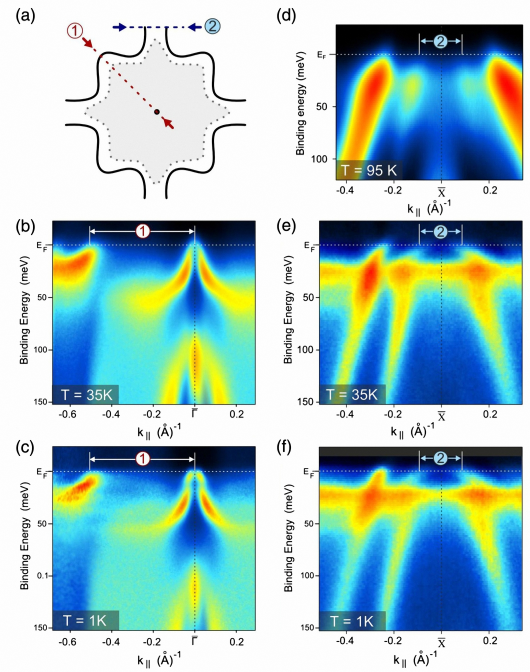}
\caption{Temperature-dependent evolution of the Fermi surface of YbRh$_2$Si$_2$. The temperature dependence is analyzed by probing two characteristic segments of the Brillouin zone, schematically shown in (a). Panels (b), (c) correspond to the “diagonal” direction, while (d)–(f) represent the “neck” direction. The respective ARPES-derived energy–momentum maps are taken at temperatures across the single-ion $T_\mathrm{K}$ and the coherence temperature $T^\ast$. Note that the data shown in (d) were recorded at a different instrument with a modified experimental geometry, resulting in altered sensitivity to $4f$ and valence states due to symmetry-related matrix-element effects\cite{Kummer_PRX}.}
\label{fig:YRS_Temp}
\end{figure}

The first ARPES experiments were planned to be performed at the ``One Cubed ARPES'' instrument at BESSY, where, due to technical constraints, the reliable temperature range extended from approximately $700$~mK to about of $45$~K. All details and the first results from this experiment are presented and discussed in Ref.~[\citeonline{Kummer_PRX}]. In figure~\ref{fig:YRS_Temp} we summarize the key experimental findings, focusing on two directions in $k$-space: (1) along the $\bar{{\rm M}}$–$\bar{\Gamma}$–$\bar{{\rm M}}$ direction (diagonal direction), and (2) along the $\bar{{\rm M}}$–$\bar{{\rm X}}$–$\bar{{\rm M}}$ ("neck" direction), where the most notable changes were anticipated based on the results for YbRh$_2$Si$_2$ and \linebreak YbCo$_2$Si$_2$.

Our analysis of the experimental data indicates that the Fermi surface does not change within the investigated temperature range, and within the sensitivity of the experiment. In particular, neither the size of the square-like part of the Fermi surface, characterized by segment~1, nor the opening of the necks, characterized by segment~2, shows any detectable variation. Likewise, no notable changes are observed in the dispersion of the quasiparticle bands crossing $E_\mathrm{F}$ (see figure~\ref{fig:YRS_2} and figure~\ref{fig:YRS_FS}c).

Note that our analysis is limited by the experimental resolution and by thermal broadening. Therefore, it is not straightforward to quantitatively address the temperature dependence of the quasiparticle linewidth close to $E_\mathrm{F}$. Qualitatively, within the studied temperature range, our data show no indication of jumps or sudden changes in the linewidth at any particular temperature. Instead, the observations suggest that the quasiparticle width increases monotonically with increasing temperature.

To summarize, the only notable temperature-depen-dent effect in our data is the increased thermal broadening, $k_\mathrm{B} T$, which becomes apparent at 35~K and, in particular, at 95~K compared to 1~K, while the overall ARPES spectra remain otherwise unchanged. We note that the ARPES measurements at 95~K were performed at the SIS beamline of SLS, which employs a different experimental geometry. As a result, the spectral appearance of the bands near the $\bar{{\rm X}}$ point looks slightly different from that obtained at BESSY. Nevertheless, the characteristic “neck” at $\bar{{\rm X}}$ is clearly observed and apparently remains open up to at least 95~K.

The already significant thermal broadening and the relatively large intensity of the surface state in the 95~K data shown in figure~\ref{fig:YRS_Temp}d may give the impression of a spectral gap forming at $E_\mathrm{F}$, which could be interpreted as a closing of the neck. However, a detailed analysis of the spectral intensity at $E_\mathrm{F}$ demonstrates~\cite{Kummer_PRX} unambiguously that no such spectral gap is present at 95~K. Thus, we do not observe the transition from a "large" to a "small" Fermi surface as predicted, for example, by simplified slave-boson mean-field treatments~\cite{Coleman1985JoMaMM}. The temperature-dependent valence change seen in angle-integrated photoemission spectrum and in resonant X-ray emission spectroscopy is therefore not accompanied by a simultaneous change of the Fermi surface.

Thus, the discussed ARPES experiments demonstrate that the Fermi surface of the prototypical Kondo lattice YbRh$_2$Si$_2$ remains remarkably stable over a wide temperature range around $T_\mathrm{K}$. In particular, we find no signatures of the transition from localized to itinerant $4f$ behavior. These experimental results appear inconsistent with simplified mean-field treatments of the periodic Anderson model.

At the same time, clear temperature-dependent variations of the Yb$^{2+}$ and Yb$^{3+}$ features are observed in angle-integrated Yb $4f$ photoemission and resonant $X$-ray emission spectroscopy spectra~\cite{Kummer_PRX}. This indicates that the relation between the local $f$-valence variation, coherent Kondo screening, and the resulting Fermi surface topology is more subtle than commonly assumed.

Our results suggest that the formation of the "large'' (low-temperature) Fermi surface occurs at a significantly higher temperature than the onset of coherence, in line with scanning-tunnelling microscopy studies on \linebreak YbRh$_2$Si$_2$~\cite{Ernst2011N}. Thus, the ARPES data do not support the prevailing picture in the community, in which the formation of the "large'' Fermi surface and the onset of coherence between Kondo singlets are assumed to occur hand in hand, i.e., around or below $T_\mathrm{K}$. The two phenomena therefore appear not to be directly linked. Further theoretical and experimental work is required to clarify how the Fermi surface evolves across the crossover from coherent to incoherent regimes in this particular Kondo lattice system and in Kondo lattices in general. We note that our observation of a large Fermi surface at near 1~K supports a spin-density wave (SDW) scenario \cite{Abrahams2012PNAS} for the quantum critical point in YbRh$_2$Si$_2$ and thus question the Kondo breakdown scenario. This scenario was proposed on the basis of the peculiar transport and thermodynamic properties observed in the vicinity of the QCP \cite{Si2001N, Coleman2001JoPCM}.

Thus, our next challenge was to experimentally explore the anticipated transition of the Fermi surface from "small" to "large" for YbRh$_2$Si$_2$, which we expected to occur at temperatures well above 100~K. Because ARPES does not reliably resolve the fine CEF-split $4f$ states and their dispersion at such elevated temperatures, we turned to Compton scattering (CS) experiments. After several measurements performed at the SPring-8 facility, high-resolution CS measurements allowed us to overcome many of the technical limitations of both ARPES and dHvA and to directly probe the Fermi surface topology of the prototypical Kondo lattice YbRh$_2$Si$_2$~\cite{Monika_Compton}.

The measurements reveal pronounced temperature-driven changes of the Fermi surface between 14~K and 300~K in zero magnetic field. As a reference for the small Fermi surface, we investigated the isoelectronic compound YbCo$_2$Si$_2$, whose electron occupation number density remains essentially unchanged over this temperature range. In contrast, YbRh$_2$Si$_2$ exhibits strong modifications of the electron density upon heating, and at room temperature its Fermi surface closely resembles that of the small-Fermi-surface reference system~\cite{Monika_Compton}. These results provide clear evidence for the long-sought Kondo crossover in YbRh$_2$Si$_2$, from itinerant $4f$ states at low temperatures associated with the coherent Kondo lattice to localized $4f$ moments far above $T_\mathrm{K}$. The characteristic temperature scale of this crossover is estimated to lie between approximately 100\,K and 300\,K~\cite{Monika_Compton}.

Discussing the relevant aspects of the “small”–“large” Fermi surface problem for YbRh$_2$Si$_2$, it was essential to identify appropriate isoelectronic reference systems corresponding to the "small" and "large" Fermi surface limits. Such reference compounds allowed us to obtain clear experimental spectral patterns for both cases, providing a solid basis for comparison and enabling a meaningful discussion of the Fermi surface topology in YbRh$_2$Si$_2$.

While a suitable reference for the "small" Fermi surface, as YbCo$_2$Si$_2$, appeared rather evident~\cite{Monika_YbCoSi_2014, Monika_Compton}, identifying an appropriate reference system for the "large" Fermi surface was a more subtle task.
In this context, Christoph Geibel proposed to consider the compound \linebreak EuRh$_2$Si$_2$, in which Eu has a stable divalent state (Eu$^{2+}$). This material crystallizes in the same structure as  \linebreak YbRh$_2$Si$_2$ and undergoes a transition from the paramagnetic (PM) to the AFM phase below $T_{\mathrm{N}} = 24.5$~K~\cite{Hossain2001JAC, Seiro2011}, where the Eu $4f$ moments, as in several other $\mathit{Ln}$Rh$_2$Si$_2$ compounds, order ferromagnetically within the $ab$ plane, while the corresponding Eu layers stack antiferromagnetically along the $c$ axis. Reflections corresponding to an incommensurate propagation vector $(0,0,\tau)$ along the $c$ axis are clearly observed in resonant magnetic X-ray scattering at the Eu $L_3$ edge below $T_{\mathrm{N}}$~\cite{Chick14}.

The idea behind this proposal followed naturally from the Luttinger Fermi surface sum rule (Luttinger theorem). Applied to the canonical heavy-fermion compound YbRh$_2$Si$_2$, it implied that in the Kondo regime the large $f$-derived Fermi surface should have approximately the same volume as that of the divalent system EuRh$_2$Si$_2$~\cite{Monika_2019}. The reasoning was straightforward. In YbRh$_2$Si$_2$, once the Kondo effect developed, the $4f$ hole became itinerant and contributed to the Fermi volume together with the conduction states. The resulting Fermi surface therefore counted $N$ conduction-band holes plus one additional $f$-derived hole. In EuRh$_2$Si$_2$, in contrast, the $4f$ electrons of Eu$^{2+}$ remained localized and did not participate in the Fermi surface. However, the conduction bands contained $N+1$ holes. Consequently, both isostructural and isoelectronic systems enclosed the same total number of holes in their PM state~\cite{Monika_2019}. According to the Luttinger theorem, this implied that their Fermi surface should have the same volume, making EuRh$_2$Si$_2$ a natural reference system for the large-Fermi-surface configuration of YbRh$_2$Si$_2$.

This simple, and perhaps audacious, proposition creates the tempting opportunity to explore experimentally with ARPES the validity of the Luttinger sum rule for the Yb-based Kondo lattice. Moreover, another exciting issue is to gain insight into how the anticipated large Fermi surface could be modified upon AFM order in a heavy-fermion Yb system. This knowledge is highly relevant for the mysterious AFM phase of YbRh$_2$Si$_2$ below 70~mK, which is a precursor of the quantum critical regime, and superconductivity below 2~mK, which was reported for this system \cite{Schuberth_2016}.

To explore the aforementioned issues, we have performed ARPES in UV and soft X-ray modes on EuRh$_2$Si$_2$ and combined our experiment with \textit{ab~initio} DFT calculations for EuRh$_2$Si$_2$. Our remarkable result is that, both in ARPES and DFT, the Fermi surfaces of both YbRh$_2$Si$_2$ and EuRh$_2$Si$_2$ materials are almost identical in size and shape, even though the masses of the dispersing electrons are heavy in the former, whereas they are light in the latter~\cite{Monika_2019}.

\begin{figure}[h!]
\centering
	\includegraphics[width=0.95\linewidth]{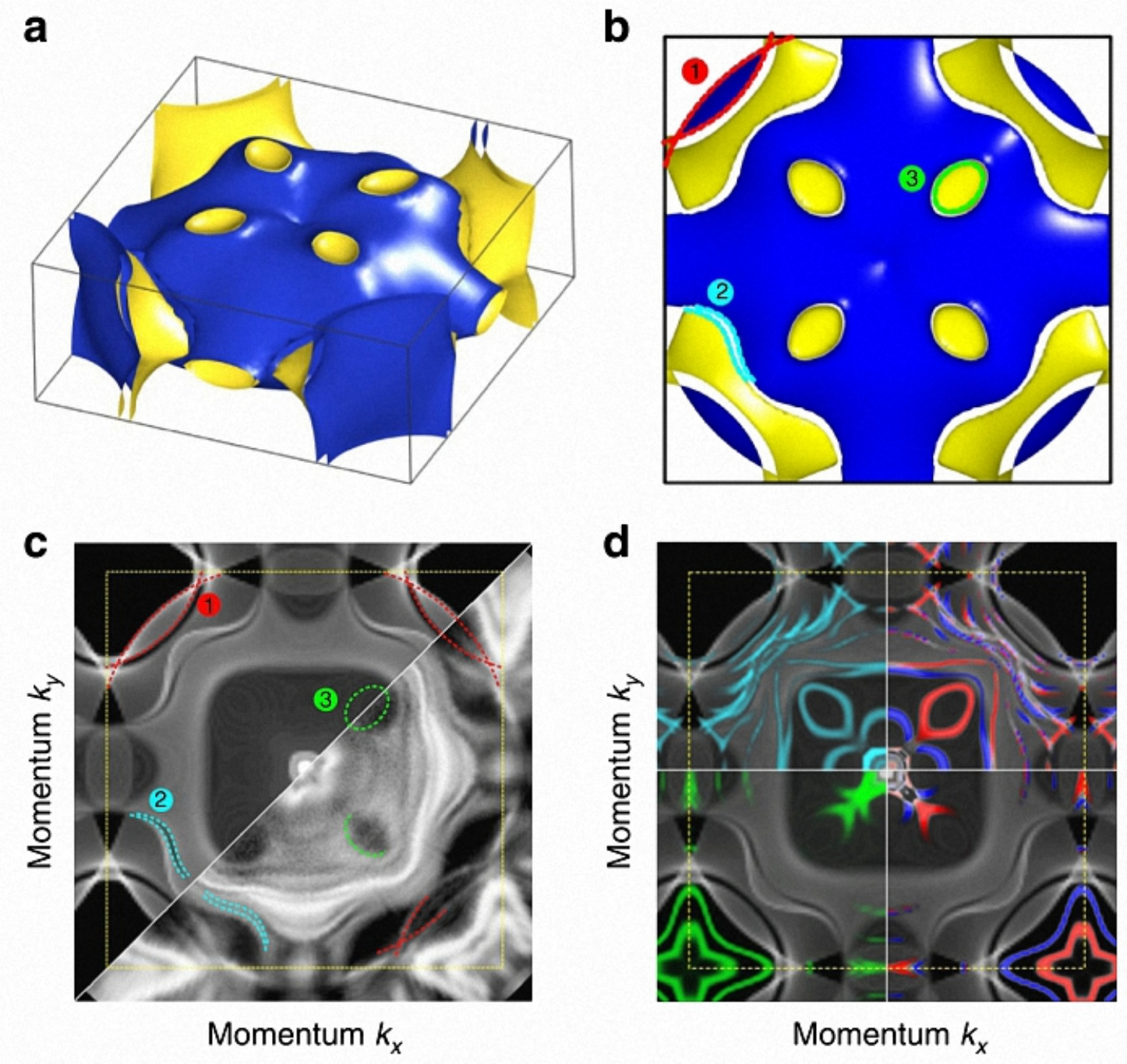}
	\caption{Fermi surface of AFM-ordered EuRh$_2$Si$_2$ derived from UV-ARPES and DFT calculations.
	(a) 3D view and (b) top view of the calculated Fermi surface along $k_z$.
	(c) Comparison between the projected bulk Fermi surface (top left panel) and the ARPES-derived Fermi surface (bottom right panel) taken at 10~K. The yellow dashed line marks the border of the Brillouin zone.
	(d) In addition to the projected bulk Fermi surface, surface states and surface resonances obtained from slab calculations are shown for Eu termination (cyan, top left panel) and Si termination (green, bottom left panel). The spin polarization of the termination-specific surface-related states is indicated in red and blue on the right side~\cite{Monika_2019}.}
	\label{fig:ERS_1}
\end{figure}

In the light of the established similarities in the magnetic correlations between both compounds, EuRh$_2$Si$_2$ becomes a relevant model system to get an idea of how the AFM order might affect the Fermi surface of YbRh$_2$Si$_2$. Both our experimental ARPES results and our calculations show that AFM order leads to significant changes in the Fermi surface of EuRh$_2$Si$_2$, both because of extensive band folding and a large band splitting (see figure~\ref{fig:ERS_1}). These large changes cannot be ignored when discussing the anomalies seen in a number of transport properties at the field induced quantum critical point, which are at the heart of the debate on the nature of this quantum critical point. It is very unlikely that the Hall effect, which has received considerable attention in the debate on this quantum critical point, does not react to such large changes in the Fermi surface. Thus, our results on EuRh$_2$Si$_2$~\cite{Monika_2019} indicate the formation of the AFM state in YbRh$_2$Si$_2$ to be very likely connected with strong changes in the Fermi surface due to band folding and band splitting, which have to be taken into account when analyzing and discussing the anomalous properties observed at the quantum critical point.

The most essential results obtained in studying this problem were the following: (i) our results support the validity of the Luttinger theorem for a Yb-based Kondo lattice and demonstrated how the large Fermi surface, including the strongly correlated hole, should look; (ii) we discussed how the large Fermi surface in YbRh$_2$Si$_2$ might change upon AFM ordering below 70~mK, an enigmatic phase that precedes quantum criticality and superconductivity in this system; and (iii) ARPES measurements revealed the curious property that the itinerant bulk-band states in the AFM phase, consisting of FM layers stacked antiferromagnetically along $c$ axis, became spin-polarized at the FM Eu-terminated surface (see figure~\ref{fig:ERS_1}c,d). This result had not been observed before, was challenging to predict, and appeared to be highly interesting and relevant for technological applications.

\section*{AFM EuRh$_2$Si$_2$: ferromagnetism at the silicide surface}

Another essential result obtained from our ARPES experiments on AFM-ordered EuRh$_2$Si$_2$ was the clear observation of spin splitting of the diamond-like Shockley surface state located around the $\bar{{\rm M}}$ point (see figure~\ref{fig:ERS_1}c,d)~\cite{Chick14}. Note that in YbRh$_2$Si$_2$ and YbCo$_2$Si$_2$ (see figure~\ref{fig:YRS_YBC}b and figure~\ref{fig:YbCoSi}), this surface state remains spin-degenerate or, more precisely, any possible splitting is too small to be resolved, either because of the limited sample quality or the experimental resolution of these particular measurements. Later, using high-quality samples, the Rashba SOC-induced spin splitting of the surface state could be clearly resolved in the PM phase, as will be discussed below.

\begin{figure}[h!]
\includegraphics[width=0.95\linewidth]{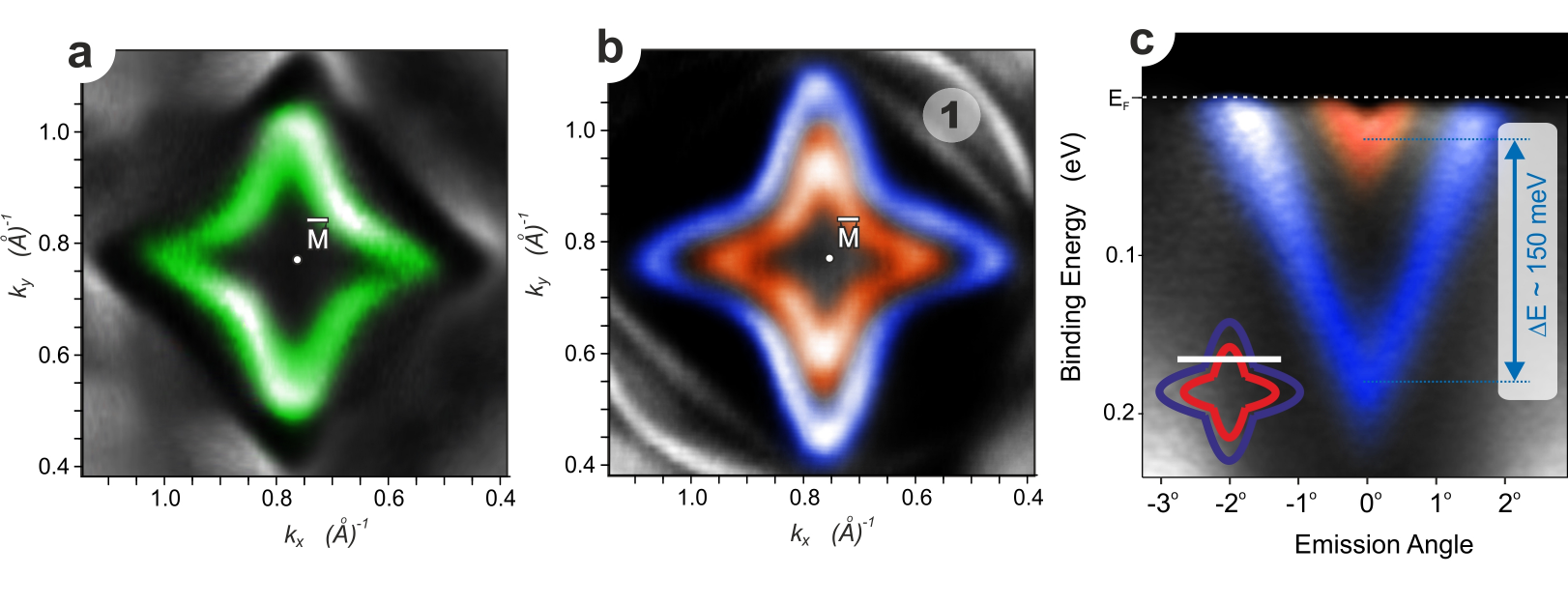}
\caption{Fermi-surface maps taken for the Si-terminated surface of EuRh$_2$Si$_2$ near the $\bar{{\rm M}}$ point at 50~K (a) and at 11~K (b), i.e., above and below the bulk AFM transition at $T_\mathrm{N} = 24.5$~K, respectively. (c) ARPES-derived band map taken at 11~K, demonstrating the maximum splitting of the surface state of 150~meV. The inset schematically indicates the measurement direction~\cite{Chick14}. Color highlighting is used to emphasize the discussed surface electron state. }
\label{fig:ERS_2}
\end{figure}

However, the observation of the aforementioned large spin splitting in AFM-ordered EuRh$_2$Si$_2$ is particularly intriguing because (i) it is evidently linked to the magnetic order originating from Eu (see in figure~\ref{fig:ERS_2}), and (ii) the surface state appears only for the Si-terminated surface, implying that its splitting is influenced by the FM ordered Eu $4f$ moments located in the fourth layer below the Si surface and hidden beneath the Si-Rh-Si surface block~\cite{Chick14}.

\begin{figure*}[t]
\centering
\includegraphics[width=0.95\linewidth]{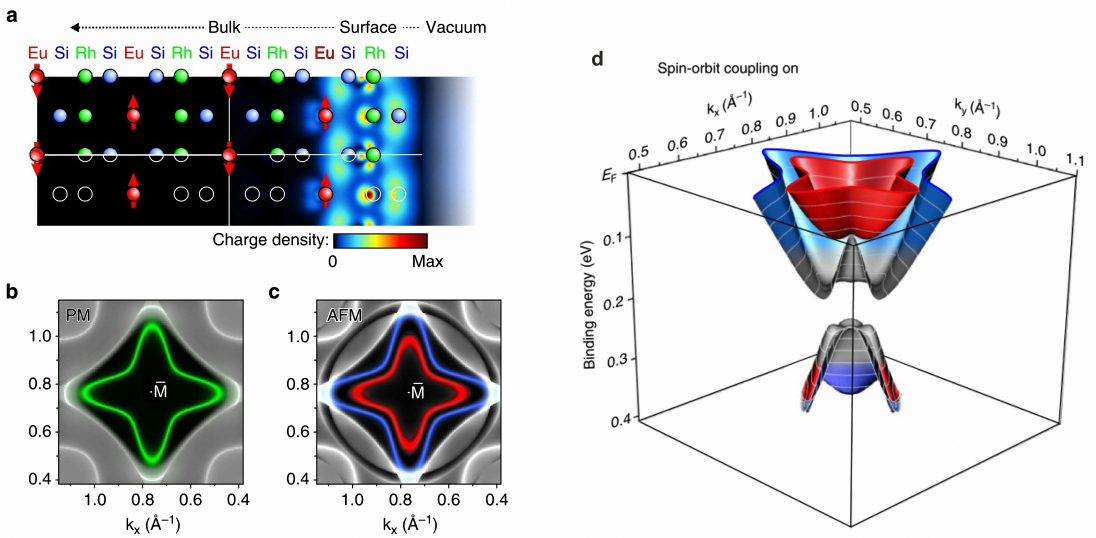}
\caption{(a) Probability-density distribution (projected onto the $ac$ plane) of the surface state at the $\bar{{\rm M}}$ point obtained for the PM phase, superimposed on the slab crystal structure used for the band-structure calculations. (b,c) Calculated Fermi surfaces obtained from bulk and slab calculations for the PM and AFM phases, respectively. The diamond-shaped surface state lies within the large gap (black) of the projected bulk bands (gray) around the $\bar{{\rm M}}$ point. In the PM phase the surface state is shown in green (unsplit), while in the AFM phase it is split and highlighted in red and blue. (d) Three-dimensional representation of the surface state calculated for the AFM phase including spin-orbit coupling. Close to the $\bar{{\rm M}}$ point, spin-orbit interaction opens a gap associated with an avoided band crossing, which leads to a change in the spin character of the bands near the gap edges; these states are indicated in gray~\cite{Chick14}.}
\label{fig:ERS_3}
\end{figure*}

\begin{figure*}[t]
\centering
\includegraphics[width=15cm]{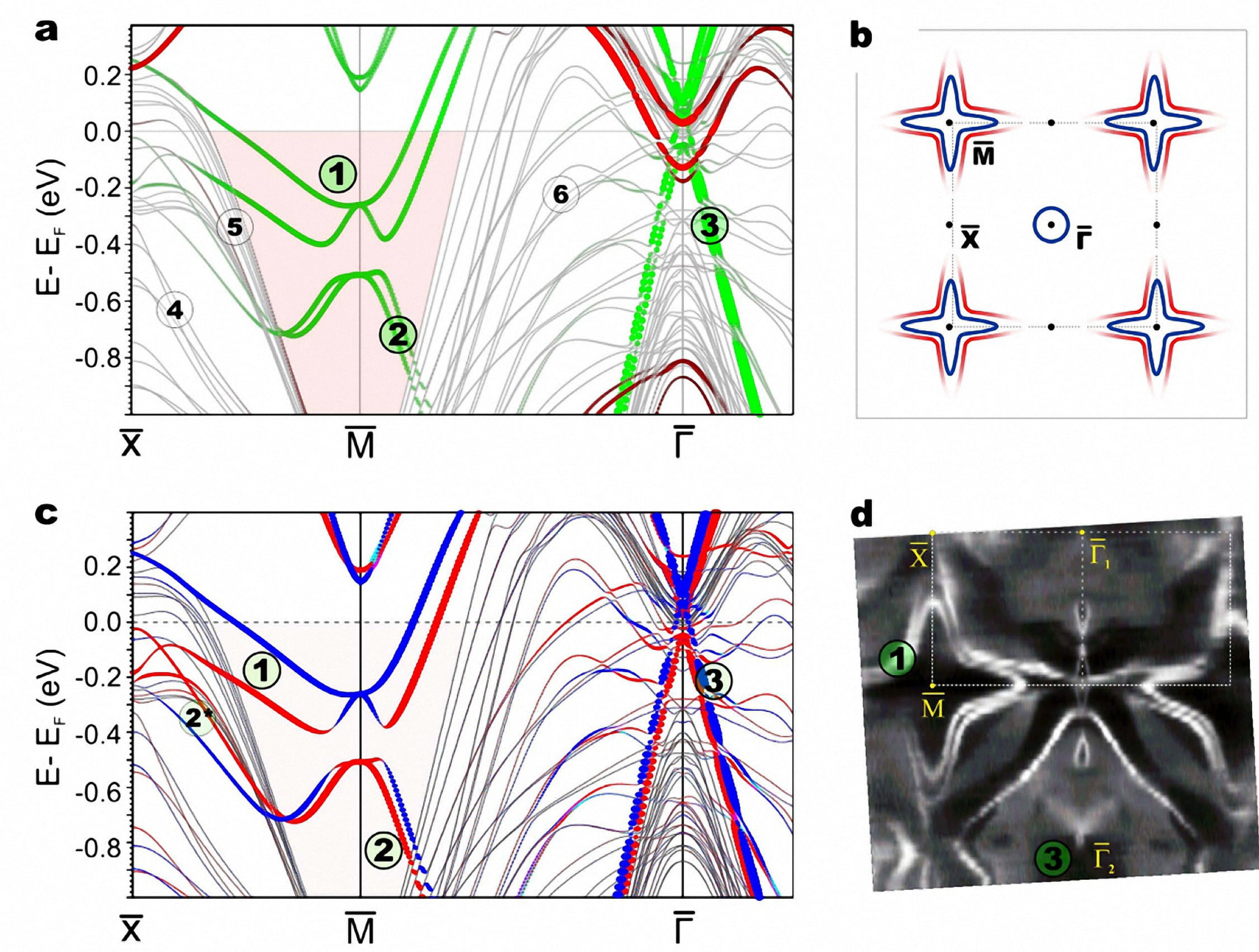}
\caption{Surface-related electronic structure of AFM-ordered GdRh$_2$Si$_2$.
(a) Calculated electronic band structure for a slab of AFM-ordered GdRh$_2$Si$_2$. Surface electronic bands are shown in red for the Gd-terminated surface and green for the Si-terminated surface. The spin-split electron- and hole-like bands of the Shockley surface state at the $\bar{{\rm M}}$-projected band gap are marked as (1) and (2), respectively, while the Dirac-cone bands observed at the $\bar{\Gamma}$ point are labeled as (3). Bulk-like projected bands are shown in gray and labeled as (4), (5), and (6). (b) Schematic view of the Fermi surface for the discussed two-dimensional electron systems (2DESs) at the center of the Brillouin zone and at the $\bar{{\rm M}}$ point. Note that the lower of the two spin-split bands (1) of the Shockley state shown in (a) does not reach the Fermi energy along the $\bar{{\rm M}}$--$\bar{{\rm X}}$ direction. (c) Calculated spin-resolved electronic band structure for the Si-terminated GdRh$_2$Si$_2$ surface. The contribution of the topmost Si--Rh--Si--Gd block to the spin vector components is shown. Majority and minority bands are displayed in red and blue, respectively. The spin-polarized 2DESs are labeled in accordance with panel~a. (d) ARPES-derived Fermi surface of AFM-ordered, Si-terminated GdRh$_2$Si$_2$, measured at a temperature of 1~K using 45~eV photons~\cite{GdRhSi}.}
\label{fig:GRS_1}
\end{figure*}

\begin{figure}[t]
\includegraphics[width=0.98\linewidth]{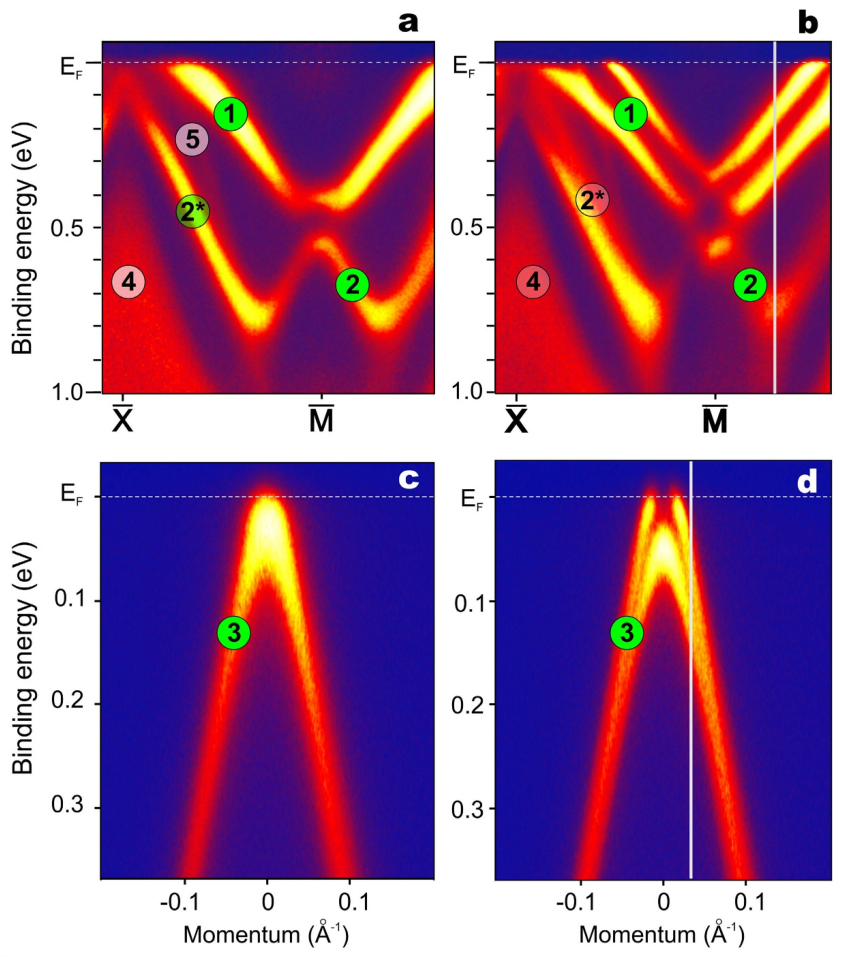}
\caption{Spin splitting of the Shockley surface state and Dirac cone. ARPES data from a Si-terminated GdRh$_2$Si$_2$ sample measured with 55~eV photons. Band maps were recorded near the $\bar{\rm{M}}$ point at 117~K (a) and 19~K (b), and near the $\bar{\Gamma}$ point at 72~K (c) and 19~K (d), along the $\bar{\rm{X}}$–$\bar{\rm{M}}$ and $\bar{\rm{X}}$–$\bar{\Gamma}$ directions, respectively. White vertical lines mark the energy-distribution curves used to analyze the temperature-dependent spin splitting. Surface- and bulk-related features are labeled according to the calculated bands shown in figure~\ref{fig:GRS_1} in Ref.~[\citeonline{GdRhSi}].}
\label{fig:GRS_2}
\end{figure}

The observation of a spin-split surface state at the $\bar{{\rm M}}$ point is in fact remarkable, and as we discuss in detail in the following sections, this state can serve as a sensitive probe of magnetic properties at the surface of the large family of $Ln$$T_2$Si$_2$ compounds. Our analysis of EuRh$_2$Si$_2$ shows that the Si-terminated surface of this AFM can exhibit pronounced FM behavior. This property is transmitted from a magnetically ordered Eu layer located several atomic planes below the surface. The transfer of ferromagnetic properties from such a well-protected subsurface layer to the surface, resulting in spin-split $\bar{{\rm M}}$ surface states, is particularly appealing from the perspective of spintronics, as it suggests the possibility of realizing robust spin-polarized surface states without requiring a perfectly ordered magnetic surface.

ARPES measurements showed that the spin splitting emerges below $\sim 41~$K and saturates at approximately 150~meV upon cooling. Here we note that in our first experiments~\cite{Chick14} the onset of the splitting was estimated to be about 32.5~K. However, subsequent measurements performed on higher-quality single crystals revealed that the splitting already appears at $\sim$41~K\cite{Generalov_NanoLett_2017}, illustrating the importance of crystal quality for such ARPES studies. In figure~\ref{fig:ERS_2} we show the ARPES patterns revealing the surface state around the $\bar{{\rm M}}$ point, taken above (panel a) and below (panel b) the bulk AFM transition at $T_{\mathrm{N}} = 24.5$~K, respectively. The giant spin splitting of this state is clearly resolved in the 11~K data (panel c). Figure~\ref{fig:ERS_3} provides $ab~initio$ insight into the surface- and bulk-related electronic states in EuRh$_2$Si$_2$. The calculations describe well the discussed surface state and reveal that it is predominantly localized within the surface Si-Rh-Si-Eu block~\cite{Chick14}.

An essential distinction of the present case from most previously reported examples of surface ferromagnetism is that the magnetic properties do not originate directly from the outermost atomic layer, which is typically highly sensitive to surface disorder or contamination. Instead, the surface state and its spin splitting are observed for the Si-terminated surface of EuRh$_2$Si$_2$, while the surface-related ferromagnetism itself is induced by the ordered Eu $4f$ moments located three atomic layers beneath the surface. These moments are effectively shielded from surface perturbations by the Si–Rh–Si overlayer. These results suggest that the mechanism responsible for the formation of surface ferromagnetism in EuRh$_2$Si$_2$ may also operate in other antiferromagnetic metallic or semiconducting compounds in which surface states exist within an energy gap at the Fermi level. In such systems, the exchange splitting of surface states could induce magnetization in functional surface layers, for example, in topological insulators or Rashba-type surface systems deposited on the antiferromagnetic substrate, thereby lifting the spin degeneracy of the corresponding surface states. More generally, the spin splitting of surface states may serve as a sensitive probe of ferromagnetic properties in the (sub)surface region of rare-earth intermetallic materials. In this context, ARPES provides a unique advantage over more conventional techniques, such as X-ray magnetic circular dichroism, which typically probe a much larger depth and may therefore be less sensitive to the magnetic order of the first buried $4f$ layer.

As we see, the spin splitting emerges already below $\sim$41~K, while the bulk becomes antiferromagnetically ordered only below $T_\mathrm{N} = 24.5$~K. This indicates the presence of pronounced two-dimensional ferromagnetic behavior at the Si-terminated surface in a temperature range where the bulk remains paramagnetic. Such a situation is particularly intriguing, as the surface exhibits clear ferromagnetic character while the bulk is either nonmagnetic due to its PM state above $T_\mathrm{N}$ or has a vanishing net magnetization in the antiferromagnetic phase. This combination makes the system especially interesting from the perspective of potential applications, where a nonmagnetic bulk coexists with robust two-dimensional ferromagnetism at the surface. Another interesting aspect is that one would normally expect the magnetic ordering temperature at the surface to be lower than in the bulk, simply because of the reduced coordination and the smaller number of neighboring Eu $4f$ moments available for exchange coupling. However, the situation here is exactly the opposite. The surface-related magnetic splitting appears at a higher temperature than the bulk AFM ordering. This observation suggests that the exchange coupling within the Eu planes is significantly stronger than along the $c$ axis, implying a pronounced two-dimensional character of the magnetic interactions in this system.

\section*{AFM GdRh$_2$Si$_2$ and HoRh$_2$Si$_2$: Surface-related ferromagnetism and its temperature scales}

Having observed the peculiar surface-related ferromagnetic properties in EuRh$_2$Si$_2$, it was natural to analyze the underlying mechanisms in more detail and to extend the investigation to the closely related $LnT$$_2$Si$_2$ AFM compounds~\cite{Kliemt_CRT_2020}, and as the next system we decided to consider GdRh$_2$Si$_2$~\cite{GdRhSi}. The expectations are rather straightforward. In EuRh$_2$Si$_2$, europium is present in the divalent Eu$^{2+}$ configuration with a half-filled $4f^7$ shell, carrying a large local magnetic moment of about $7\,\mu_{\mathrm{B}}$ per atom. In contrast, Gd$^{3+}$ possesses an additional $5d$ conduction electron compared with Eu$^{2+}$. This difference suggests that the exchange interaction between the localized $4f$ moments and the itinerant electrons may be stronger in GdRh$_2$Si$_2$. Consequently, one may expect that the magnetic effects observed at the surface of EuRh$_2$Si$_2$ could become more pronounced in GdRh$_2$Si$_2$, potentially leading to stronger or more robust magnetic signatures. This expectation is already supported by the comparison of their bulk properties: the Néel temperature of GdRh$_2$Si$_2$ is about 107~K~\cite{GdRhSi}, i.e., almost four times higher than that of the Eu-based system~\cite{Chick14}. Note that in the AFM phase, below 107~K, the Gd $4f$ moments become ferromagnetically ordered within the $ab$ plane, while the corresponding Gd layers stack antiferromagnetically along the $c$ axis\cite{GdRhSi,Kristin_GRS,Vyazovskaya_2019,Windsor2020}.

In figure~\ref{fig:GRS_1}, we show the surface electronic band structure obtained from \textit{ab~initio} calculations for AFM ordered GdRh$_2$Si$_2$~\cite{GdRhSi}. To separate surface-related states from bulk electronic bands, we used a thick slab terminated by Gd on one side and Si on the other, allowing us to identify bulk-like bands, projected band gaps, and surface states for both terminations. The two-dimensional states associated with the Si-terminated surface are shown in green, while those for the Gd-terminated surface are highlighted in red.

Figure~\ref{fig:GRS_1}b schematically illustrates the Fermi surface of the relevant two-dimensional states. The calculated spin texture for the Si-terminated surface is shown in figure~\ref{fig:GRS_1}c, while figure~\ref{fig:GRS_1}d presents the ARPES-derived Fermi surface of Si-terminated GdRh$_2$Si$_2$ measured at 1\,K. The non-symmetrized ARPES data reveal the evolution of photoemission intensity caused by matrix-element effects when moving from the first to the second Brillouin zone.

The spectral features are similar to those observed in EuRh$_2$Si$_2$. For the Si termination, pairs of strongly dispersive bands appear within the large projected band gap around the $\bar{{\rm M}}$ point, labeled as (1) and (2) in figure~\ref{fig:GRS_1}a,c. Both majority and minority states are degenerate at $\bar{{\rm M}}$ and open a gap due to spin-orbit coupling. These features originate from Shockley surface states confined to the topmost atomic layers of the Si-terminated crystal. A similar spin splitting of this surface state is observed as in the case of AFM EuRh$_2$Si$_2$, indicating that this behavior is intrinsic and appears to be a common feature of the Rh$_2$Si$_2$ surface electronic structure. As expected, the Shockley surface state is absent for the Gd-terminated surface.

Next, we summarize the ARPES results before turning to the analysis of the temperature scales derived from the spin splitting of the Shockley surface state and the Dirac cone. We then compare these scales with those obtained from the XMLD experiment to distinguish between surface- and bulk-related magnetic behavior.

Figure~\ref{fig:GRS_2} shows temperature-dependent ARPES spectra of the Si-terminated GdRh$_2$Si$_2$ surface near the $\bar{\Gamma}$ and $\bar{{\rm M}}$ points of the surface Brillouin zone. The bulk Néel temperature $T_\mathrm{N} \approx 107$~K marks the onset of in-plane ferromagnetic alignment of the Gd $4f$ moments, which stack antiferromagnetically along the $c$ axis~\cite{GdRhSi}. To ensure a paramagnetic starting point, the crystal was cleaved at 120~K, well above $T_\mathrm{N}$. The Si termination is readily identified by the intense Shockley surface state at the $\bar{{\rm M}}$ point, which is absent for Gd termination. The high-temperature band map in the PM state in figure~\ref{fig:GRS_2}a shows the Shockley surface state without spin splitting. Upon cooling, a clear splitting develops. Monitoring an energy-distribution curve at $k_{\parallel}$ near one third of the $\bar{{\rm X}}$--$\bar{{\rm M}}$ distance (figure~\ref{fig:GRS_2}b) reveals the onset of the splitting around $\sim 90$~K, which increases and reaches about 160~meV at low temperature. A similar analysis for the Dirac-cone states at the $\bar{\Gamma}$ point shows that the splitting emerges around $\sim 60$~K. As seen, it is not yet resolved at 72~K (figure~\ref{fig:GRS_2}c) and becomes clearly visible at 19~K with a maximum value of about 70~meV (figure~\ref{fig:GRS_2}d).

\begin{figure}[h!]
\centering
\includegraphics[width=0.80\linewidth]{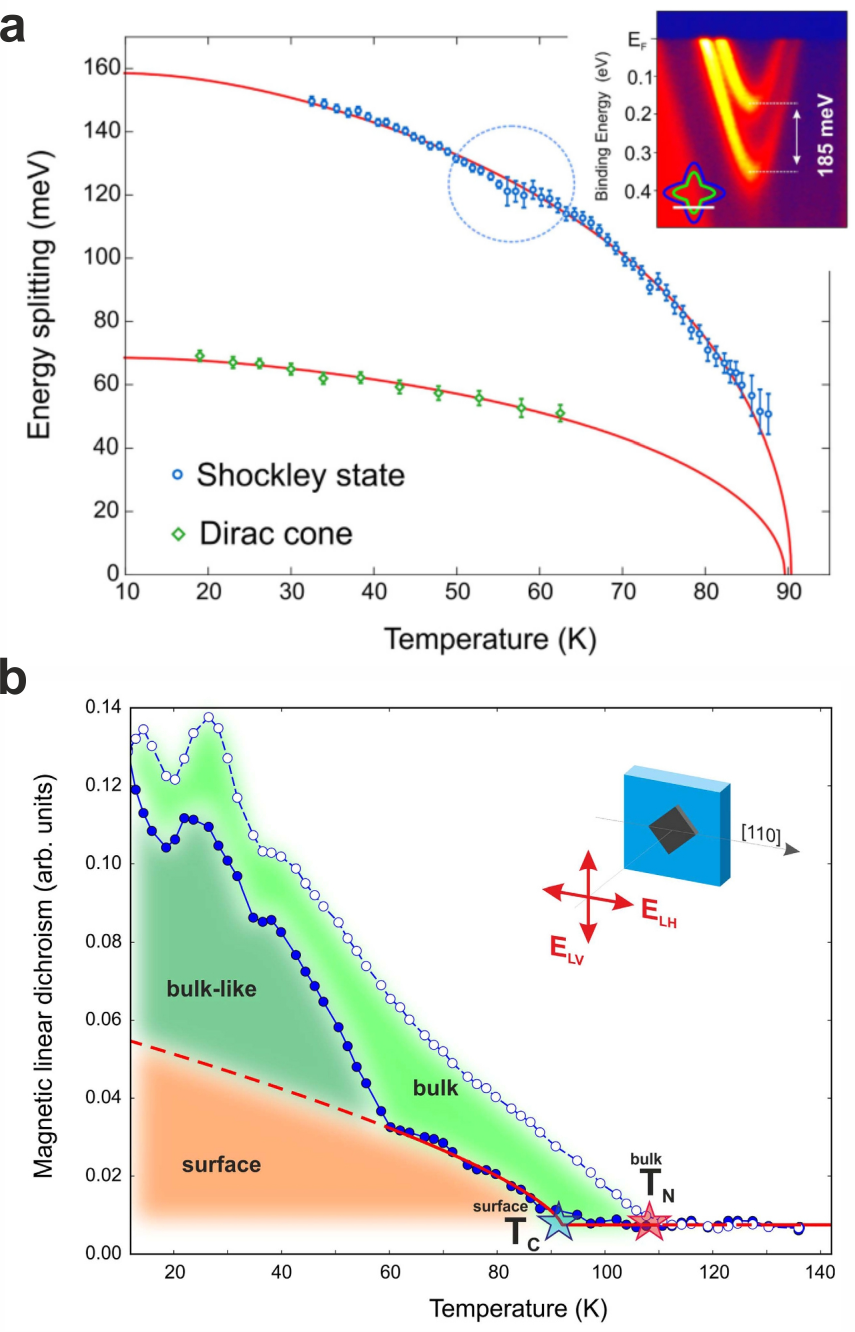}
\caption{Temperature scales from ARPES and XMLD.
(a) ARPES-derived temperature evolution of the spin splitting of the Shockley surface state and the Dirac cone. The solid lines represent fits to both data sets obtained using the Weiss molecular-field approximation to the Heisenberg model~\cite{Getzlaff2008}. The inset shows the maximum spin splitting of the Shockley state, reaching 185~meV, and schematically illustrates the measurement direction.
(b) Temperature dependence of the XMLD signal measured in fluorescence yield (TFY, open symbols) and total electron yield (TEY, solid symbols). TFY probes the bulk magnetization, whereas TEY is more sensitive to the surface region. The dashed lines serve as guides to the eye. The red solid line shows the fit obtained using the Weiss molecular-field approximation to the Heisenberg model~\cite{Getzlaff2008}. The inset schematically illustrates the experimental geometry of the XMLD experiment~\cite{GdRhSi}.}
\label{fig:GRS_3}
\end{figure}

Figure~\ref{fig:GRS_3}a summarizes the detailed temperature dependence of the spin splitting of the Dirac cone and the Shockley surface state derived from ARPES. For the Shockley surface state, the spin splitting emerges at $\sim 90$~K and rapidly increases upon cooling, reaching about 160~meV. At higher temperatures the splitting becomes smaller than the lifetime and instrumental broadening and therefore cannot be resolved. The temperature dependence shows a noticeable kink around 60~K. The splitting is anisotropic, ranging from about 160~meV up to a maximum of 185~meV along the $\bar{\rm X}$--$\bar{\rm M}$ direction slightly away from the $\bar{{\rm M}}$ point. The experimentally observed values are consistent with our theoretical calculations. On average, the splitting is somewhat larger than that reported for EuRh$_2$Si$_2$ ($\sim150$~meV~\cite{Chick14}), although the difference is smaller than might be expected. As we mentioned, the splitting of the Dirac cone becomes clearly resolved only below $\sim 60$~K and remains less than half of that observed for the Shockley state. Since the Dirac cone has mainly Rh $4d$ character, spin-orbit coupling contributes to the observed splitting. Calculations estimate the SOC contribution near the $\bar{\Gamma}$ point to be about 40–45~meV, comparable to the measured splitting around 60~K. As SOC is temperature independent, it may dominate the observed splitting above $\sim 65$~K and even persist above $T_\mathrm{N}$. Nevertheless, fitting the data within the Weiss molecular-field framework yields a critical temperature of about 90~K, similar to that obtained for the Shockley state. This indicates that the ARPES-observed splitting is primarily driven by magnetic exchange interactions, while SOC partially masks the effect in the Dirac cone.

\begin{figure}[t]
\centering
\includegraphics[width=0.98\linewidth]{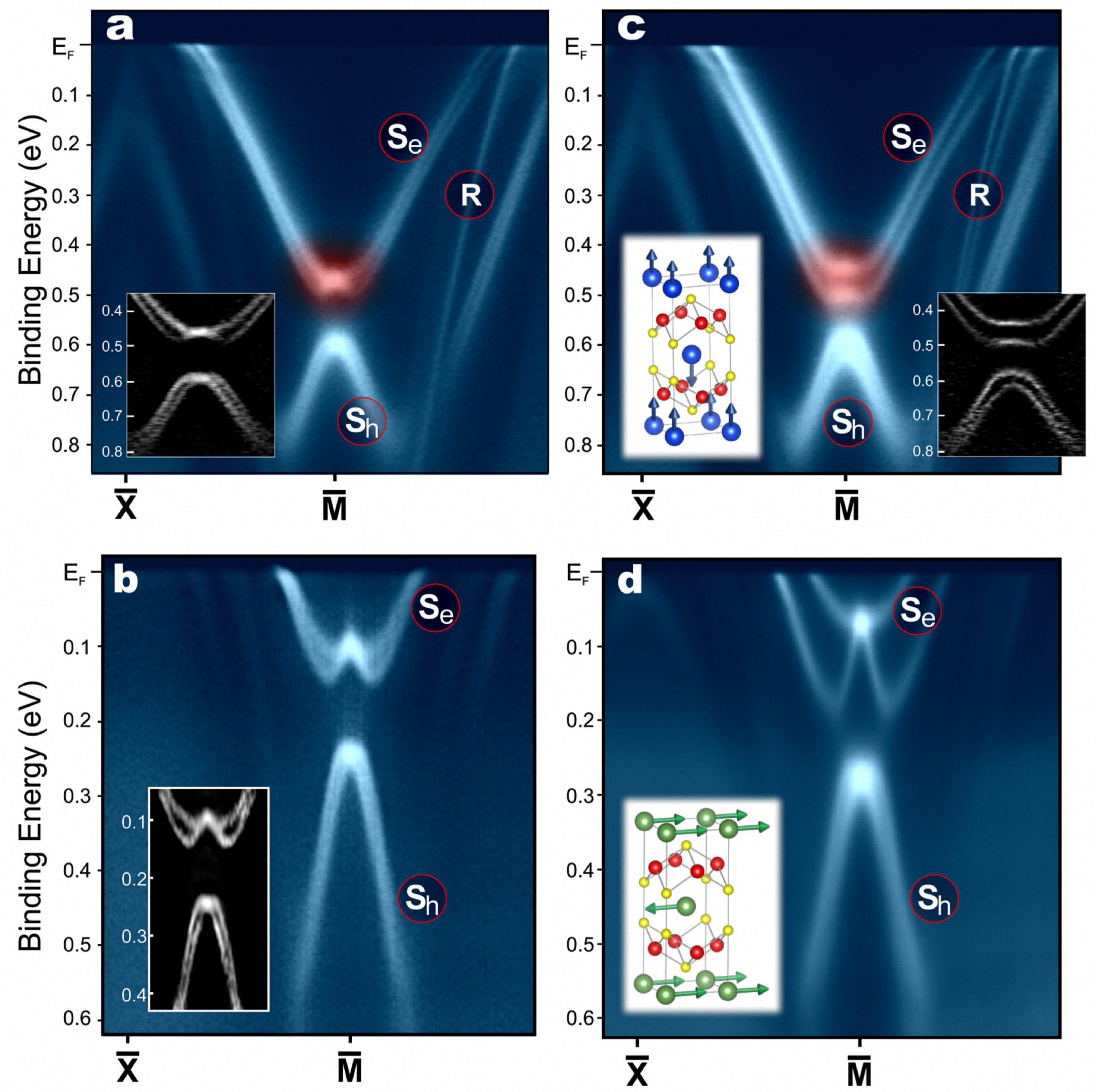}
\caption{Comparison of ARPES data from EuRh$_2$Si$_2$ and HoRh$_2$Si$_2$. The key observation is the spin splitting of the surface state at the $\bar{{\rm M}}$ point in the AFM-ordered phase of HoRh$_2$Si$_2$. ARPES intensity maps measured along the $\bar{{\rm X}}-\bar{{\rm M}}-\bar{{\rm X}}$ direction for HoRh$_2$Si$_2$ (a,c) and EuRh$_2$Si$_2$ (b,d), recorded above (a,b) and below (c,d) the transition into the AFM ordered state. Insets in panels (a) and (b) show the second derivatives of the respective ARPES maps in the vicinity of the $\bar{{\rm M}}$ point, highlighted by red-like shading. The data for HoRh$_2$Si$_2$ were taken at 40~K (a) and 15~K (c), and for EuRh$_2$Si$_2$ at 55~K (b) and 10~K (d). The insets in the lower left corners of panels (c) and (d) show the corresponding unit cells with the orientation of the $4f$ magnetic moments on the rare-earth atoms indicated by arrows. Yellow (red) spheres represent Si (Rh) atoms~\cite{Generalov_NanoLett_2017}.}
\label{fig:HRS_1}
\end{figure}

To further clarify the origin of the temperature-depen- dent spin splittings, we performed XMLD measurements at the Gd $M_5$ edge (see the summary presented in figure~\ref{fig:GRS_3}b). XMLD is sensitive to both FM and AFM order, as the signal is defined by $\langle M_J^2\rangle$, which is independent of the sign of the moment. By simultaneously recording the total electron yield (TEY) and total fluorescence yield (TFY), we probed the magnetic order near the surface and in the bulk, respectively.

The TFY data reveal the onset of antiferromagnetic order in the bulk at $\sim107$~K, consistent with the bulk Néel temperature. In contrast, the TEY signal indicates that magnetic order at the surface emerges only around $\sim 90$~K, in excellent agreement with the onset of the Shockley-state spin splitting observed by ARPES. Both TEY and ARPES curves exhibit a distinct kink near ${\sim 60}$~K.

Thus, the combined ARPES and XMLD results on GdRh$_2$Si$_2$  reveal distinct temperature scales for magnetic order at the surface and in the bulk of this model AFM system. Once the Si-terminated surface becomes FM ordered, it remains largely decoupled from the bulk over a finite temperature range. Coupling sets in only when an additional exchange channel becomes active. Our results suggest that this channel is possibly opened by the Dirac-cone state, which strongly hybridizes with the bulk states (see in figure~\ref{fig:GRS_1}a,c) and can mediate exchange coupling between surface and bulk magnetic moments~\cite{GdRhSi}.

These findings establish the ferromagnetic Si-termi- nated surface of GdRh$_2$Si$_2$ as a promising model platform for engineering magnetically active 2D electron systems and exploring novel electronic and magnetic functionalities in 2D materials~\cite{GdRhSi}.

Additional magnetic functionalities of the Si–Rh–Si–$Ln$ structural block emerge from ARPES experiments on the related AFM compound HoRh$_2$Si$_2$ (see figure~\ref{fig:HRS_1}) \cite{Generalov_NanoLett_2017}. The essential difference from the previously considered materials is that in the AFM phase the Ho $4f$ magnetic moments are ordered along the $c$ direction and exhibit a small, gradual canting away from the $c$ axis below a certain temperature~\cite{Kliemt_CRT_2020, Generalov_NanoLett_2017}, whereas in the Gd- and Eu-based materials they are aligned within the $ab$ plane.

The results obtained from our ARPES experiments may be of interest for the community working on the realization of spin- or magnetically active transistors, whose implementation remains a challenge. A common strategy is to create highly mobile 2D electron systems at surfaces or interfaces and to control their spin and energy. In many existing systems, however, spin-dependent properties require the application of an external magnetic field. HoRh$_2$Si$_2$ can serve as a model compound for such studies, since at the Si-terminated surface the interplay between spin-orbit coupling and exchange magnetic interaction of the 2D states can be tuned solely by temperature.

As shown for Bi$_2$Te$_3$ in Ref.~[\citeonline{Chulkov_2012}], the Dirac surface state is protected by time-reversal symmetry and therefore remains degenerate at the $\Gamma$ point in the nonmagnetic case. When the magnetic moments are oriented perpendicular to the surface (${\bf M} \parallel {\bf e}_z$), the exchange field breaks time-reversal symmetry and lifts the degeneracy of the two spin states at this high-symmetry point, opening a gap at the Dirac point. In contrast, for an in-plane magnetization the Dirac cone is mainly shifted in momentum space rather than gapped. A similar behavior is observed by ARPES for the $\bar{{\rm M}}$ surface states of AFM HoRh$_2$Si$_2$ (see figure~\ref{fig:HRS_1}c).

\begin{figure*}[t]
\centering
\includegraphics[width=0.98\linewidth]{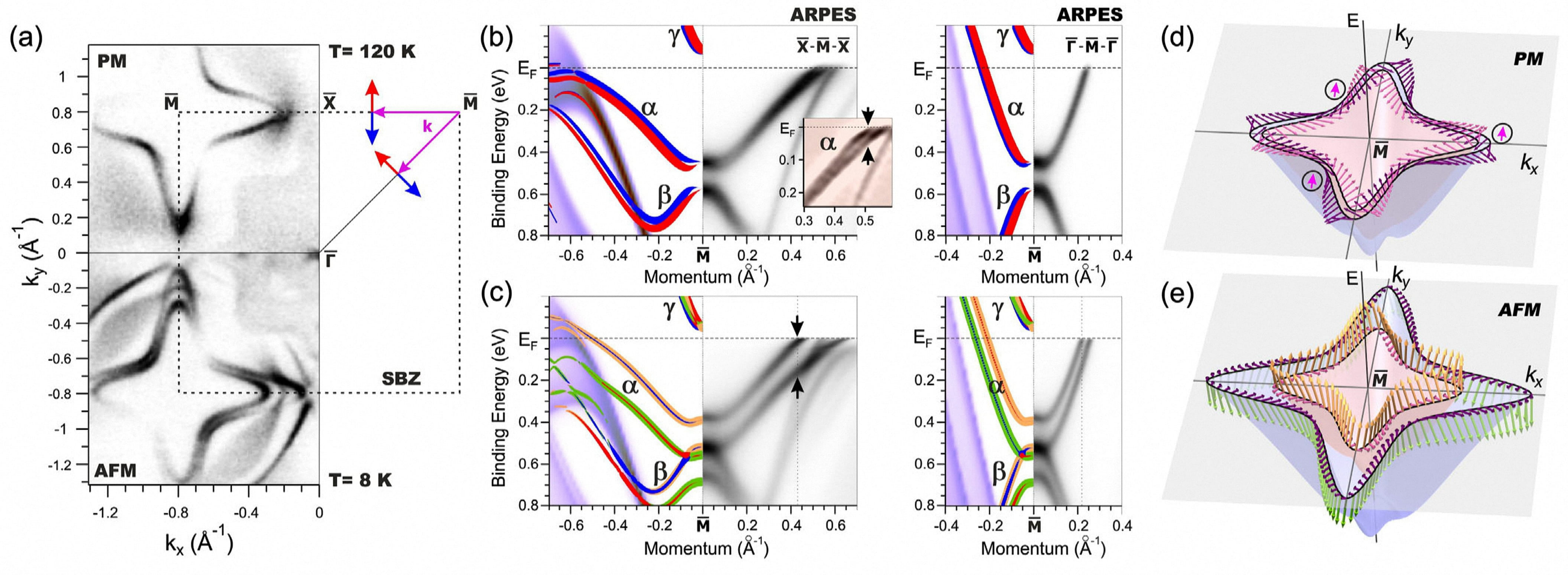}
\caption{Key results obtained for TbRh$_2$Si$_2$.
(a) Fermi contours obtained by ARPES for the PM and AFM phases. Red and blue arrows indicate the color scheme used for the in-plane spin components derived from DFT calculations shown in panels (b) and (c). Calculated and measured band structures for the Si-terminated TbRh$_2$Si$_2$ in the PM (b) and AFM (c) phases are displayed. The inset in (b) shows the spin-orbit splitting of the $\alpha$ state; black arrows indicate the 35\,meV splitting. Black arrows in (c) mark the 140\,meV splitting of the $\alpha$ state in the AFM phase. The violet–brown palette represents the calculated surface-projected bulk states. For the surface states, the in-plane spin polarization is shown in red and blue, while yellow and green denote the positive and negative $S_z$ components, respectively. Spin-resolved \textit{ab~initio} constant-energy contours for the $\alpha$ state in the PM (d) and AFM (e) phases, calculated at a binding energy of 0.23\,eV. The in-plane spin orientation $S_k$ is indicated by pink (purple) arrows for the inner (outer) contour. In the AFM phase, the color of the spin vector $\mathbf{S}=S_k+S_z\hat{z}$ represents the sign of $S_z$, as in figure~1c~\cite{Usachov_PRL_2020}.}
\label{fig:TbRS_1}
\end{figure*}

Thus, by tuning the temperature across the magnetic phase transitions (AFM ordering and the onset of the $4f$ moment canting), the spin configuration of the 2D electrons can be controlled~\cite{Generalov_NanoLett_2017}. In the high-temperature phase the spins are confined within the surface plane by the Rashba SOC, whereas in the AFM phase they tend to align along the direction of the ordered $4f$ moments, i.e., perpendicular to the surface or tilted with respect to it. Temperature-dependent ARPES experiments~\cite{Generalov_NanoLett_2017} show how the corresponding level splittings evolve with temperature. In combination with \textit{ab~initio} calculations and modeling, these splittings were related to different spin textures. Depending on whether Rashba SOC or the exchange magnetic interaction dominates, the spins either remain within the $ab$ surface plane or tilt toward the out-of-plane ordered $4f$ moments~\cite{Generalov_NanoLett_2017}.

\section*{Cubic Rashba effect disclosed for the AFM compound TbRh$_2$Si$_2$}

It is worth noting that for the $Ln$Rh$_2$Si$_2$ materials discussed above, where $Ln$ = Eu, Gd, and Ho, spin-resolved ARPES measurements have not been performed. The information on the spin properties of the two-dimensional states, including the Dirac cone at $\bar{\Gamma}$ and the $\bar{{\rm M}}$ surface state, has therefore been inferred from a combination of \textit{ab~initio} calculations and conventional ARPES measurements. Already at an early stage, our analysis of the DFT-derived spin structure of the $\bar{{\rm M}}$ surface state in GdRh$_2$Si$_2$ appeared puzzling. The theoretical results showed that the spin texture of this surface state in the PM phase, driven by spin-orbit coupling, differs markedly from that expected for the classical Rashba effect. In particular, the spin structure exhibits an unusual texture characterized by a triple winding of the electron spin along the constant-energy contour. In contrast, for the conventional Rashba effect the spin rotates only once along the Fermi contour, forming the well-known helical spin texture with spins oriented tangentially to the momentum and confined within the surface plane~\cite{LaShell_1996}.

We explored this unusual property further in the\linebreak intermediate-valence material EuIr$_2$Si$_2$~\cite{Susanne2019, Usachov_PRB.101.245140}, which will be discussed in detail later, and in the AFM compound TbRh$_2$Si$_2$~(see ARPES and DFT results in figure~\ref{fig:TbRS_1} and the full story in Ref.~[\citeonline{Usachov_PRL_2020}]). In this system, similar to the Ho case, the Tb $4f$ moments order along the $c$ direction in the AFM phase. The experimental results, including spin-resolved ARPES measurements, together with detailed theoretical analysis by our collaborators based on a $\boldsymbol{k}\cdot\boldsymbol{p}$ model~\cite{Nechaev_2018, Usachov_PRL_2020} and calculations of the spin photocurrent within the \textit{ab~initio} one-step theory of photoemission, consistently confirmed the “triple-winding” spin structure of the $\bar{{\rm M}}$ surface state~\cite{Usachov_PRL_2020} and provided a natural explanation in terms of a cubic Rashba effect. In this scenario, the spin splitting originates from higher-order terms in $k$ in the effective $\boldsymbol{k}\cdot\boldsymbol{p}$ model Hamiltonian, giving rise to a characteristic spin texture with a threefold winding of the electron spin along the constant-energy contour~\cite{Usachov_PRL_2020}.

An important outcome of the work on TbRh$_2$Si$_2$ is that the unusual “triple-winding” spin structure proves to be remarkably robust. It persists when the system undergoes the transition from the PM to the AFM phase, where the Tb $4f$ moments become ordered perpendicular to the surface and generate a strong exchange field. The “triple-winding” spin texture remains intact even though the spin-orbit coupling at the Rh atoms is substantially weaker than the out-of-plane exchange field produced by the Tb $4f$ moments.

Since TbRh$_2$Si$_2$ also belongs to the large family of materials crystallizing in the tetragonal ThCr$_2$Si$_2$ structure~\cite{Kliemt_CRT_2020}, we therefore anticipate that the discovered cubic Rashba effect can be further explored within this broad class of compounds, which provides a versatile platform for investigating its properties in detail. For instance, the strength of the spin-orbit coupling can be tuned by varying the transition-metal element. The orientation and magnitude of the exchange magnetic field may be manipulated through the rotation of the $4f$ moments via coupling to the CEF, for example by employing Dy or Ho as the rare-earth element. As we know, several members of the ThCr$_2$Si$_2$ family exhibit Kondo physics, heavy-fermion behavior, or intermediate-valence properties. The presence of these interactions offers additional degrees of freedom to control the unusual surface magnetism at the Si-terminated surface and, in particular, to further explore phenomena related to the cubic Rashba effect~\cite{Usachov_PRL_2020, Manchon_2022}.

\begin{figure*}[t]
\centering
\includegraphics[width=16cm]{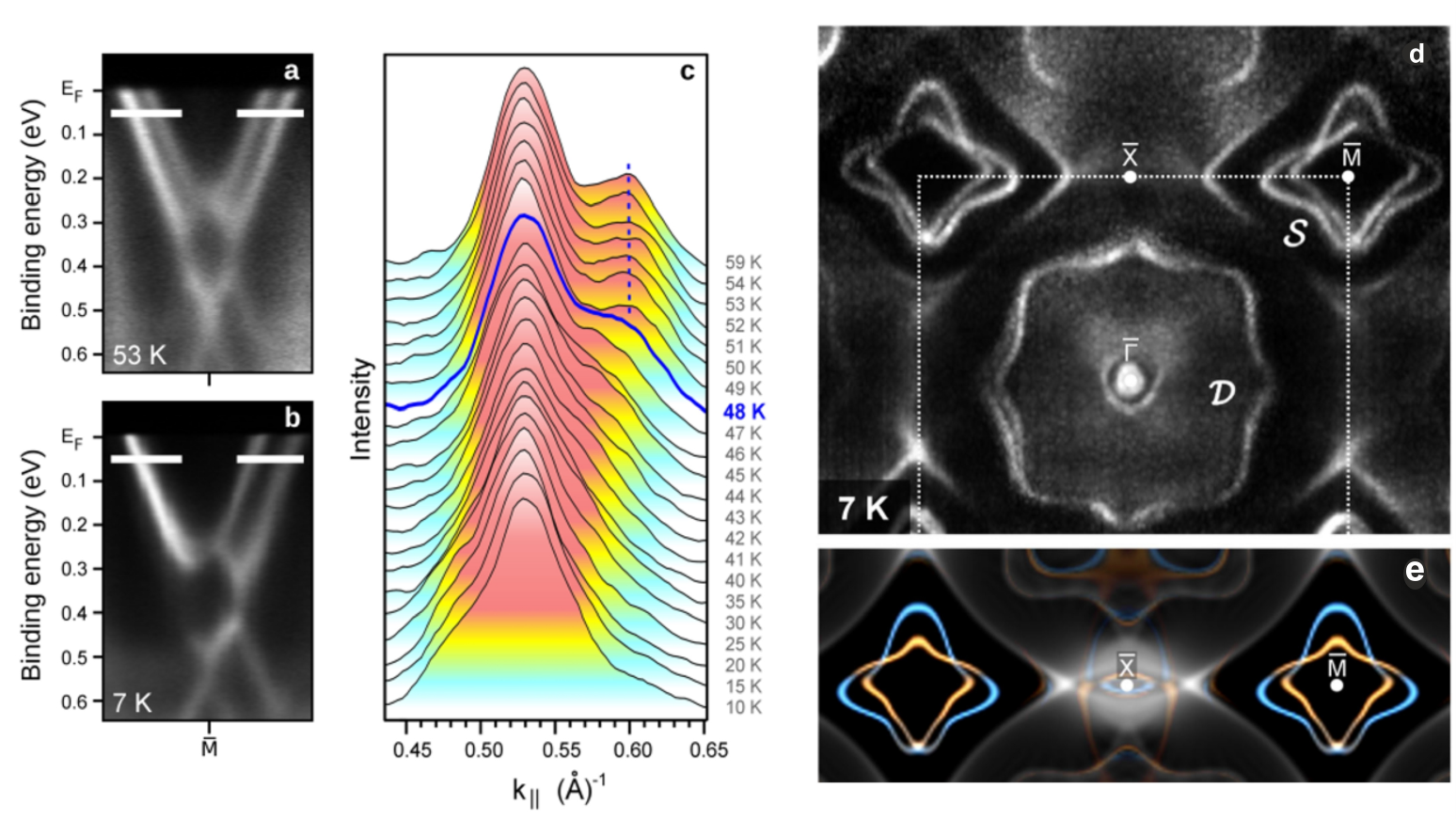}
\caption{Insight into ferromagnetism at the Si-terminated EuIr$_2$Si$_2$ surface. ARPES band maps of the $\bar{{\rm M}}$ surface state taken at (a) 53~K and (b) 7~K along the $\bar{{\rm X}}-\bar{{\rm M}}$ direction, where the asymmetry develops at low temperatures. White bars indicate the momentum distribution curves that were tracked as a function of temperature to extract the magnetic transition temperature. The temperature evolution of the momentum distribution curves is shown for (c) the left branch of the electron-like band. The emergence of asymmetry in the surface state is reflected by a drop in intensity and a change in peak separation at 48~K, marking the onset of magnetic order. (d) ARPES-derived Fermi surface of EuIr$_2$Si$_2$ taken from the Si-terminated surface at 7~K in the ferromagnetically ordered phase of the Si surface. (e) Calculated Fermi surface shown as a superposition of projected bulk and slab band structures. Color highlighting is used to emphasize the surface-related electronic states. For the bulk, the mean $4f$ occupancy was set to $n_{4f}=6.2$, corresponding to the experimentally determined valence $\nu = 2.83$ at 7~K~\cite{Susanne2019}.}
\label{fig:EIS_1}
\end{figure*}

\section*{Strong spin-orbit coupling, 2D ferromagnetism, and Kondo physics in $\boldsymbol{Ln}$Ir$_2$Si$_2$}

In discussing the unusual surface-related properties in which exchange magnetic interactions and spin-orbit coupling are both involved and closely intertwined, we considered that it would be particularly interesting to enhance the spin-orbit coupling by introducing a heavier element into the system. Following this line of thought, we turned to the $Ln$Ir$_2$Si$_2$ family and decided to explore in detail the surface-related properties of the intermediate-valence compound EuIr$_2$Si$_2$~\cite{Susanne2019, EIS_2020}, the Kondo lattice system YbIr$_2$Si$_2$~\cite{Generalov_SISY_2018}, and the antiferromagnet GdIr$_2$Si$_2$~\cite{Schulz_GdIr_2021}. This approach allowed us to investigate the behavior of two-dimensional states in situations where strong spin-orbit coupling is combined either with 2D ferromagnetism or Kondo physics.

\section*{Intermediate-valent EuIr$_2$Si$_2$: 2D ferromagnetism and strong spin-orbit coupling}

We now turn to EuIr$_2$Si$_2$~\cite{Susanne2019}, a compound discovered in 1986~\cite{Chevalier1986}, whose bulk properties have been extensively investigated and are well established. In the bulk, EuIr$_2$Si$_2$ exhibits a valence-fluctuating Eu state and therefore remains nonmagnetic down to the lowest tempera- \linebreak tures~\cite{Chevalier1986,Seiro2011,Seiro2019}.

Our ARPES measurements reveal that the Si-termi-nated surface of EuIr$_2$Si$_2$, similarly to the $Ln$Rh$_2$Si$_2$ systems discussed above, also hosts a well-defined $\bar{\rm M}$ surface state residing in the topmost Si–Ir–Si–Eu layer. Remarkably, ARPES reveals that Eu atoms in this surface-layer block are not in a valence-fluctuating state, unlike Eu in the bulk, but instead demonstrate magnetically active divalent-configuration behavior. Temperature-dependent ARPES measurements demonstrate that the $4f$ magnetic moments of this Eu layer order ferromagnetically and at a surprisingly high temperature of about 48~K, with a large magnetic moment consistent with the Eu$^{2+}$ state~\cite{Susanne2019}. At this point, we would like to note that photoelectron-diffraction (PED) results obtained for Si-terminated \linebreak EuIr$_2$Si$_2$, together with their analysis, indicate that Eu in the Si–Ir–Si–Eu surface block exhibits an effective valence of about 2.1~\cite{EIS_2020}. This implies a finite hybridization of the valence states, which nevertheless does not suppress the peculiar long-range ferromagnetic order of the Eu $4f$ moments and their apparent quasi-two-dimensional behavior.

We note here that, for the EuRh$_2$Si$_2$ system discussed above, a curious hybridization of the Dirac cone with Eu $4f$ states was experimentally observed, coexisting with the AFM order below 24.5~K~\cite{Hoeppner2013}.

Coming back to EuIr$_2$Si$_2$, the revealed FM ordering below 48~K leads to a strong exchange polarization of the surface-state electrons. Because Ir atoms contribute significantly to the $\bar{\rm M}$ surface state, the broken inversion symmetry at the surface already produces a strong Rashba-type spin splitting in the PM phase at high temperatures. Below $T \approx 48$ K, the ordering of the Eu $4f$ moments introduces an additional exchange field, resulting in a competition between two fundamental spin-splitting mechanisms: spin-orbit coupling and magnetic exchange interaction. This interplay gives rise to a momentum-dependent and temperature-tunable spin splitting that can be clearly resolved experimentally. The observed behavior is well reproduced by \textit{ab~initio} calculations as well as by a relativistic $\boldsymbol{k}\cdot\boldsymbol{p}$ model describing the interplay between exchange interaction and the Rashba effect~\cite{Nechaev_2018, Susanne2019, Usachov_PRB.101.245140}. It was also established that the temperature-dependent intermediate Eu valence in the bulk allows tuning of the energy and momentum extent of the projected bulk band gaps that confine the two-dimensional carriers. This provides an additional degree of freedom for controlling spin-polarized carriers at the surface of a crystal that remains nonmagnetic in the bulk. Figure~\ref{fig:EIS_1} shows the most interesting ARPES data, which allow us to gain insight into the 2D ferromagnetic properties of the Si-terminated surface of EuIr$_2$Si$_2$~\cite{Susanne2019}.

While the properties of the subsurface Eu layer beneath the Si-terminated surface were addressed in detail, several important questions remained open and required a more comprehensive analysis. In particular, what are the magnetic and valence properties of the next deeper Eu layer, located in the eighth atomic plane below the Si-terminated surface? Does this layer already exhibit the non-integer valence state characteristic of the bulk? Furthermore, what is the behavior of Eu ions at an Eu-terminated surface? A divalent surface configuration can be reasonably expected, as commonly observed in non-integer-valent and trivalent Eu systems, where the reduced coordination at the surface leads to a shift of the $4f$ states towards higher binding energies. However, it remained unclear whether the outermost Eu layer also develops magnetic order, at which temperature this occurs, and how both the valence state and magnetic ordering evolve in the subsurface region of an Eu-terminated crystal.

\begin{figure}[t]
\centering
\includegraphics[width=0.98\linewidth]{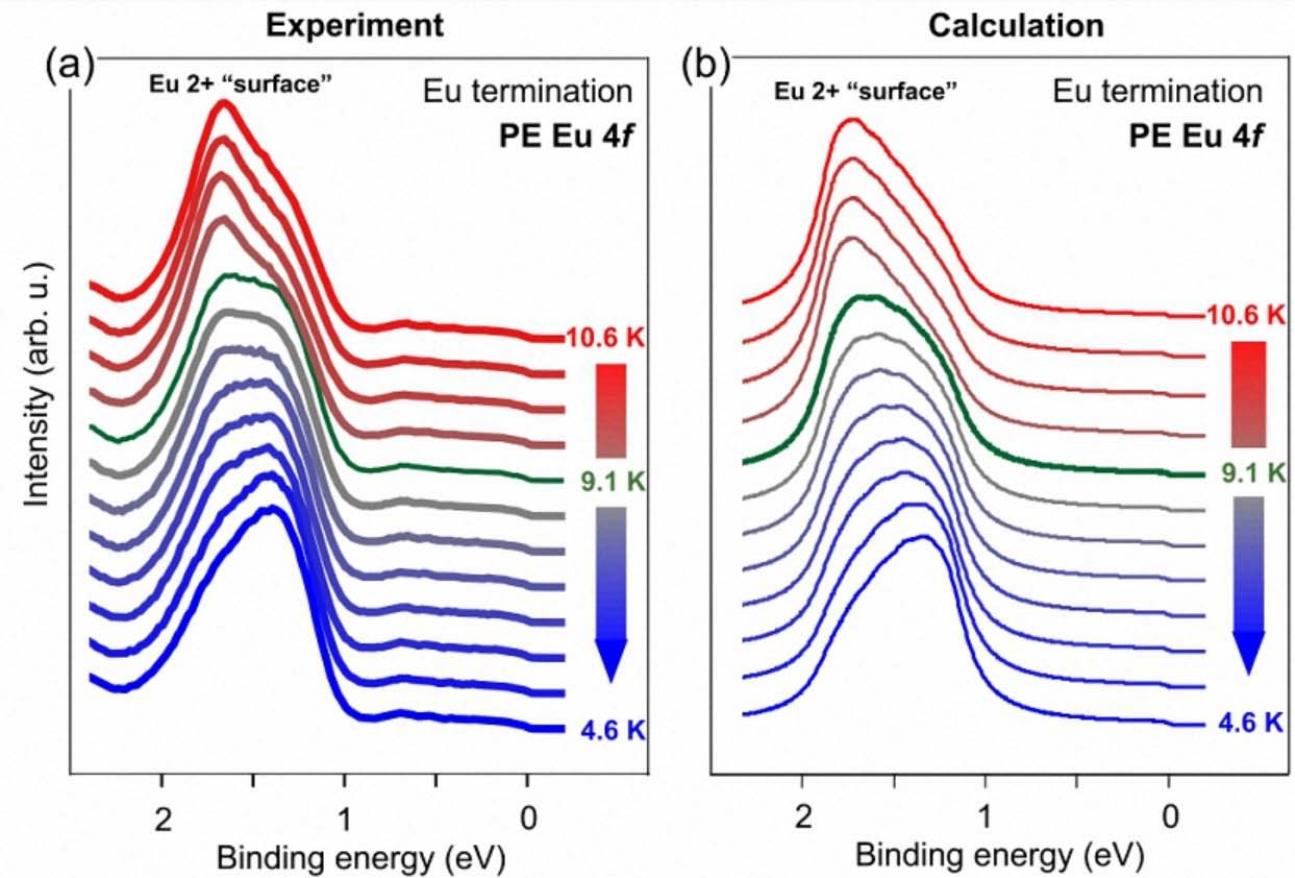}
\caption{FM properties of the Eu-terminated EuIr$_2$Si$_2$.
(a) Measured and (b) calculated normal-emission PES spectra of the Eu$^{2+}$ 4$f$ multiplet for the Eu-terminated surface as a function of temperature, with spectra taken using circularly polarized photons of 110 eV. In the calculations, a critical temperature of 9.5~K and magnetization oriented along the [100] direction were assumed~\cite{EIS_2020}.}
\label{fig:EIS_1A}
\end{figure}

A detailed discussion of these issues is provided in Ref.~[\citeonline{EIS_2020}]. There, we describe a theoretical framework for modeling PED patterns in systems with a localized open shell such as rare-earth intermetallic compounds. We then present experimentally derived PED patterns for EuIr$_2$Si$_2$ obtained from different surface terminations and analyze them within this modeling approach. On this basis, conclusions are drawn regarding the magnetic properties of both Si- and Eu-terminated surfaces, as well as the valence states of Eu atoms in successive subsurface layers. In particular, we demonstrate that the Eu-terminated surface also undergoes magnetic ordering below approximately 10~K. Figure~\ref{fig:EIS_1A} shows temperature-dependent Eu 4$f$ spectra, which display pronounced modifications of the 4$f$ multiplet structure upon crossing the transition temperature~\cite{EIS_2020}.

Taken together, the experimental and theoretical results reveal EuIr$_2$Si$_2$ as a remarkable type of material in which a valence-fluctuating and nonmagnetic bulk coexists with robust 2D ferromagnetism at the surface. The resulting highly spin-polarized surface states, whose properties can be tuned through the interplay of Rashba spin-orbit coupling and exchange interaction, make this system particularly attractive for spintronic applications. \linebreak More broadly, these findings highlight valence-fluctuating rare-earth compounds as a promising platform for designing materials with unconventional magnetic surface properties. They call for renewed attention to a wide class of systems in which heavy $5d$ elements are combined with rare-earth or transition-metal atoms, where similar or even more complex surface-confined ferromagnetic states may have remained unnoticed. Particularly in this context, we suggest beginning with well-known intermediate-valent Eu-based materials such as EuPd$_2$Si$_2$~\cite{Sampathkumaran1981VFS} and \linebreak EuNi$_2$P$_2$~\cite{EuNiP_1985, EuNiP_ARPES_2009}.

\section*{YbIr$_2$Si$_2$: A 2D Kondo lattice at its surface with strong spin-orbit coupling}

We next turn to the Kondo-lattice compound YbIr$_2$Si$_2$, whose Si–Ir–Si–Yb surface layer provides a particularly intriguing platform where strong spin-orbit coupling, broken inversion symmetry, and Kondo interactions coexist~\cite{Generalov_SISY_2018}. While this system is often discussed in connection with the quantum-critical properties of YbRh$_2$Si$_2$, here we focus primarily on its surface electronic structure, probed by ARPES and analyzed using \textit{ab~initio} calculations. Based on the knowledge gained from studies of the $Ln$Rh(Ir)$_2$Si$_2$ materials discussed above, one may anticipate the presence of a $\bar{\rm M}$ surface state at the Si-terminated surface, CEF-split states originating from the Yb $4f$ levels, a strong Rashba spin–orbit coupling leading to a spin splitting of the surface state, as well as interactions between Yb $4f$ states and spin-split surface states.

\begin{figure}[t]
\centering
\includegraphics[width=8cm]{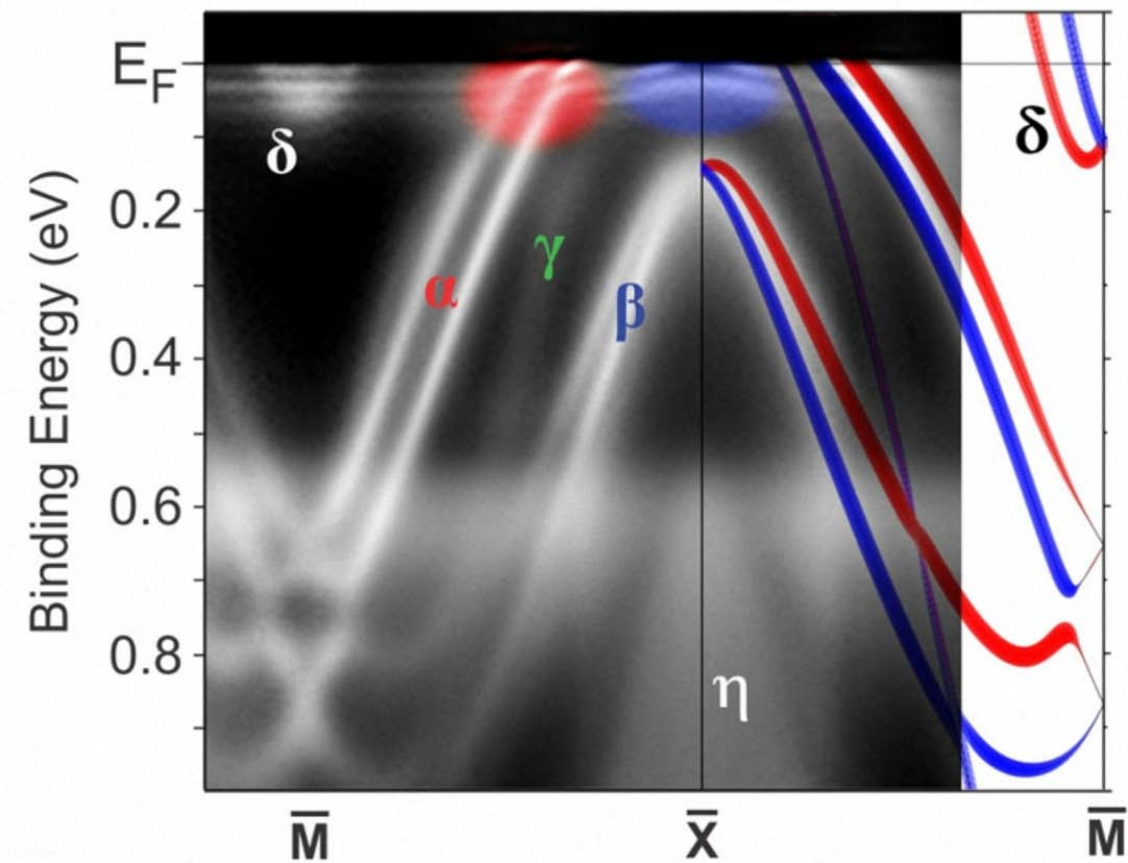}
\caption{ARPES data taken from the Si-terminated surface of YbIr$_2$Si$_2$ at a temperature of 1~K along the $\bar{\rm M}$–$\bar{\rm X}$ direction, superimposed with itinerant bands calculated by DFT. The spin polarization of the calculated bands is indicated by red and blue colors. Regions where the interplay between $4f$ and $spd$ states occurs near $E_\mathrm{F}$ are highlighted by red and blue patches~\cite{Generalov_SISY_2018}.}
\label{fig:YIS_1}
\end{figure}

In addition to these anticipated features, clearly visible in figure~\ref{fig:YIS_1}, ARPES measurements performed on the Si–Ir–Si–Yb surface reveal spin-polarized 2D $4f$ states that coexist and interact with strongly spin-polarized itinerant $\bar{\rm M}$ surface states. The results show that within the surface block the strong spin-orbit coupling of the Ir atoms induces spin polarization of the itinerant electrons, while the 2D lattice of Yb $4f$ moments forms the basis for coherent Kondo interaction at the surface. As a consequence, the spin-polarized $4f$ electrons interact in a temperature-dependent manner with the highly mobile spin-polarized itinerant states, thereby influencing both their dispersion and spin texture. At elevated temperatures, the strongly spin-polarized 2D states are highly mobile, while at lower temperatures they begin to couple coherently to the $4f$ moments, leading to a reduction of their group velocity. This implies that the Kondo interaction offers a route to tune the velocities of itinerant carriers via many-body renormalization.

A key aspect of YbIr$_2$Si$_2$ is that the $4f$ electrons near the surface become effectively 2D and acquire spin polarization as a result of the broken inversion symmetry combined with strong spin-orbit interaction. This creates an unusual situation in which localized $4f$ states participate in a 2D spin-polarized electronic environment while simultaneously engaging in Kondo interactions with itinerant carriers~\cite{Generalov_SISY_2018}.

The observation of interactions between spin-polarized itinerant states and $4f$ electrons at the surface establishes an interesting link between several central directions of modern condensed-matter physics, including strongly correlated electron systems, spin-orbit–driven phenomena, and low-dimensional spin-polarized electronic states. Systems of this type therefore provide a promising platform for exploring the interplay between Kondo physics, spin-orbit coupling, and reduced dimensionality, as well as for designing materials in which spin-polarized electronic states can be manipulated in strongly correlated environments~\cite{Generalov_SISY_2018}.

\section*{AFM GdIr$_2$Si$_2$: Classical and cubic Rashba effect}

Exploring EuIr$_2$Si$_2$ and YbIr$_2$Si$_2$, it was also naturally logical to study the surface-related FM properties in the AFM system GdIr$_2$Si$_2$, with particular focus on its Si termination~\cite{Schulz_GdIr_2021}. Note that, because the DFT results discussed above for TbRh$_2$Si$_2$~\cite{Usachov_PRL_2020}, which revealed the spin structure of the $\bar{\rm M}$ surface state with the characteristic “triple-winding” spin texture, were found to be fully consistent with the predictions of the $\boldsymbol{k}\cdot\boldsymbol{p}$ model, with the calculations of the spin photocurrent performed within the \textit{ab~initio} one-step theory of photoemission, and with spin-resolved ARPES measurements, we proceeded to investigate related systems such as GdIr$_2$Si$_2$ primarily on the basis of conventional ARPES experiments combined with DFT calculations.

Investigating GdIr$_2$Si$_2$, as anticipated, we found on the basis of combined ARPES measurements and \textit{ab~initio} calculations that three well-defined surface electron states are formed within the pronounced projected bulk band gap in the vicinity of the $\bar{\rm M}$ point. In the PM phase of GdIr$_2$Si$_2$, these states exhibit sizeable spin splittings originating from the Rashba spin-orbit interaction associated with the heavy Ir atoms. The calculations further show that two of the observed states display a spin texture characteristic of the conventional linear Rashba effect, whereas the third state demonstrates the previously discussed cubic Rashba behavior with the corresponding triple-winding spin structure. Below the Néel temperature, ARPES data reveal a complex photoemission intensity pattern indicative of the formation of magnetic domains. Analysis shows that both the dispersion and the spin structure of the surface states are governed by the interplay between the Rashba spin-orbit field and the in-plane exchange field arising from the FM ordered Gd 4$f$ moments in the near-surface layer~\cite{Schulz_GdIr_2021}.

\section*{CeIrIn$_5$: Rashba splitting and surface/bulk distinction of Ce atoms}

On the basis of the results obtained for $Ln$Ir$_2$Si$_2$ materials, we were motivated to explore systems containing both $Ln$ atoms and heavy elements such as Ir, since in such compounds one may anticipate similar or even more advanced surface-related properties and characteristic temperature scales. In that regard, we turned to the CeIrIn$_5$ system, which belongs to the widely studied material family $Ln$$T$In$_5$ (with $T=\mathrm{Co},\mathrm{Rh},\mathrm{Ir}$)~\cite{Fisk_2_2012, Fisk_1_2014}. This family has attracted considerable attention since the early 2000s and continues to be intensively investigated due to its remarkable bulk electronic and magnetic properties.

We focus on CeIrIn$_5$ and perform ARPES measurements combined with DFT analysis of its electronic structure, with particular emphasis on (i) surface-related properties, including the surface reconstruction anticipated from the crystallographic structure~\cite{Mende_2022}, and (ii) the search for a pronounced Rashba effect, as well as for differences in the behavior of Ce atoms in the bulk and in the near-surface region~\cite{Mende_2021}. We have mentioned above that such a property of Ce was already identified in early studies of Ce compounds by the Kaindl group~\cite{Laubschat_PRL_1990,Weschke_PRB_1991,Laubschat_SS_1992}, and it was therefore reasonable to anticipate a similar behavior in CeIrIn$_5$ as well~\cite{Mende_2021}. In the latter, we demonstrated, based on straightforward DFT calculations, how the likely surface terminations upon cleaving of the CeIrIn$_5$ crystal can be anticipated from the analysis of the spatial distribution of the charge density across the lattice. In particular, both CeIn- and In-terminated surfaces are expected. Subsequently, slab geometries were constructed and the corresponding electronic structures were calculated for slab and bulk configurations, treating the Ce $4f$ states as frozen core states. This approach allowed us to compute in detail the surface- and bulk-derived contributions to the band structure, which in turn helped us to quickly navigate the ARPES measurements and identify the surface terminations of the freshly cleaved sample~\cite{Mende_2021, Mende_2022}.

\begin{figure}[h!]
\includegraphics[width=0.98\linewidth]{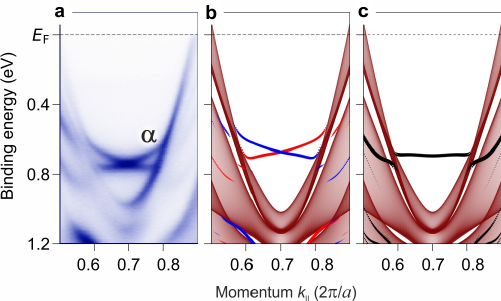}
\caption{Insight into the surface state located near the $\bar{\rm M}$ point, as revealed by ARPES data (a) taken along the $\bar{\rm M}$--$\bar{\Gamma}$--$\bar{\rm M}$ direction for the CeIn-terminated surface of CeIrIn$_5$, and by full-relativistic DFT calculations (b), where red and blue denote positive and negative values of the spin expectation $\langle S_y \rangle$, respectively. Scalar-relativistic DFT results are shown in (c) for comparison~\cite{Mende_2021}.}
\label{fig:CeIrIn_1}
\end{figure}

Essentially, the calculations predicted for the CeIn-terminated surface the presence of a surface state near the $\bar{{\rm M}}$ point exhibiting spin splitting driven by the conventional (linear) Rashba effect, associated with a trivial spin texture characteristic of the classical (non-cubic) Rashba scenario. Indeed, ARPES measurements showed such a state at binding energies closely matching the theoretical predictions, confirming it as an intrinsic feature of the CeIn-terminated surface of CeIrIn$_5$. Notably, this surface state displays a heavy, nearly linear dispersion. These findings are summarized in figure~\ref{fig:CeIrIn_1}, taken from Ref.~[\citeonline{Mende_2021}]. At this point, we note that although the discussed surface state in CeIrIn$_5$ is located well below the Fermi level, our analysis indicates that in related 218 compounds such as Ce$_2$IrIn$_8$ a similar Rashba spin-split surface state is expected. In this case, it should appear closer to the Fermi level and therefore provide an interesting opportunity to study its coupling to the Kondo-related spectral feature. Since both states are anticipated to reside within the $\bar{\rm M}$-point gap, one may expect a well-defined spectral signature of the spin-dependent interaction between the Kondo peak and the spin-split surface state, which can be explored in a temperature-dependent manner.

\begin{figure*}[h!]
\centering
\includegraphics[width=15cm]{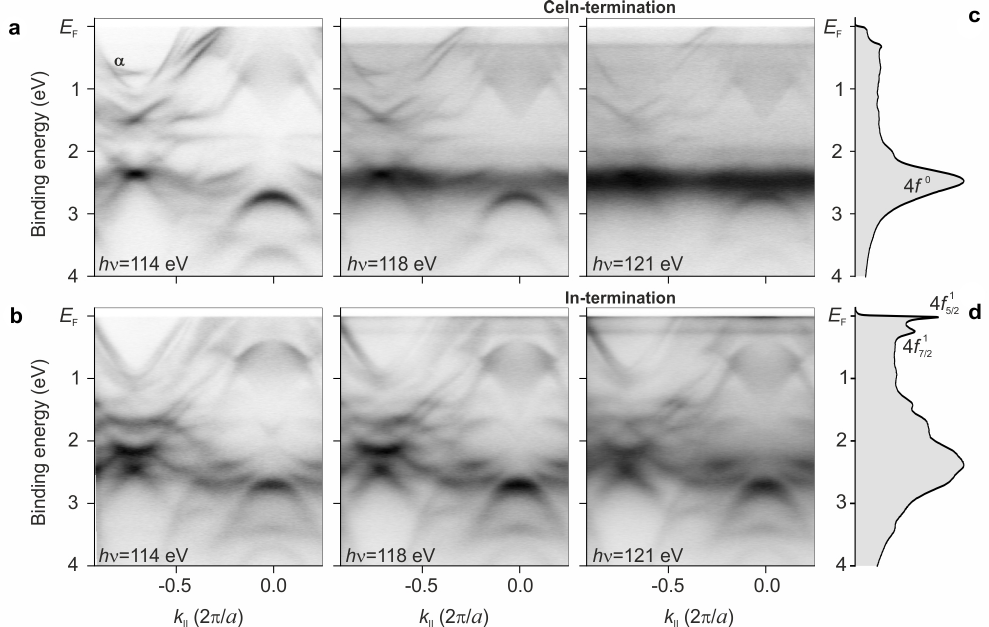}
\caption{Insight into the evolution of momentum-resolved Ce $4f$ emission across the Ce $4d$–$4f$ resonance threshold. ARPES spectral patterns measured along the $\bar{\rm M}$–$\bar{\Gamma}$ direction for (a) the CeIn-terminated and (b) the In-terminated surfaces of CeIrIn$_5$, taken at photon energies of 114, 118, and 121~eV. Angle-integrated spectra recorded at 121~eV for both surface terminations are shown in (c) and (d)~\cite{Mende_2021}.}
\label{fig:CeIrIn_2}
\end{figure*}

In figure~\ref{fig:CeIrIn_2}, we show the essentially different electronic responses of Ce atoms at the surface and in the bulk-like environment of CeIrIn$_5$ as revealed by ARPES measurements. The previously discussed surface state can also be clearly distinguished in the spectra taken from the CeIn-terminated surface. By tuning the photon energy across the Ce $4d$–$4f$ resonance, the momentum-resolved data show a pronounced modification of the $4f$ spectral line shape for the two surface terminations. In particular, the CeIn-terminated surface exhibits a considerably weaker and less well-defined $4f$ intensity near the Fermi level, whereas the In-terminated (bulk-like) surface reveals a strong and sharp resonance-enhanced feature. The obtained results are consistent with earlier observations by the Kaindl group, indicating a more $\gamma$-like character, i.e., significantly weaker hybridization of Ce atoms at the surface compared to the more $\alpha$-like behavior in the bulk~\cite{Laubschat_PRL_1990,Weschke_PRB_1991,Laubschat_SS_1992}.
The data therefore provide direct spectroscopic evidence that the $4f$ electronic landscape in CeIrIn$_5$ is strongly layer-dependent, with important implications for the development of correlated quasiparticle states in the near-surface region~\cite{Mende_2021}.

We believe that these results also call for detailed investigations of the electronic and magnetic properties emerging at the surfaces of strongly correlated Ce-, Eu- and Yb-based materials, several of which have already been discussed above, particularly in systems with layered, quasi-two-dimensional structures.

They also have important implications for the design of novel functional and quantum materials based on thin layers of $f$-electron compounds as building blocks. We continuously emphasize this general message throughout our review.

In such architectures, different fundamental interactions, including strong spin-orbit coupling, exchange magnetism, heavy-fermion behavior, and unconventional superconductivity, may originate from distinct atomic layers and coexist within a single structure. The controlled combination of these interactions therefore provides a \linebreak promising route toward predicting and realizing materials with new and tunable functionalities~\cite{Mende_2021, Mende_2022}.

\begin{figure*}[h!]
\includegraphics[width=1.00\linewidth]{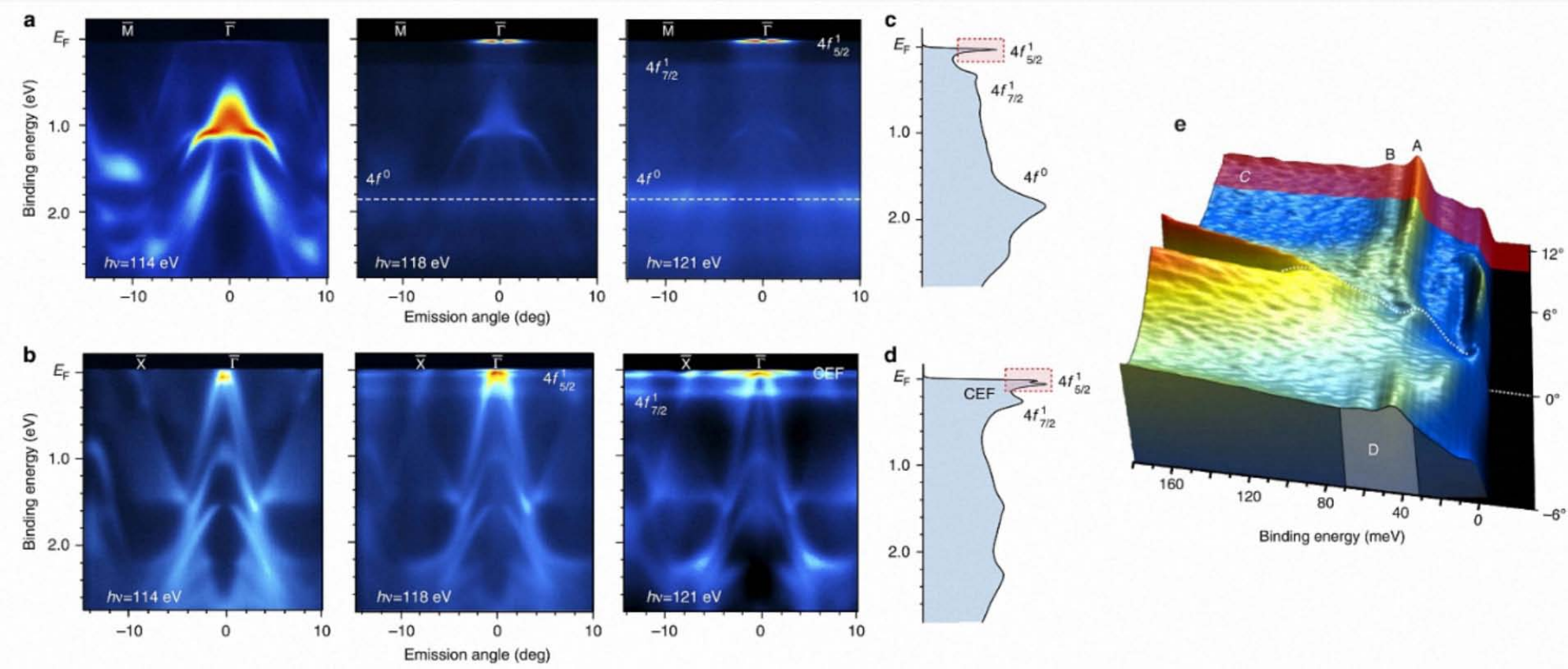}
\caption{AFM Kondo lattice CeRh$_2$Si$_2$. ARPES patterns taken from (a) the Ce-terminated surface along the $\bar{\Gamma}$–$\bar{\rm M}$ direction and (b) the Si-terminated surface along the $\bar{\Gamma}$–$\bar{\rm X}$ direction. The data illustrate the evolution of momentum-resolved Ce-4$f$ emission across the Ce $4d \rightarrow 4f$ resonance threshold. Angle-integrated spectra obtained at a photon energy of 121~eV for both surface terminations are shown in (c) and (d); dotted rectangles highlight the spectral weight associated with the $4f^{1}_{5/2}$ states. (e) Fine spectral structure measured on the Si-terminated surface of CeRh$_2$Si$_2$ using 40~eV photons. The visible asymmetry of the ARPES intensity with respect to the $\bar{\Gamma}$ point arises from the use of circularly polarized light~\cite{patil_16}.}
\label{fig:CRS_1}
\end{figure*}

\section*{CeRh$_2$Si$_2$: Distinct Kondo-related temperature scales at the surface and in the bulk}

On the next example of an AFM Kondo lattice, \linebreak CeRh$_2$Si$_2$~\cite{patil_16, Georg_2020}, we continue the discussion of the considerably different properties of Ce atoms at the surface and in the bulk of a Kondo material. Similarly to the above-discussed 122 systems, CeRh$_2$Si$_2$ cleaves between $Ln$ and Si atomic planes, leaving behind a crystal surface composed of a mosaic of Ce- and Si-terminated regions. This property makes the compound particularly suitable for ARPES measurements addressing two central questions: (i) how the Ce-4$f$ spectral function differs for surface and bulk Ce atoms~\cite{patil_16}, and (ii) how it evolves for both types of Ce sites as a function of temperature~\cite{Georg_2020}. High-resolution ARPES measurements performed at the Ce $4d \rightarrow 4f$ resonance, which enhances the Ce-4$f$ photoemission signal, provide direct insight into how the hybridization between itinerant and 4$f$ states shapes the spectroscopic response of surface and bulk Ce in this Kondo lattice. The analysis demonstrates that two well-defined and qualitatively different types of spectra can be clearly distinguished and unambiguously associated with Ce- and Si-terminated surfaces. The Ce-terminated surface, representative of weakly hybridized Ce atoms, exhibits a pronounced and sharp $4f^{0}$ ionization peak at about 1.9~eV binding energy together with a rather structureless spectral weight at the Fermi level (see figure~\ref{fig:CRS_1}).

\begin{figure*}[t]
\includegraphics[width=1.00\linewidth]{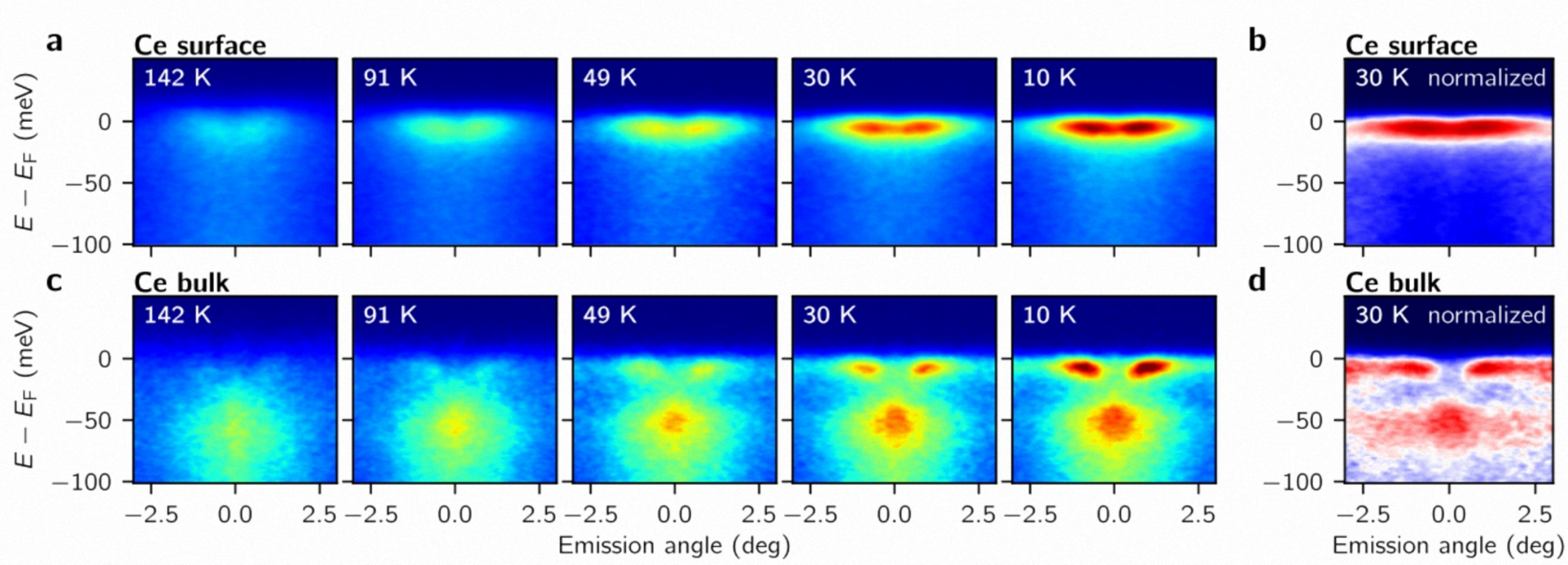}
\caption{Temperature-dependent studies of AFM Kondo lattice CeRh$_2$Si$_2$: Evolution of the $k$-resolved Ce-4$f$–derived states in the vicinity of the $\bar{\Gamma}$ point measured for (a) the Ce-terminated and (c) the Si-terminated surfaces of CeRh$_2$Si$_2$. The spectra were recorded along the $\bar{\rm M}$–$\bar{\Gamma}$–$\bar{\rm M}$ direction of the surface Brillouin zone. (b, d) Corresponding data after normalization of each angular slice to the same integrated intensity of the 30~K spectrum, allowing weak contributions of the CEF-related sidebands to be enhanced~\cite{Georg_2020}.}
\label{fig:CRS_2}
\end{figure*}

In contrast, spectra obtained from Si-terminated regions, which reflect bulk-like Ce atoms, reveal a strong hybridization between Ce-4$f$ and valence states, manifested by a pronounced hybridized $4f^{0}$ feature, similarly to the case of CeIrIn$_5$, and by the emergence of a momentum-dependent fine spectral pattern near the Fermi level. This structure originates from the CEF splitting of the magnetic $4f^{1}$ configuration and directly evidences the formation of hybridized quasiparticle states. In particular, the bulk-sensitive spectra indicate the presence of a weakly hybridized excited crystal-field level at about 48~ meV, a second excitation around 62~meV whose spectral weight is distributed over a broader energy range due to hybridization, as well as a strongly dispersive quasiparticle band at the Fermi level related to the $\Gamma_{7}$ ground state of Ce. Remarkably, these hybridization-driven features persist deep in the AFM phase, indicating that, while the Kondo interaction essentially quenches the Ce-4$f$ moments through strong $f–d$ hybridization, the remaining magnetic degrees of freedom still maintain long-range AFM order. These results represent a prototypical example of the ARPES response expected for a Ce-based magnetic Kondo lattice. They provide essential insight into the entangled localized–itinerant character of Ce-4$f$ electrons and establish a robust spectroscopic basis for evaluating distinct energy scales, including the effective Kondo interaction, distinctly at the surface and in the bulk of Ce Kondo systems.

The latter point became the focus of our subsequent work~\cite{Georg_2020}, which was aimed at a direct visualization of how the Kondo-lattice electronic structure evolves with temperature simultaneously at the surface and in the bulk under otherwise similar experimental conditions. By carefully selecting Ce- and Si-terminated surface domains, we were able to follow the temperature dependence of the Ce-4$f$ spectral function over a wide temperature range and to trace particularly the evolution of the fine electronic structure near the Fermi level (see figure~\ref{fig:CRS_2}). The obtained ARPES data provided a clear view of how the low-energy electronic structure of a Kondo lattice develops with temperature at the surface and in the bulk, and the most essential findings of this work can be summarized as follows:

\smallskip
\noindent
(i) Measurements performed over a broad temperature range allowed us to estimate the characteristic Kondo temperatures at the surface and in the bulk, and these values are found to differ substantially, enabling a comprehensive discussion of the microscopic origin of this unexpected difference.

\smallskip
\noindent
(ii) Performed analysis shows that the momentum dependence of the Kondo peak deviates significantly between surface and bulk regions of the Kondo lattice. This behavior is linked to the different sets of itinerant bands available for hybridization in the two environments and is likely also influenced by modified CEF splittings and corresponding changes in the 4$f$ ground-state degeneracy.

\smallskip
\noindent
(iii) The high-quality ARPES data, in conjunction with DFT band-structure calculations, enabled the development of an improved fitting procedure that provides more accurate insight into the properties of electronic states above the Fermi level than the standard approach based on division by a resolution-convolved Fermi–Dirac function. The presented methodological aspect offers an alternative route for analysing fine electronic structures in Ce systems and, in particular, for investigating the momentum dependence of the Kondo peak above the Fermi level.

\smallskip
\noindent
(iv) The presented study highlights the essential role of surface properties in ultraviolet ARPES investigations of $f$-electron materials. Without reliable knowledge of the crystal termination and the corresponding electronic environment, the measured spectra may represent a superposition of contributions from surface, subsurface, and bulk rare-earth atoms, which can lead to misleading conclusions regarding bulk electronic behavior. This observation is of broad relevance for spectroscopic studies of strongly correlated electron systems in general and of Ce compounds in particular.

\smallskip
\noindent
(v) Finally, the present study allows for a direct insight into the broader problem of Kondo lattices with two different local moment sublattices, providing some reason for the cross talking between the two Kondo effects being weak~\cite{Georg_2020}.

\begin{figure*}[h!]
\centering
\includegraphics[width=14 cm]{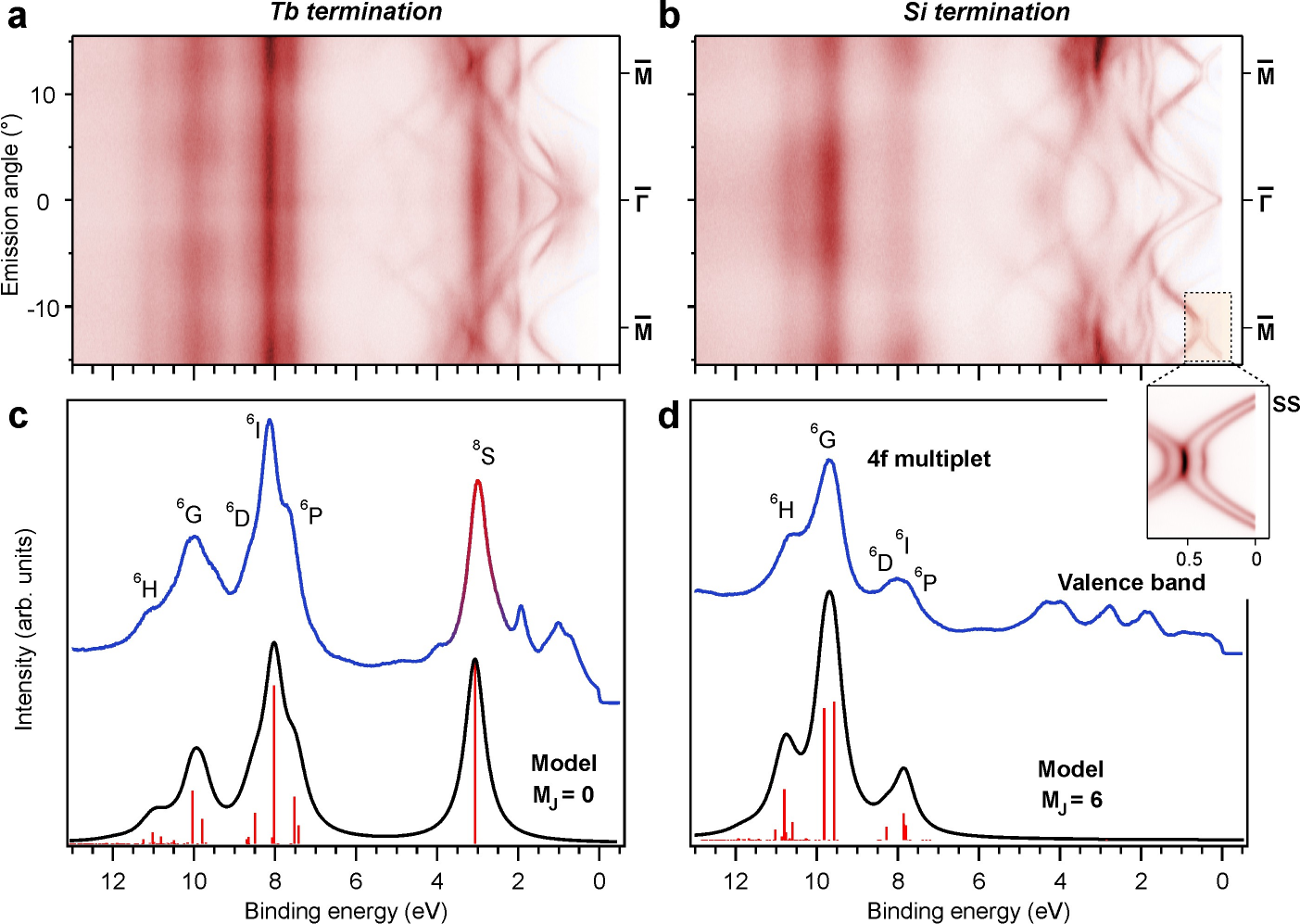}
\caption{Evaluation of Tb 4$f$ moments orientation from ARPES. Band map patterns taken along the $\bar{\rm M}-\bar{\Gamma}-\bar{\rm M}$ direction for (a) Tb- and (b) Si-terminated surfaces of TbRh$_2$Si$_2$ in the AFM phase at $T=21$~K. The inset shows surface states taken at $T=8$~K.
(c,d) Angle-integrated photoemission spectra obtained from the data in panels (a,b) by integrating over an angular range of $\pm5^\circ$ around the $\bar{\Gamma}$ point (blue curves), compared with calculated 4$f$ multiplet spectra for single-$M_J$ ground states $|0\rangle$ and $|6\rangle$ (black curves)~\cite{Usachov_JPCL_2022}.}
\label{fig:TRS_5}
\end{figure*}

\section*{Assessing the orientation of 4$\boldsymbol{f}$ moments by ARPES and XAS}

In our systematic studies of the $Ln$Rh$_2$Si$_2$ materials, where the electronic and magnetic properties were accurately analyzed at both $Ln$- and Si-terminated surfaces, we observed that the line shape of the $Ln$ 4$f$ multiplets changes substantially depending on the surface termination~\cite{Usachov_JPCL_2022, Tarasov_PRB_2022}. As exemplified in figure~\ref{fig:TRS_5}, a direct comparison of the ARPES patterns obtained from Tb- and Si-terminated surfaces of the AFM TbRh$_2$Si$_2$, discussed above, clearly reveals pronounced differences in the spectral shape of the nondispersive Tb $4f$ emission features. To gain deeper insight into this behavior, we performed modeling of the $4f$ multiplet structure, following the approach used by F. Gerken in 1983~\cite{Gerken_1983}, but extending it to the calculation of photoemission spectra for individual $M_J$ ground states~\cite{Usachov_JPCL_2022}.

A comparative analysis of the experimentally measured multiplet structures with modeled spectra calculated for different $|M_J\rangle$ ground states enables reliable identification of the magnetic ground state and the corresponding direction of the total angular momentum $\langle \mathbf{J} \rangle$ in individual atomic layers~\cite{Usachov_JPCL_2022}. This concept was explicitly verified experimentally for several representatives of the layered $Ln$Rh$_2$Si$_2$ family. The resulting variations of the 4$f$ spectral line shapes ultimately indicate the pronounced reorientation of 4$f$ magnetic moments at the RE surface with respect to the bulk due to the change of the CEF.

A central outcome of this work is the elaboration of a comprehensive database of calculated 4$f$ photoemission spectra for all trivalent lanthanides and for the full set of $|M_J\rangle$ ground states~\cite{Usachov_JPCL_2022}. This spectral library enables quick qualitative determination of the ground state already in the PM phase and allows prediction of the orientation of magnetic moments in magnetically ordered states, including FM and AFM, without the need for an externally applied magnetic field. The methodology is applicable on equal footing in laboratory-based (UV) as well as synchrotron-based photoemission experiments and is particularly advantageous for layered and quasi-2D 4$f$ systems, as well as for $Ln$-based molecules and supramolecular complexes. We believe that this approach will provide a solid basis for probing and controlling the magnetic anisotropy of rare-earth-based functional materials, including multilayers and interfaces relevant for device engineering and technological applications. By linking measurable 4$f$ photoemission line shapes to the orientation of local magnetic moments, classical 4$f$ photoelectron spectroscopy emerges as a conceptually simple yet powerful probe of layer-resolved magnetism, complementary to established techniques such as magnetic dichroism and resonant X-ray scattering.

\begin{figure*}[h!]
\includegraphics[width=1.00\linewidth]{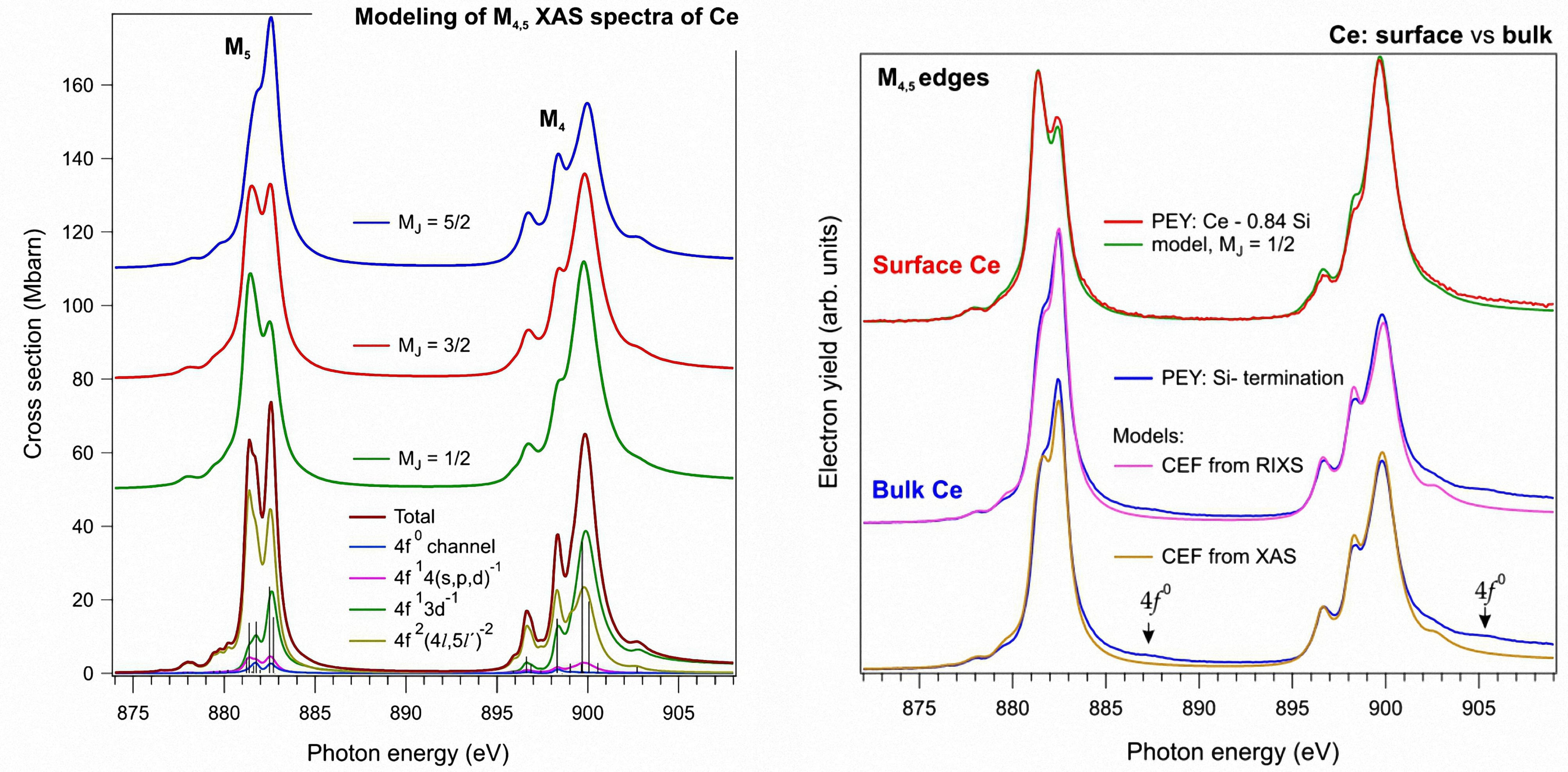}
\caption{Left panel: Modeling of the Ce $M_{4,5}$ XAS spectra for individual $\lvert M_J\rvert$ ground states. The calculated unpolarized Ce 3$d$ XAS spectrum with partial contributions from the individual decay channels is shown at the bottom.
Right panel: Insight into surface and bulk Ce properties based on Ce 3$d$ XAS spectra. Further details can be found in Ref.~[\citeonline{Usachov_PRB_2025}].}
\label{fig:CRS_3}
\end{figure*}

Continuing this research line, and using the layered AFMs HoRh$_2$Si$_2$ and DyRh$_2$Si$_2$ as model systems, we obtained detailed insight into the temperature-dependent canting of the 4$f$ moments in the near-surface region through analysis of the photoemission $4f$-multiplet line shape~\cite{Usachov_JPCL_2023, Kliemt_PRB_2023}. Bulk-sensitive experiments show that both compounds exhibit a progressive canting of the 4$f$ moments away from the crystallographic $c$-axis upon changing temperature, which provides a suitable platform to explore this property by means of 4$f$-PES measurements. Focusing on well-defined Si-terminated surfaces, we probed lanthanide 4$f$ moments located below the topmost \linebreak Si–Rh–Si block. Temperature-dependent 4$f$-ARPES spectra reveal systematic and reproducible modifications of characteristic multiplet features~\cite{ Usachov_JPCL_2023}. By applying an elaborated theoretical framework that incorporates CEF effects together with exchange magnetic interactions, we show that these subtle spectral variations are directly related to the temperature-driven canting of the 4$f$ moments. The analysis further demonstrates that the canting behavior is layer-dependent, with pronounced differences between lanthanide layers in the near-surface region and in the bulk, primarily arising from modified CEF conditions at the crystal surface. In the specific case of the Ho-terminated surface of HoRh$_2$Si$_2$, fundamental core-level chemical shifts in the 4$f$ emission enable a clear distinction between the magnetic response of surface, subsurface, and bulk layers. Thus, the combination of 4$f$-sensitive ARPES and quantitative modeling establishes high-precision photoelectron spectroscopy as an effective probe for monitoring and controlling the orientation and subtle canting of lanthanide moments~\cite{Usachov_JPCL_2023, Kliemt_PRB_2023}.

Having established from 4$f$ photoemission measurements that the orientation of $Ln$ 4$f$ magnetic moments can be quantitatively accessed through analysis of the 4$f$-multiplet line shape, it is natural to expect that this information should also be reflected in X-ray absorption spectra. Therefore, we investigated the same materials, such as the antiferromagnet TbRh$_2$Si$_2$ discussed above, by performing XAS measurements at the $Ln$ 4$d$ edge, where spectra were taken in both surface-sensitive partial- \linebreak electron-yield (PEY) and bulk-sensitive total-electron- \linebreak yield (TEY) modes~\cite{Usachov_RPED_2024}. As before, particular attention was paid to $Ln$- and Si-terminated surfaces. A comparison between experimentally obtained XAS spectra and results of detailed modeling reveals a consistent pattern of modified $4f$-moment orientation in the near-surface region, in full agreement with the conclusions derived from the above $4f$-PES results. Complementing the findings from photoemission, the study~\cite{Usachov_RPED_2024} establishes a quantitative framework for modeling XAS spectra at the lanthanide 4$d$ edge, including contributions from individual $M_J$ states. The Supplementary Information~\cite{Usachov_RPED_2024} provides a comprehensive description of the computational approach, together with calculated XAS spectra across the lanthanide series at the $4d \rightarrow 4f$ absorption threshold.

As a next step, we decided to apply our methodology to Ce-based systems, and in particular to CeRh$_2$Si$_2$, which has been discussed extensively above. In analogy to other $Ln$Rh$_2$Si$_2$ compounds, modifications of the CEF environment are naturally expected in the surface region. This should affect the magnetic ground state and lead to different properties of the Ce $4f$ moments at the surface compared to the bulk. Such considerations clearly motivated a systematic study by means of XAS at both the Ce $3d$ and $4d$ absorption edges, with spectra taken in bulk- and surface-sensitive detection modes (TEY and PEY, respectively)~\cite{Usachov_PRB_2025}.

By comparing the experimental spectra with the modeled spectra for different Ce ground states, we were able to disentangle surface and bulk contributions to the Ce XAS signal (see figure~\ref{fig:CRS_3} taken from Ref.~[\citeonline{Usachov_PRB_2025}]). The results reveal a clear reorientation of the Ce $4f$ magnetic moments at the Ce-terminated surface relative to their bulk alignment, driven by surface-induced changes of the CEF. These findings emphasize that surface effects must be explicitly considered when interpreting XAS and related photon-in/electron-out spectroscopies in strongly correlated Ce materials~\cite{Usachov_PRB_2025}. In particular, even Ce $3d$ XAS spectra, which are commonly treated by the community as predominantly reflecting bulk properties, contain significant information about the surface region. With a careful line-shape analysis, one can therefore extract the relevant spectral contribution and make reliable statements about the ground-state properties and orientation of Ce $4f$ moments at the surface~\cite{Usachov_PRB_2025}.

\section*{Resonant 4$\boldsymbol{f}$-PED on YbRh$_2$Si$_2$: \\ Determination of the ground state}

The application of PED to strongly correlated $4f$ systems has already been demonstrated in our studies of EuIr$_2$Si$_2$, where this approach allowed us to unveil how the mean valence of Eu evolves from the surface to the bulk and how it depends on the specific surface termination~\cite{EIS_2020}. Although PED is not yet widely used within the community, our results demonstrate that it represents a highly efficient method for probing the electronic and magnetic properties of lanthanide-based materials~\cite{Usachov_RPED_2024, Usachov_RPED_2025}.

Since its introduction in the 1970s, PED has developed into a versatile spectroscopic and imaging technique capable of providing detailed information on local atomic structure, chemical environment, magnetic ordering, and electronic states in crystalline solids and low-dimensional systems. The method relies on the coherent multiple scattering of photoelectrons emitted from selected atomic sites, giving rise to characteristic angular diffraction patterns that encode structural and electronic information. More recently, PED has also opened new perspectives for investigating ultrafast processes, where diffraction imaging enables access to electronic dynamics on femtosecond and picosecond timescales.

In strongly correlated lanthanide compounds, resonant PED (RPED) offers further opportunities by exploiting the enhancement of photoemission intensity at core-level absorption thresholds. The interference between direct photoemission and autoionization channels substantially increases the sensitivity of diffraction measurements, making RPED particularly suitable for studies of deep atomic layers, dilute species, buried interfaces, and complex surface terminations. In Ref.~[\citeonline{Usachov_RPED_2025}], we demonstrate that a consistent theoretical approach explicitly incorporating the resonant PE process enables reliable modeling of RPED patterns and can be successfully applied to representative Yb-based systems such as the heavy-fermion compound YbRh$_2$Si$_2$ and its trivalent counterpart YbCo$_2$Si$_2$. In particular, a conclusive determination of the ground-state properties of YbRh$_2$Si$_2$, together with a discussion of the limitations of RPED in the case of trivalent YbCo$_2$Si$_2$, demonstrates that RPED is a powerful spectroscopic tool for studies of heavy-fermion phenomena~\cite{Usachov_RPED_2024}. Moreover, this approach can be extended quite naturally to other strongly correlated materials and applied at other resonant excitations such as $3d \rightarrow 4f$. Surprisingly, in the case of YbRh$_2$Si$_2$, RPED provides, among other applied techniques, a particularly conclusive and straightforward answer to the widely debated problem of the ground-state determination~\cite{Usachov_RPED_2024}.

\section*{Insight into the electronic structure of the centrosymmetric skyrmion magnet GdRu$_2$Si$_2$}

While discussing the $Ln$$T_2$Si$_2$ family, we should also note the particularly intriguing material GdRu$_2$Si$_2$~\cite{Khanh_NatNano_2020, Yasui_NatComm_2020, Eremeev_NA_2023}. The discovery of a square magnetic skyrmion lattice in this antiferromagnet, realized without the presence of geometrical frustration, has placed GdRu$_2$Si$_2$ at the focus of considerable current research efforts~\cite{Khanh_NatNano_2020, Yasui_NatComm_2020}. Essentially, GdRu$_2$Si$_2$ hosts the shortest-period skyrmion lattice reported to date, which makes it an attractive platform for the development of next-generation high-density magnetic memory devices with ultralow energy consumption.

Note that despite the experimental discovery of \linebreak skyrmions in GdRu$_2$Si$_2$ in 2020, there was \linebreak debate~\cite{Khanh_NatNano_2020, Yasui_NatComm_2020, Nomoto_PRL_2020, Bouaziz_2022} over the dominant magnetic interaction and the mechanism that stabilizes the formation of skyrmions in this material. To conclusively address this problem, detailed information about the electronic structure of this material was essential. Since a peculiar helical AFM order of the Gd 4$f$ magnetic moments below 46~K at zero magnetic field is the precursor of the skyrmion phase at higher fields, it was important to determine in detail the electronic structure and its changes upon this phase transition.

All these aspects were comprehensively addressed in our ARPES studies, complemented by DFT calculations performed for both the PM and AFM phases~\cite{Eremeev_NA_2023}. In particular, surface and bulk electronic states were disentangled. Subtle modifications of the bulk-projected band structure, as well as of surface-resonant states emerging across the transition from the PM to the AFM phase, were observed experimentally and reproduced theoretically. These changes are associated with the stabilization of the helical magnetic order.

\begin{figure*}[h!]
\centering
\includegraphics[width=15 cm]{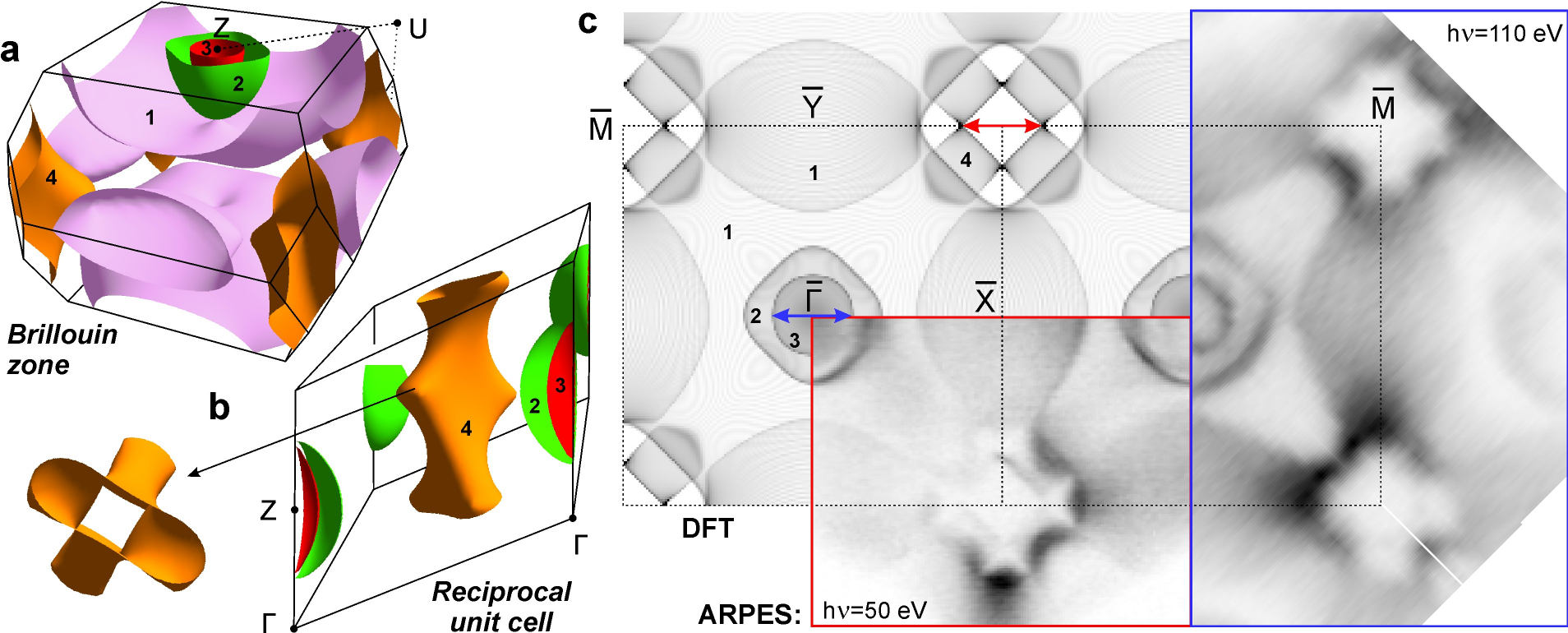}
\caption{Insight into the origin of the skyrmion phase in GdRu$_2$Si$_2$~\cite{Eremeev_NA_2023}. Calculated Fermi surface of GdRu$_2$Si$_2$ in the PM phase shown for the first Brillouin zone (a) and in reciprocal-lattice units (b), together with its projection onto the (001) surface (c). Selected regions highlight the comparison with ARPES-derived Fermi-surface maps. The red arrow indicates the proposed nested region in the 3D Fermi surface while the blue arrow shows the region near the center of the Brillouin zone~\cite{Bouaziz_2022}.}
\label{fig:GdRuS_1}
\end{figure*}

To clarify the origin of this spiral-AFM state, we analyzed possible nesting conditions of the Fermi surface (see figure~\ref{fig:GdRuS_1}). The good agreement between theory and ARPES experiment allows us to conclude that the dominant Fermi surface nesting occurs near the corners of the Brillouin zone, rather than at its center as proposed in Ref.~[\citeonline{Bouaziz_2022}], and shown as blue arrow in figure~\ref{fig:GdRuS_1}. This conclusion is further supported by an orbital analysis revealing a pronounced Gd-$5d$ contribution to the nested states, as well as by theoretical studies of unit-cell distortions. We were thus able to identify the electronic states that play a key role in stabilizing the helical magnetic order via the RKKY interaction, which ultimately gives rise to the emergence of the skyrmion phase in GdRu$_2$Si$_2$ at elevated magnetic fields.

\section*{Photoemission studies of $\boldsymbol{Ln}$Co$_2$P$_2$ systems}

So far, we have provided a rather detailed overview of the $Ln$$T_2$Si$_2$ materials and demonstrated how essential it is to maintain a systematic approach in such studies. The logic of moving from one system to another is quite straightforward, and each subsequent, clearly motivated experiment has delivered new and sometimes astonishing results, revealing new phenomena and their characteristic temperature scales.

As the next example, we turn to another interesting family of materials, namely $Ln$Co$_2$P$_2$~\cite{Poelchen_2022, Usachov_APR_2025}. Our initial interest in these systems was motivated by their relatively high magnetic ordering temperatures. For instance, in AFM CeCo$_2$P$_2$, the Néel temperature associated with the ordering of the Co magnetic moments exceeds room temperature and reaches approximately 440~K~\cite{Reehuis_JAC_1998}. In light of our previous findings on ferromagnetic surface states in compounds such as GdRh$_2$Si$_2$ and EuRh$_2$Si$_2$, such elevated ordering temperatures suggested that these antiferromagnets might also host ferromagnetic surface states with comparably high characteristic temperatures, making them attractive from both fundamental and appli- cation-oriented perspectives. An analysis of the bulk properties of CeCo$_2$P$_2$ indicates that the Ce ions behave rather passively and are not magnetically active, which is consistent with the high Néel temperature governed by the Co sublattice.

At the same time, we anticipated that Ce ions in the near-surface region could exhibit behavior distinct from that in the bulk, becoming more magnetically active and potentially giving rise to quasi-two-dimensional Kondo-lattice states that may interact with the ordered Co moments at the surface. Such expectations were natural and based on the experience gained from the $Ln$$T_2$Si$_2$ systems discussed above. To address this problem, we performed a detailed investigation of the P–Co–P–Ce surface termination of CeCo$_2$P$_2$, and ARPES measurements reveal that in fact Ce sublattice becomes magnetically active near the surface. This behavior originates from symmetry breaking, reduced coordination, and the effective magnetic field generated by the uncompensated ferromagnetic Co layer within the surface block, which promotes partial occupation of the Ce $4f$ shell~\cite{Poelchen_2022}. Further, ARPES reveals a pronounced admixture of Ce $4f$ states with itinerant electronic bands in the vicinity of the Fermi level, including exchange-split states associated with the Co layer. Temperature-dependent ARPES spectra further reveal significant variations of the $4f$ spectral weight near $E_\mathrm{F}$, consistent with a Kondo-lattice scenario. These findings indicate that the Ce sublattice within the topmost P–Co–P–Ce block develops quasi-two-dimensional Kondo-lattice behavior that couples ferromagnetically to the ordered Co lattice. These results show the remarkable diversity of $f$-electron-driven phenomena that may emerge at the surfaces of materials whose bulk electronic structure exhibits only weak or negligible $f$-derived character.

Based on the analysis of our ARPES results together with bulk-sensitive experiments and evaluation of the band structure from first-principles calculations, we concluded that CeCo$_2$P$_2$ may host collective magnetic excitations originating from the Co sublattice. Owing to the unusual electronic structure in the vicinity of $E_\mathrm{F}$, these magnons were expected to exhibit unconventional scattering behavior. This understanding provided the basis for subsequent resonant inelastic X-ray scattering experiments, which confirmed this anticipation and clearly demonstrated the existence of long-lived magnons in \linebreak CeCo$_2$P$_2$ even in the terahertz regime, an astonishing result for a metallic system~\cite{Poelchen_2023}.

\begin{figure*}[h!]
\includegraphics[width=1.00\linewidth]{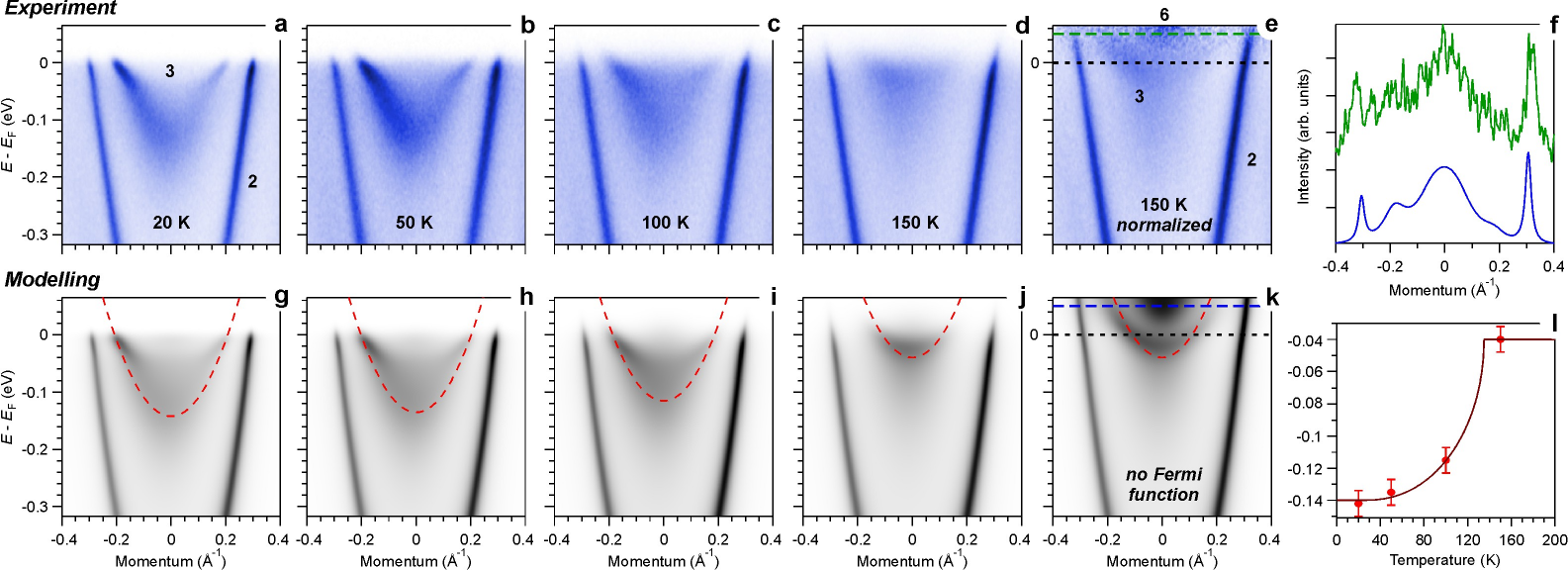}
\caption{ARPES insight into the electron-boson interactions in LaCo$_2$P$_2$. Temperature evolution of spin-polarized Co 3$d$ band near the $\bar{\rm M}$ point from ARPES measurements taken along the $\bar{\Gamma}$–$\bar{\rm M}$–$\bar{\Gamma}$ direction (a–e), together with simulations (g–k). (f) Profiles of the measured and simulated PE intensity at 50~meV above $E_F$. l) Temperature-dependent changes in the energy position of the bottom of band 3 without renormalization (corresponding to the red dashed lines in (g–k), which represent the bare band). Further details can be found in Ref.~[\citeonline{Usachov_APR_2025}].}
\label{fig:CeCoP}
\end{figure*}

We next turn to LaCo$_2$P$_2$, which represents an instructive non-$4f$ member of the $Ln$Co$_2$P$_2$ family~\cite{Usachov_APR_2025}. In contrast to the Ce-based compound discussed above, \linebreak  LaCo$_2$P$_2$ is a weak itinerant ferromagnet with a Curie temperature of about 135~K. Remarkably, it is the only representative of this series in which the Co $3d$ moments order ferromagnetically, whereas compounds with $Ln$ = Ce, Pr, Nd, and Sm exhibit AFM ordering involving both the lanthanide and Co sublattices. Moreover, LaCo$_2$P$_2$ appears to lie close to a magnetic instability, since the neighboring BaCo$_2$P$_2$ is nonmagnetic while other $Ln$Co$_2$P$_2$ compounds are magnetically ordered. This particular position in the magnetic phase diagram makes LaCo$_2$P$_2$ a convenient reference system for addressing the electronic and magnetic properties of itinerant transition-metal $3d$ states in layered pnictides. The absence of $4f$ electrons in LaCo$_2$P$_2$ enables a more direct investigation of electron–boson interactions involving both phonons and magnetic excitations. The ferromagnetic ordering of the Co sublattice lifts the spin degeneracy of the electronic states and leads to characteristic splittings of Co-derived bands near the $\bar{\rm M}$ point of the Brillouin zone, providing direct access to spin-polarized quasiparticles and their coupling to bosonic modes, which can be explore temperature dependently (see the details in figure~\ref{fig:CeCoP} taken from Ref.~[\citeonline{Usachov_APR_2025}]).

Understanding such effects in reference systems is essential for clarifying how electron–boson interactions renormalize transition-metal $d$ states in layered pnictides and may influence the formation of superconducting ground states in related materials. ARPES indeed reveals a substantially renormalized electronic structure of LaCo$_2$P$_2$, reflecting the interplay of electron–phonon and electron- magnon interactions. A careful analysis allows us to disentangle bulk and surface contributions. In the bulk, large Co-derived electron pockets around the Brillouin-zone corners exhibit pronounced kinks and reduced quasiparticle lifetimes, indicative of strong many-body renormalization. In contrast, at the P-terminated surface the magnetic order of the Co moments is suppressed, while Rashba-type spin splitting emerges due to inversion- symmetry breaking. Obtained results show that the low-energy electronic structure of LaCo$_2$P$_2$ is governed by the competition between electron-phonon and electron- magnon interactions, which strongly renormalize the spin-polarized Co $3d$ states. In this respect, LaCo$_2$P$_2$ serves as a convenient reference system for studying itinerant magnetism and its coupling to bosonic excitations in layered pnictides. Such insight is important for understanding related many-body effects in transition-metal pnictides more generally, including those that are believed to be relevant for the superconducting properties of the iron-based compounds~\cite{Usachov_APR_2025}.

\section*{Time-resolved studies on 4$\boldsymbol{f}$ systems}

The ARPES results discussed above provided a solid basis for extending these studies to time-resolved experiments aimed at clarifying the microscopic mechanisms governing quasiparticle dynamics in systems with localized 4$f$ moments coupled to itinerant electrons. The first femtosecond pump–probe ARPES experiment addressed this problem for the heavy-fermion compound YbRh$_2$Si$_2$~\cite{Kummer_PRB_2012}. These measurements enabled momentum-resolved access to relaxation processes in unoccupied electronic states above $E_\mathrm{F}$ and revealed strongly energy- and band-dependent quasiparticle lifetimes. In particular, an enhanced lifetime was observed near a dispersive band located about 0.2~eV above $E_\mathrm{F}$, indicating the importance of hybridization between valence states and localized Yb 4$f$ levels for ultrafast relaxation pathways. Several microscopic scenarios were proposed in order to describe the experimental observations~\cite{Kummer_PRB_2012}.

Later on, time-resolved studies were extended to magnetically ordered lanthanide systems of the $Ln$T$_2$Si$_2$ family in order to explore ultrafast spin dynamics and the coupling between electronic and magnetic degrees of freedom~\cite{Lee_ADMI_2022}. By combining time- and angle-resolved PE spectroscopy with time-resolved resonant soft-X-ray diffraction, the femtosecond evolution of the electronic distribution, the exchange splitting of the $\bar{\rm M}$ surface state of the Si-terminated surface, and the resonant magnetic diffraction amplitude of the (001) reflection were studied in AFM GdRh$_2$Si$_2$~\cite{Lee_ADMI_2022}. The simultaneous sensitivity to in-plane surface FM order and out-of-plane bulk AFM order enabled a detailed picture of the ultrafast magnetic-order dynamics in this quasi-two-dimensional system. The similar temporal evolution of exchange splitting and magnetic diffraction amplitude indicates strong exchange coupling between itinerant conduction electrons and localized Gd 4$f$ moments. The observed dynamics can be qualitatively described within a magnetic three-temperature framework based on the Landau-Lifshitz- Bloch equation. However, the transient recovery of magnetic order at electronic and lattice temperatures exceeding the equilibrium transition temperatures points to a nonthermal enhancement of magnetic interactions and deviations from mean-field behavior that increase with excitation fluence. These findings demonstrate that a quantitative description of ultrafast magnetization dynamics in 4$f$ systems requires theoretical approaches beyond simple thermal models~\cite{Lee_ADMI_2022}.

Time-resolved investigations were further extended to a systematic study of ultrafast magnetization dynamics in a series of isostructural lanthanide antiferromagnets $Ln$Rh$_2$Si$_2$, enabling direct comparison of nonequilibrium spin dynamics for different 4$f$ occupations. Using femtosecond resonant soft-X-ray diffraction, complemented by first-principles calculations, pronounced variations in the transient magnetic response were observed across the lanthanide series~\cite{Windsor_NatMater_2022}. Analysis of the time-dependent magnetic diffraction amplitude allowed the extraction of angular-momentum transfer rates, which were found to correlate with the de Gennes factor of the respective rare-earth ion. The results were discussed in terms of exchange-mediated transfer of angular momentum between antiparallel 4$f$ moments via conduction electrons, thereby providing direct experimental constraints on the microscopic mechanisms governing ultrafast spin dynamics in lanthanide antiferromagnets~\cite{Windsor_NatMater_2022}.

We note that these results on ultrafast dynamical processes obtained in the time-resolved experiments discussed above emphasize the essential role of prior static ARPES and X-ray absorption studies performed on well-characte- rized high-quality samples. Such measurements establish the electronic structure and magnetic ground-state properties of the investigated systems and thereby provide a solid basis for the design and interpretation of nonequilibrium experiments. This strategy is expected to remain highly valuable for future studies aimed at exploring ultrafast $4f$-driven magnetic phenomena and the dynamics of quasiparticles formed through $4f$ hybridization in novel crystalline systems as well as in engineered $4f$ architectures.

\section*{Crystal growth}

Before turning to the summary and outlook, we will provide an overview of crystal-growth procedures, including specific strategies and practical know-how developed by our team throughout our systematic studies of lanthanide-based compounds. To reliably investigate materials using synchrotron-based UV techniques, particularly ARPES, they must be available in high-quality crystalline form, either as thin films or bulk single crystals. The experimental results discussed above were obtained by our team using exclusively single crystals, which were cleaved in ultra-high vacuum immediately prior to investigation. To ensure the purity and single-crystalline form of the compounds, which contained air-sensitive lanthanides as well as elements with high vapor pressure, the crystal growth was performed under a protective atmosphere.

\subsection*{Flux method: Experimental methods and setups}

\begin{figure}[h!]
    \centering
    \includegraphics[width=\linewidth]{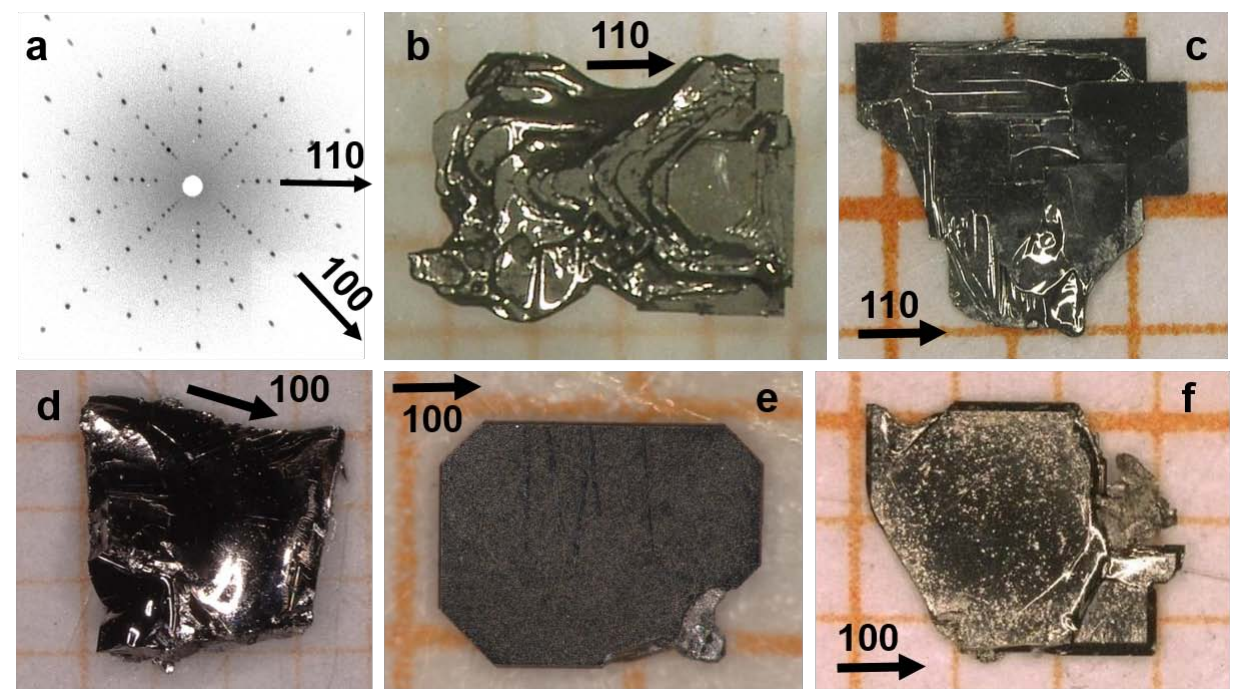}
    \caption{Laue pattern of GdRh$_2$Si$_2$ (a) and oriented GdRh$_2$Si$_2$ (b), HoRh$_2$Si$_2$ (c), CeIrIn$_5$ (d), GdIr$_2$Si$_2$ (e), and CeCo$_2$P$_2$ (f) single crystals on mm grid. While the largest surface of the crystals is always oriented perpendicular to the crystallographic $[001]$ direction, the direction of the longest edges differ for the different compounds.}
    \label{fig:samples}
\end{figure}

Vapor pressure refers to the equilibrium pressure exerted by a substance's vapor coexisting with its liquid or solid phase in a closed system. It rises with temperature, as increasing thermal energy enables more particles to vaporize. Each substance exhibits distinct vapor pressure behavior, with some, such as Eu, Yb, or P (e.g., $p(650\,^{\circ}\mathrm{C}) = 133\,\mathrm{bar}$), displaying exceptionally high vapor pressures at elevated temperatures, necessitating specialized techniques like flux methods or precursor synthesis. Ytterbium and europium readily oxidize to stable Yb$_2$O$_3$ or Eu$_2$O$_3$ upon exposure to oxygen, necessitating strict oxygen-free handling conditions. In practice, these elements are stored and manipulated in an argon-filled glove box. For crystal growth, closed crucible systems, typically made of metal or quartz, are employed, either under vacuum or argon. These setups using quartz ampules are widely used due to their simplicity and compatibility with nested crucibles featuring a sieve \cite{Canfield_2016}, enabling flux-crystal separation post-growth. Remarkably, we recently demonstrated that EuPd$_3$Si$_2$ crystals can be grown at temperatures beyond quartz's stability limit using a staggered crucible design (see figure~1 in Ref.~[\citeonline{Ocker_2025_132_arxiv}]).
To enable experiments in a box furnace at temperatures $>1100^\circ$C in air, we designed this multilayer crucible setup. Key design choices include:
(i) A graphite inner crucible to prevent melt reactions with niobium, avoiding aluminum/oxygen contamination of the melt from Al$_2$O$_3$. (ii) A flexible, sealed Nb crucible to withstand high vapor pressures from evaporating elements. (iii) A quartz outer ampule to protect Nb from oxidation.
Despite quartz's nominal $1100^\circ$C limit, we observe minimal deformation and air-tight sealing up to 1230$^\circ$C, with additional tests confirming ampule integrity under experimental conditions.
Historically, Eu-based single crystals were grown in closed crucible using the Bridgman method~\cite{Onuki_2017, Hedo_2014}. However, Eu's high vapor pressure at high temperatures and low boiling point ($T_\text{b} = 1439^\circ$C at 1~bar) cause evaporation, altering stoichiometry. The use of flux and sealed ampules mitigates this effect by dissolving the volatile substance in the flux and by confining vapor losses to cold areas within the crucible. For the growth of high-vapour-pressure materials like the Eu(Rh$_{1-x}$Co$_x$)$_2$Si$_2$ system, a newly developed screwable graphite crucible further reduced evaporation~\cite{Walther_2025}.
Closed crucible systems for crystal growth experiments introduce several important considerations.
First, the interaction between the melt and crucible material can lead to crucible degradation, which in turn contaminates both the melt and the resulting crystals.
For instance, it was observed that when growing YbRh$_2$Si$_2$ from indium flux in Al$_2$O$_3$ crucibles, the crucible was attacked by the melt and Al was released into the melt leading to an incorporation of Al in the crystals.
Second, the crucible walls may act as heterogeneous nucleation sites, promoting simultaneous crystal growth from multiple locations.
Sometimes, this results in polycrystalline samples rather than single crystals, with possible inclusions of residual flux. In other cases, this can be advantageous, as many faceted single crystals can grow freely in the melt in the flux. Such a scenario was found when growing platelet-shaped 122 materials in preparation of ARPES experiments. The lateral crystal sizes of the platelet-shaped crystals varied between the systems from $\approx$ 1\,mm$\times$1\,mm for EuIr$_2$Si$_2$ \cite{Susanne2019} up to $\approx$ 4\,mm$\times$4\,mm in case of HoRh$_2$Si$_2$ \cite{Generalov_NanoLett_2017} with thicknesses of some tenths to hundreds of micrometers and  with individual masses between 1 and 100 mg. Typical examples are shown in figure~\ref{fig:samples}.
Perhaps the most critical limitation upon using closed crucibles is the inability to visually monitor the growth process in real time.
Since closed crucibles are typically opaque and growth often occurs within sealed furnaces lacking observation windows, direct observation and adjustment of growth parameters during the experiment is generally impossible.
This lack of visual feedback significantly complicates  the ability to optimize experimental conditions and achieve desired outcomes, thus, typically many crystal growth experiments need to be done to optimize a crystal growth process.
In contrast, the Czochralski pulling technique can be used in some cases. It allows for direct monitoring and adjustment of parameters of the crystal growth during the growth process. We recently demonstrated that this method can be even used for growing high-vapour-pressure materials under enhanced Ar overpressure \cite{Kliemt2016, Kliemt2022a}.

\subsection*{Experimental conditions}

Single crystals of the compounds $Ln$Rh$_2$Si$_2$ ($Ln$ = Ce, Eu, Gd, Ho, Tb), $Ln$Ir$_2$Si$_2$ ($Ln$ = Eu, Gd, Yb), $Ln$Co$_2$P$_2$ ($Ln$ = La, Ce), and CeIrIn$_5$  were grown using different crystal growth methods and experimental parameters which are described in the following. CeRh$_2$Si$_2$ was grown by the Czochralski method using a tetra-arc furnace where a procedure similar to that described in \linebreak Refs.~[\citeonline{Onuki_2003,Knafo_2010}] was applied. The single crystalline grains were extracted from a large Czochralski-grown specimen after growth. In contrast, YbRh$_2$Si$_2$ was grown in closed crucibles using a high-temperature indium-flux technique. Optimal growth was achieved after optimization by adjusting the initial stoichiometry of the elements within \linebreak the flux (Yb\,:\,Rh\,:\,Si\,:\,In = 3\,:\,2.2\,:\,1.8\,:\,24). The elements were placed in an inner graphite crucible and sealed in an outer tantalum crucible. For crystal growth, a modified Bridgman technique in a high-temperature furnace (GERO HTRV 70-250/18) was used. After heating to a maximum temperature of $T_{\rm max}=1550^{\circ}$C and a homogenization time of 1\,h, the crucible was cooled by slowly moving the hot zone of the furnace away from the sample platform at very low velocities producing a temperature gradient of 1-4\,K/h \cite{Krellner2012, Kliemt_CRT_2020}. In a similar way, single crystals of YbIr$_2$Si$_2$ and YbCo$_2$Si$_2$ were obtained \cite{Krellner2012, Klingner_2011}. Also similarly, GdRh$_2$Si$_2$, HoRh$_2$Si$_2$, TbRh$_2$Si$_2$ and many other $Ln$Rh$_2$Si$_2$ compounds \cite{Kliemt_2015, Generalov_NanoLett_2017, Kliemt_CRT_2020} were grown by using a stoichiometry of the initial melt of $Ln$ : Rh : Si : In = 1 : 2 : 2 : 24. In case of EuRh$_2$Si$_2$ and EuIr$_2$Si$_2$ the indium flux method from a melt with the initial stoichiometry Eu : Rh(Ir) : Si : In = 1 : 2 : 2 : 20 was used. The elements were directly sealed in a tantalum crucible under an argon atmosphere and a maximum temperature of $T_{\rm max}=1400^{\circ}$C was used during the growth by Bridgman method \cite{Seiro2011, Seiro_2014, Susanne2019}.
GdIr$_2$Si$_2$ exists in a high-temperature (P-type phase, space group No. 129, $P4/nmm$) and in a low-temperature phase (I-type phase, space group No. 139, $I4/mmm$)  \cite{Kliemt_CRT_2020, Kliemt_2022}. To obtain single crystals in the I phase, the elemental ratio of Gd : Ir : Si : In = 1 : 2 : 2 : 49 and a maximum temperature of the furnace of $T_{\rm max}=1550^{\circ}$C was used.
All $Ln$Rh$_2$Si$_2$ and $Ln$Ir$_2$Si$_2$ compounds are stable in HCl, therefore the excess flux from the growth process was subsequently removed with diluted HCl. The well faceted crystals of the 122 compounds in the ThCr$_2$Si$_2$ structure type grow as platelets with the $c$ axis perpendicular to the largest facet. The respective in-plane orientation of the single crystals needs to be determined by Laue method for each compound individually, figure~\ref{fig:samples}a. For some compounds, we found a preferred alignment of the basal plane facets as shown in Tab.~3 in Ref.~[\citeonline{Kliemt_CRT_2020}]. CeCo$_2$P$_2$ and LaCo$_2$P$_2$ \cite{Poelchen_2022, Poelchen_2023} were grown from the tin flux similarly to the procedure described in Ref.~[\citeonline{Reehuis_1990}] with an optimized initial stoichiometry of $Ln$ : Co : P : Sn = 1.6 : 2 : 2 : 30 using the modified Bridgman technique. For the phosphorus-based compounds, the maximum furnace temperature was $T_{\rm max}=1300$ to $1400^{\circ}$C and therefore lower than for the Rh and Ir compounds and the initial heat-up rate was done with 100 K/h with an additional hold time of 10 h at $450^{\circ}$C, to ensure a complete solution of P in Sn \cite{Kliemt_CRT_2020}. After crystal growth, the Sn flux was removed by etching in HCl. Single crystals of CeIrIn$_5$ \cite{Mende_2021} were grown by an In-self-flux method in evacuated quartz ampules in the homogeneous temperature field of a box furnace. The elements were used in a ratio of Ce : Ir : In = 1 : 1 : 20, as described previously \cite{Moshopoulou_2001}. After heating to $T_{\rm max}=1100^{\circ}$C, the crystals were grown during a slow cooling to $T_{\rm fin}=800^{\circ}$C at a rate of 3 K/min where the residual indium flux was removed by centrifugation \cite{Moshopoulou_2001}.

\subsection*{Characterization of the single crystals}

Following the growth process, all samples were routinely characterized by powder X-ray diffraction (PXRD) using a Bruker D8 diffractometer equipped with Cu-K$_{\alpha}$ radiation operating in Bragg-Brentano geometry. The chemical composition of the single crystals was determined by energy-dispersive x-ray spectroscopy (EDX) using a Zeiss-DSM 940A scanning electron microscope with an EDAX detector. Prior to the ARPES measurements, the orientation of the single crystals was determined using the Laue backscattering method in a M\"uller Micro instrument with white X-rays from a tungsten anode. To investigate the physical properties of the materials, physical characterization was performed with the commercial measurement options of a Quantum Design Physical Property Measurement System (PPMS). For many of the $Ln$-based materials, the primary focus was on the precise determination of their magnetic transition temperatures in preparation for ARPES studies.

\section*{Summary and concluding remarks}

The studies discussed in this review demonstrate that modern photoemission-based spectroscopies have enabled a qualitatively new level of insight into the physics of lanthanide-based crystals. A central conclusion emerging from systematic ARPES and XAS investigations is that the surface regions of these materials constitute electronically and magnetically distinct environments, characterized by their own properties and related temperature scales, which in some cases are even richer and more compelling than those in the bulk~\cite{Susanne2019, Poelchen_2022, Georg_2020}. Their distinct behavior originates from a combination of factors, including reduced coordination, modified crystal-electric-field environments~\cite{Usachov_JPCL_2022, Usachov_JPCL_2023}, broken inversion symmetry together with spin–orbit coupling~\cite{Usachov_PRL_2020, Manchon_2022}, as well as surface relaxation, reconstruction~\cite{Mende_2022}, and the emergence of surface states and resonances~\cite{Chick14, Hoeppner2013}. In many cases, surface properties are governed not solely by the outermost atomic layer, but by a few topmost layers, where each layer within this near-surface block fulfills a distinct function and, collectively, defines the resulting surface behavior~\cite{Susanne2019, Schulz_GdIr_2021, Generalov_SISY_2018, Usachov_PRL_2020}.

For example, the ferromagnetic silicide surface formed by the Si–Ir–Si–Gd block of AFM GdIr$_2$Si$_2$~\cite{Schulz_GdIr_2021} provides a model in which a strong \textit{cubic} Rashba effect~\cite{Usachov_PRL_2020}, arising from spin–orbit interaction on Ir together with the symmetry and orbital composition of the surface states at the BZ corner, is combined with 2D FM from the Gd layer. This unique coexistence makes such surfaces exceptional platforms for studies of electronic and magnetic phenomena induced in nanostructures deposited on top. Surface-stabilized ferromagnetism over a wide temperature range provides a unique opportunity to couple the underlying magnetic order to functional overlayers, including topological systems, molecular assemblies, low-dimensional metallic and semiconducting systems. More generally, the properties discussed above point to a straightforward route for realizing magnetically coupled 2D systems and thereby exploring novel functionalities at the surfaces of correlated-electron materials.

Systematic investigations of high-quality Yb- and Ce-based single crystals have enabled a reliable separation of bulk and surface contributions in ARPES spectra, direct visualization of momentum-dependent $f$–$spd$ hybridization for both surface and bulk states of the Kondo lattice~\cite{Vyalikh_PRL_2010, Danzenbacher_PRL_2011, Generalov_SISY_2018}, as well as the determination of the $f$-derived Fermi surface and its properties~\cite{Danzenbacher_PRL_2011, Kummer_PRX, Monika_Compton}. These studies have explicitly demonstrated that surface regions exhibit their own temperature scales, including magnetic ordering and Kondo behavior. The pronounced differences between surface and bulk properties indicate that quantum-critical phenomena may also assume a distinct character in the reduced-dimensional environment of the surface.

An important methodological advance emphasized in this review is the development of spectroscopic approaches that provide direct access to the magnetic properties of $4f$ electrons. In particular, a detailed analysis of the 4$f$ multiplet line shape enables determination of the orientation of local magnetic moments in distinct atomic layers, as well as detection of their subtle temperature-driven canting~\cite{Usachov_JPCL_2022, Usachov_JPCL_2023, Usachov_PRB_2025}, information that is difficult or often impossible to assess using conventional bulk-sensitive probes. The systematic application of this approach transforms the analysis of complex $4f$ multiplet features observed in photoemission spectra into a quantitative diagnostic tool for exploring magnetic anisotropy and surface magnetism.

The obtained results also call for systematic exploration of surface-induced states in layered strongly correlated systems based on Ce, Eu, or Yb. Quasi-2D building “Lego-like” blocks such as P–Co–P–$Ln$ or Si–T–Si–$Ln$ and their combinations~\cite{Shvets2025JETPLett} provide a promising route toward designing novel functional and quantum materials in which competing interactions, including spin–orbit coupling, exchange magnetism, crystal-field effects, heavy-fermion behaviour, and unconventional superconductivity, can be selectively combined and systematically \linebreak tuned~\cite{Bobkova2024, Bakurskiy2024, Mironov2025}.

Looking ahead, further progress will rely on extending these systematic approaches to a broader range of lanthanide materials and to experimentally unexplored regimes. In particular, exploring topological phenomena in $4f$ systems, continuing time-resolved studies of correlated electronic and magnetic degrees of freedom, and performing deeper analyses of electron--boson interactions in lanthanide compounds, including thickness-dependent low-dimensional systems, will unveil new phenomena together with the characteristic energy scales governing the properties of these systems. The continued integration of advanced spectroscopic techniques with controlled crystal growth and modern theoretical modelling will remain essential for unveiling new physics and for establishing design principles for functional materials based on lanthanide building blocks. Additionally, it would be highly desirable to establish a similarly systematic and coherent research line on $5f$ materials (e.g., U- and Pu-based compounds) in order to explore bulk and surface temperature scales governed by $5f$-driven physics.

\section*{Acknowledgements}

The authors express their appreciation to Clemens Laubschat and the members of his group, who have been deeply involved in the numerous experiments reviewed here since the early 2000s, including Sergei Molodtsov, Steffen \linebreak Danzenb\"{a}cher, Matthias Holder, Mark H\"{o}ppner, Alla Chikina, Monika G\"{u}ttler, Susanna Schulz, Alexander Generalov, Swapnil Patil, Max Mende, Yuri Dedkov and Kurt Kummer. The authors also acknowledge the support of collaborators at the synchrotron radiation facilities \linebreak BESSY~II, SLS, MAX~IV, SPring-8, HiSOR, and Diamond. Time-resolved experiments on $Ln$Rh$_2$Si$_2$ systems were driven by Laurenz Rettig and Will Windsor.

Particular appreciation is expressed to Christoph \linebreak Geibel and the members of his group for their enthusiastic participation in the growth of high-quality samples, their comprehensive analysis and characterization, and their active involvement in the interpretation of spectroscopic data, as well as in the development of experimental programs. In addition, we strongly acknowledge the technical support of Klaus-Dieter Luther and Tim F\"orster to sustain the crystal-growth apparatuses in Frankfurt.

The authors further benefited from many useful discussions with Gertrud Zwicknagl, Yuri Kucherenko, Peter Riseborough, Steffen Wirth, Sebastian Burdin, Sergey Borisenko, Akihisa Koizumi, Shin-ichi Fujimori and Jim Allen, as well as Misha Otrokov, Arthur Ernst, Sergey Eremeev, Ilya Nechaev, Eugene Krasovskii and Evgeny Chulkov. Additionally, the authors thank Alexander Fedorov, Rui Lou, Taichi Okuda, Milan Radovic, Timur Kim, Craig Polley, Khadiza Ali, and Alexei Preobrajenski.

Much of the work reviewed here was supported by DIPC, IKERBASQUE, the Deutsche Forschungsgemeinschaft, St.~Petersburg State University, and MIPT. The authors acknowledge Saint-Petersburg State University for research Project No. 125022702939-2.
The review research was particaly supported by the Russian Science Foundation Project No. 23-72-30004.

\section*{Contact information}

\noindent Corresponding author: Denis V. Vyalikh,

\noindent orcid.org/0000-0001-9053-7511,

\noindent e-mail denis.vyalikh@dipc.org

\section*{Competing Interests}

The authors declare no competing financial or non-financial interests.

\section*{Data availability}

The authors declare that the data supporting the findings of this study are available within the paper and its supplementary information files. Other data are available from the authors upon reasonable request.

\end{document}